\documentclass{aa}  
\usepackage{graphicx}
\usepackage{txfonts}
\usepackage{tikz}
\usepackage{amssymb,amsmath}
\usepackage{dsfont}
\usepackage{lscape}
\usepackage{longtable}
\usepackage{float}	
\usepackage{natbib}
\usepackage{color}
\usepackage{pdflscape}
\begin{document} 
%
%
\title{The shapes of spiral arms in the S$^4$G survey and their connection with stellar bars 
}
  \author{S. Díaz-García\inst{1,2}
          \and        
          H. Salo\inst{3}	
          \and
          J. H. Knapen\inst{1,2,4}
          \and
		  M. Herrera-Endoqui\inst{5}
          }
  \institute{Instituto de Astrof\'isica de Canarias, E-38205, La Laguna, Tenerife, Spain \\
              \email{simondiazgar@gmail.com}
         \and
             Departamento de Astrof\'isica, Universidad de La Laguna, E-38205, La Laguna, Tenerife, Spain
         \and 
             Astronomy Research Unit, University of Oulu, FI-90014 Finland
         \and
           	 Astrophysics Research Institute, Liverpool John Moores University, IC2, Liverpool Science Park, 146 Brownlow Hill, Liverpool, L3 5RF, UK           
         \and
         	 Instituto de Astronomía, Universidad Nacional Autónoma de México, Apdo. Postal 877, Ensenada, Baja California 22800, México  
          }
  \date{Received 31 May 2019 / Accepted 12 August 2019}
  \abstract
  {
\emph{Context.} 
Spiral galaxies are very common in the local Universe, but their formation, evolution, and interplay with bars remain 
poorly understood after more than a century of astronomical research on the topic.\\
\emph{Aims.} 
We use a sample of 391 nearby galaxies from the S$^4$G survey to 
characterise the winding angle and amplitude of spiral arms as a function of disc properties, 
such as bar strength, in all kinds of spirals (grand-design, multi-armed, and flocculent).\\
\emph{Methods.}
We derived global pitch angles in 3.6~$\mu$m de-projected images from 
i) average measurements of individual logarithmic spiral segments, and 
ii) for a subsample of 32 galaxies, from 2-D Fourier analyses. 
The strength of spirals was quantified from the tangential-to-radial force ratio and from the normalised $m=2$ Fourier density amplitudes.\\
\emph{Results.} 
In galaxies with more than one measured logarithmic segment, the spiral pitch angle varies on average by $\sim 10^{\circ}$ between segments, 
but by up to $\gtrsim 15-20^{\circ}$. 
The distribution of the global pitch angle versus Hubble type ($T$) is very similar 
for barred and non-barred galaxies when $1 \lesssim T \lesssim 5$. 
Most spiral galaxies ($>90\%$) are barred for $T>5$. 
The pitch angle is not correlated with bar strength, and only weakly with spiral strength. 
The amplitude of spirals is correlated with bar strength (and less tightly, with bar length) for all types of spirals. 
The mean pitch angle is hardly correlated with the mass of the supermassive black hole (estimated from central stellar velocity dispersion), 
with central stellar mass concentration, or with shear, questioning previous results in the literature using smaller samples.
\\
\emph{Conclusions.}
We do not find observational evidence that spiral arms are driven by stellar bars or by invariant manifolds. 
Most likely, discs that are prone to the development of strong bars are also reactive to the formation of prominent spirals, 
explaining the observed coupling between bar and spiral amplitudes.
}
\maketitle
%
%
%
\section{Introduction}
%
%
The most conspicuous stellar structures of spiral galaxies are the arms that 
sweep out from the centre of the disc in unbarred galaxies, or from near the ends of the stellar bar. 
Spiral arms are sites rich in gas with intense star formation, H{\sc\,ii} regions, and 
dust \citep[e.g.][]{1976ApJS...31..313S,2011ApJ...737...32E}. 
This makes them more prominent in blue bands of the spectrum, but their backbone is typically composed of 
old stars \citep[e.g.][]{1981ApJS...47..229E,1989ApJ...342..677E,1996MNRAS.283..251K,2002ApJS..143...73E}. 
Roughly two thirds of all massive galaxies are spirals \citep[e.g.][]{2011MNRAS.410..166L,2013MNRAS.435.2835W}. 

There are three main types of spiral arms \citep[e.g.][]{1990NYASA.596...40E}: grand-design, flocculent, and multi-armed. 
The fraction of each of these classes in disc galaxies is $\sim 18\%$, $\sim 50\%$, and $\sim 32\%$, respectively, 
based on the morphological classifications made by \citet[][]{2011ApJ...737...32E} and \citet[][]{2015ApJS..217...32B} to the 
Spitzer Survey of Stellar Structure in Galaxies \citep[S$^4$G;][]{2010PASP..122.1397S}, 
that comprises over 2000 nearby galaxies. 
Grand-design galaxies \citep[][]{1966PNAS...55..229L} are characterised by two long and well-defined arms. 
Flocculent spirals \citep[][]{1981ApJS...47..229E} present short and fragmented spiral arm sections. 
They are more common in faint galaxies \citep[][]{1985ApJ...288..438E}. 
The fairly symmetric multi-armed spirals constitute a category between grand-design and flocculent, closer to the former, 
characterised by a central two-armed pattern that develops long ramifications in the outer parts of the optical disc. 
In this work we analyse the shapes of these three types of spirals in the S$^4$G survey.

How spiral arms are formed in galactic discs remains a matter of debate. 
One of the most widely accepted explanations of the formation and maintenance of spirals is the density wave theory 
\citep[][]{1964ApJ...140..646L}. 
The arms are considered as density waves that rotate with a radius-independent angular speed 
through the stars and gas in a shearing, differentially rotating galactic disc. 
In this scenario, the stars in the inner parts of the disc rotate faster than 
the spiral and can overtake it, whereas in the outer parts they lag with respect to the spiral pattern. 

Spirals might also be tidally triggered in interactions with companion galaxies \citep[][]{1979ApJ...233..539K}, 
as it is the case of the Whirlpool Galaxy (or M$\,$51) \citep[see e.g.][and references therein]{2000MNRAS.319..377S,2015ApJ...800L...3W}, 
the first identified spiral nebula in history \citep[][]{1850RSPT..140..499R} 
using the \emph{Leviathan of Parsonstown} telescope (1.8 m aperture).

In the swing amplification mechanism \citep[][]{1965MNRAS.130...97G,1965MNRAS.130..125G,1981seng.proc..111T}, 
shear and self-gravity transform initially leading waves into trailing ones, 
so that the wave is amplified when the stabilising effects of random motions are negligible. 
A constraint on the theory of the formation of spirals is the number of arms expected by swing amplification. 
\citet[][]{2015ApJ...808L...8D} showed that the total number of spiral arms anticorrelates with 
the disc-to-total mass fraction at 2.2 disc scalelengths: submaximal discs are expected to be multi-armed galaxies. 
High resolution $N-$body simulations by \citet[][]{2018MNRAS.477.1451F} confirm that the disc-mass fraction 
controls the number of spirals.

Grand-design spiral arms and the symmetric inner parts of 
multi-armed galaxies have been interpreted as long-lived spiral density waves, 
while flocculent galaxies and the outer parts of multi-armed galaxies have been associated to 
random local gravitational instabilities in the disc, 
swing amplified into transient spiral arms \citep[e.g.][and references therein]{2011ApJ...737...32E}. 
According to numerical models \citep[e.g.][and references therein]{2011MNRAS.410.1637S}, 
the spiral arms in flocculent galaxies are short-lived 
(of the order of $\sim 100$ Myr), whereas grand-design spirals exhibit arms that last longer ($\sim 1$ Gyr). 
High-resolution $N-$body simulations by \citet[][]{2013ApJ...766...34D} showed that 
ragged spiral structure can survive long after the original perturbing influence has been removed.

How tightly wound the spiral arms are can be quantified by measuring the pitch angle \citep[][]{1987gady.book.....B}, 
defined as the angle subtended by the tangent to the spiral arm, 
relative to the tangent to a circle at a point at a given galactocentric radius. 
By definition, it lies between $-90^{\circ}$ and $90^{\circ}$. 
In theoretical modelling, positive (negative) pitch angles indicate trailing (leading) pattern with respect to rotation.

The pitch angle is not necessarily constant, regardless of the class of the spiral arms \citep[e.g.][]{1981AJ.....86.1847K}. 
It has been claimed that pitch angles depend on the central mass concentration and atomic gas density 
\citep[e.g.][]{1981AJ.....86.1847K,1994A&A...288..365B,2015ApJ...802L..13D,2019ApJ...871..194Y}, 
on the galactic shear rate \citep[e.g.][]{2006ApJ...645.1012S,2013A&A...553A..77G}, 
or on the steepness of the rotation curves \citep[][]{2005MNRAS.359.1065S,2014ApJ...795...90S}: 
spiral arms have been found to be more open in galaxies with rising rotation curves, 
whereas those with falling rotation curves are generally tightly wound. 
Recent work by \citet{2017MNRAS.471.2187D} reports a very tight scaling relation between 
the mass of the central supermassive black hole and the spiral pitch angle 
which, if true, hints at a surprisingly intimate link between the large-scale structure of discs and the mass in the nucleus 
\citep[see also][]{2008ApJ...678L..93S,2013ApJ...769..132B}. 

Spiral arms are known to contribute to the rearrangement of gas that leads to the formation of disc-like bulges 
\citep[e.g.][]{2004ARA&A..42..603K}. This makes them important agents for the secular evolution of disc galaxies, 
a process in which bars also play a fundamental role. 
Whether spirals are bar-driven or not has been debated over the past decades, 
ever since \citet[][]{1979ApJ...233..539K} suggested that stellar bars can lead the formation of spiral density waves. 
The latter has been mainly tested from the correlations between gravitational torques and density amplitudes associated to bars and the spirals 
\citep[e.g.][]{2003MNRAS.342....1S,2004AJ....128..183B,2005AJ....130..506B,2009AJ....137.4487B,2009MNRAS.397.1756D,2010ApJ...715L..56S}. 
%
%
\begin{figure}
\centering
\includegraphics[width=0.5\textwidth]{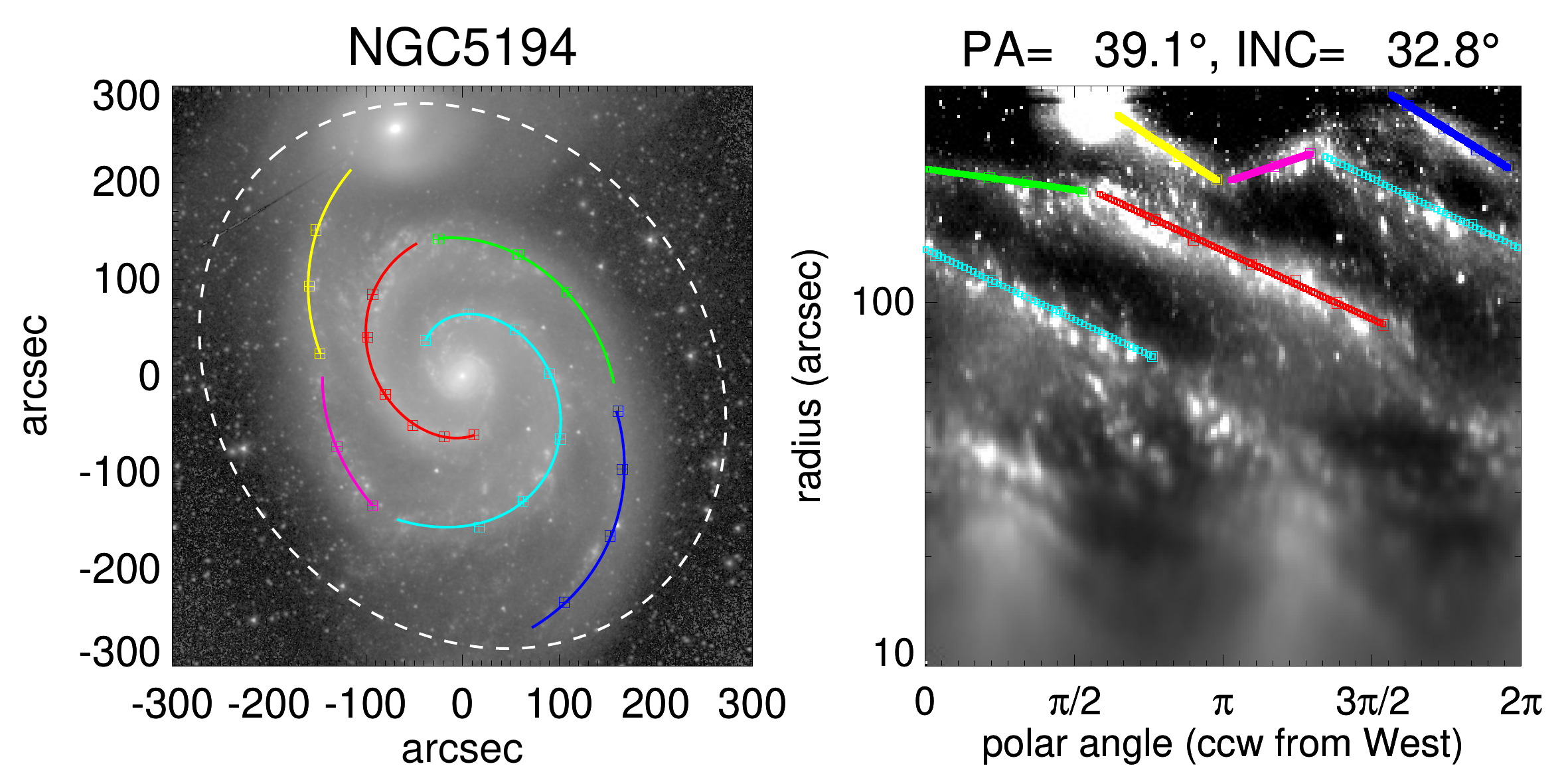}\\
\includegraphics[width=0.5\textwidth]{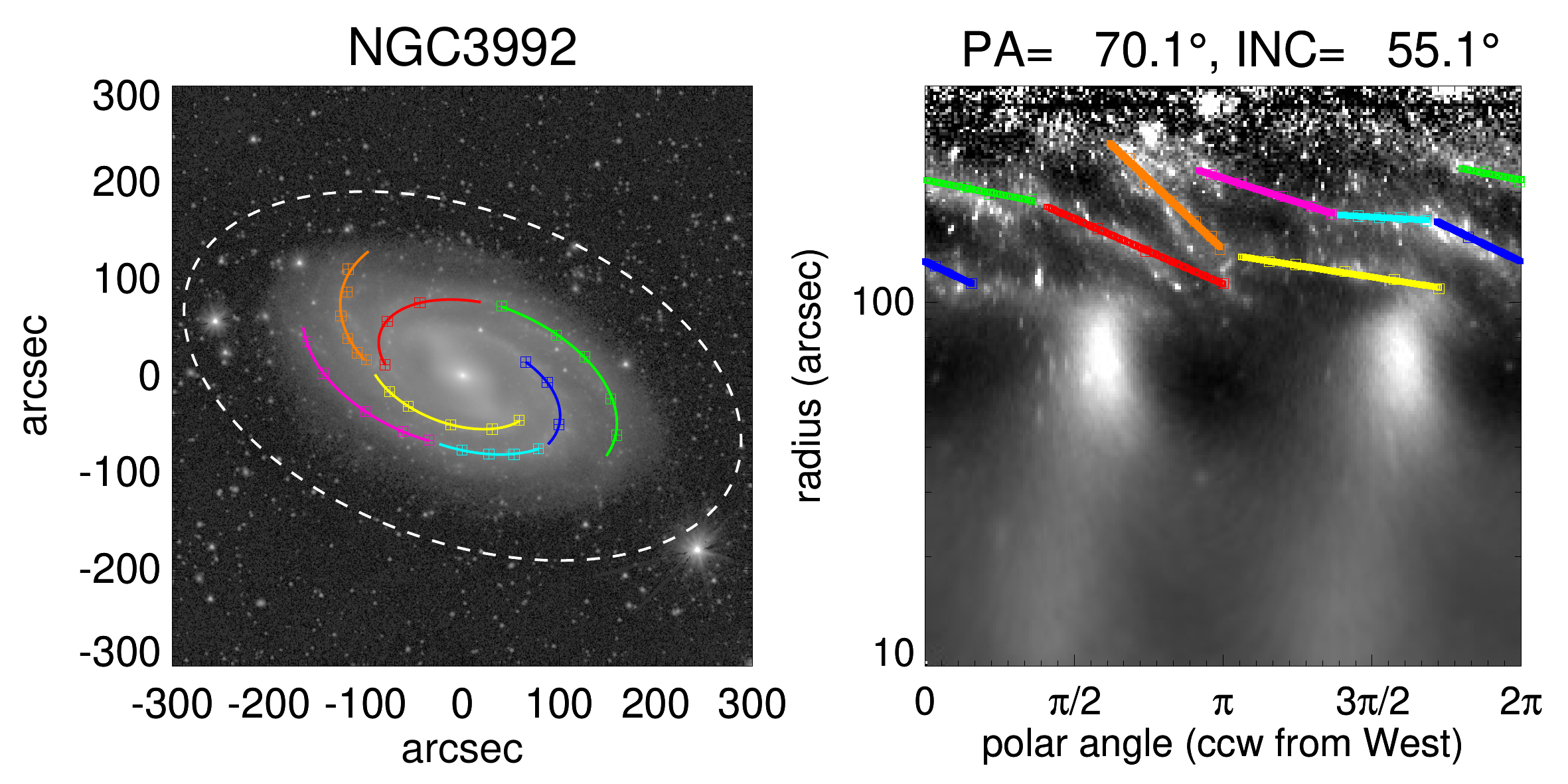}
\caption{
3.6 $\mu$m image of NGC$\,$5194 (\emph{upper panels}) and NGC$\,$3992 (\emph{lower panels}) in sky plane (\emph{left}), 
with fitted spiral arm segments, 
and logarithmic polar plot of the same galaxies de-projected to the disc plane (\emph{right}).
Images and measurements are taken from \citet[][]{2015A&A...582A..86H}.
}
\label{Figpitch}
\end{figure}

A strong coupling between bars and spirals is also expected from the manifold theory, 
which is a relatively new paradigm for the formation of spirals and rings 
\citep[e.g.][]{2006A&A...453...39R,2006MNRAS.369L..56P,2006MNRAS.373..280V,2007A&A...472...63R,2009MNRAS.394...67A}. 
Numerical simulations show that galactic material gets confined in tubes (invariant manifolds) 
that extend from the two unstable Lagrangian points at the end of the bar ($L_1$ and $L_2$). 
This theory predicts a dependence of the pitch angle of spirals on the bar perturbation strength \citep[][]{2009MNRAS.400.1706A}. 

The main goals of this paper are 
i) to reassess the existence of scaling relations involving the pitch angle of the arms in the S$^4$G, 
ii) to analyse the coupling between the amplitudes of bars and spiral arms 
in all kinds of spirals galaxies with unprecedented sample size, 
iii) to test the predicted correlation between bar strength and pitch angle, and 
iv) to check if there is any difference in the distribution of the pitch angles for barred and non-barred galaxies.

The paper is organised as follows: 
in Sect.~\ref{data1} we present the sample and the data used in this work. 
In Sect.~\ref{pitch_met}~and~\ref{non_axis} we explain our methodology to derive the global pitch angle 
of the galaxies as well as the strength of spirals and bars. 
In Sect.~\ref{pitches_hubble} we study the distribution of the pitch angle across the Hubble sequence, 
as well as the distribution of spiral types and bar frequency in our sample versus $T-$type. 
In Sect.~\ref{pitches_rad} we assess the radial variation of the pitch angle, 
in Sect.~\ref{spiral_bar} we test whether the pitch angle and amplitude of spirals are controlled by the strength of the bar, 
and in Sect.~\ref{BH_mass_sec} we analyse the dependence of the pitch angle of the spirals on the central stellar mass concentration and 
on the mass of the central supermassive black hole. 
In Sect.~\ref{Discussion} we discuss the properties of spiral arms, their dependence on disc fundamental parameters, 
and their hypothetical coupling with bars. 
Sect.~\ref{summarysection} summarises the most important results of this work. 
%
%
\section{Data and sample selection}\label{data1}
%
\subsection{The S$^4$G survey}
%
%
The S$^{4}$G survey \citep[][]{2010PASP..122.1397S} consists of 2352 nearby galaxies (distance $\lesssim 40$ Mpc) 
observed in the 3.6~$\mu$m and 4.5~$\mu$m bands with the InfraRed Array Camera \citep[IRAC;][]{2004ApJS..154...10F} 
installed on board the \emph{Spitzer Space Telescope} \citep{2004ApJS..154....1W}. 
Data taken from HyperLEDA were used to define the S$^4$G sample, which is composed of bright and large galaxies 
(extinction-corrected total blue magnitude $m_{B_{\rm corr}} \textless 15.5$ mag and 
blue light isophotal angular diameter $D_{25} \textgreater 1^{\prime}$) located away from the Milky Way plane 
(Galactic latitude $|b|\textgreater 30^{\circ}$). 

The S$^{4}$G is not complete in any quantitative sense, but it is comprised of 
galaxies of all Hubble types ($T$) with stellar masses spanning $\sim$~5 orders of magnitude. 
Due to selection based on H{\sc\,i} recessional velocities, it is deficient in gas-poor early-type galaxies: 
this is not a serious drawback for the current study because we are not interested in S0s, that lack spiral arms. 
In addition, the S$^{4}$G missed around 400 late-type galaxies without 21 cm systemic velocity 
measurements listed in HyperLEDA: we are currently obtaining $i$-band photometry with ground-based telescopes 
for the $\sim 50\%$ of those galaxies that lack high-resolution deep near-IR and optical archival imaging. 
The current work is based on the original S$^4$G.

Bulge-to-total mass ratios ($B/T$) and disc scale-lengths ($h_{\rm R}$) of S$^{4}$G 
galaxies are available from the 2-D photometric 
decomposition models by \citet[][]{2015ApJS..219....4S}. 
We take the $3.6\,\mu$m isophotal radii at the surface brightness 
25.5~mag~arcsec$^{-2}$ ($R_{25.5}$) from \citet[][]{2015ApJS..219....3M}.
%
%
\begin{figure}
\centering
\includegraphics[width=0.5\textwidth]{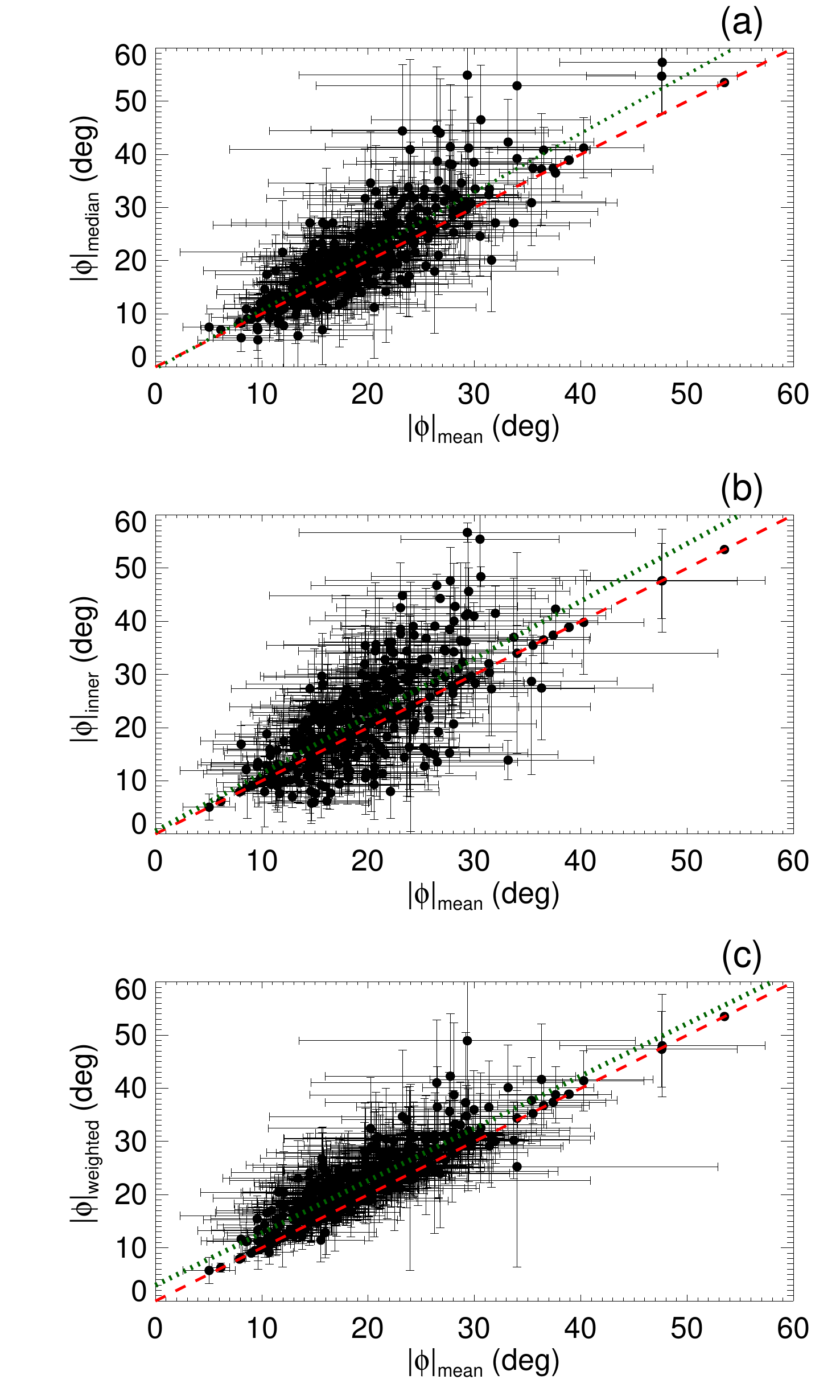}
\caption{
Comparison between mean and a) median pitch angle, 
b) mean of the innermost logarithmic segments, and c) mean weighted by the arc length of the arms. 
The $y=x$ straight line is shown in red, 
and the dotted green line shows the linear fit to the cloud of points. 
Error bars indicate the standard deviation of the mean calculated from the segments fitted in each galaxy.
}
\label{mean_vs_median}
\end{figure}
%
%
\begin{figure*}
\centering
\includegraphics[width=0.49\textwidth]{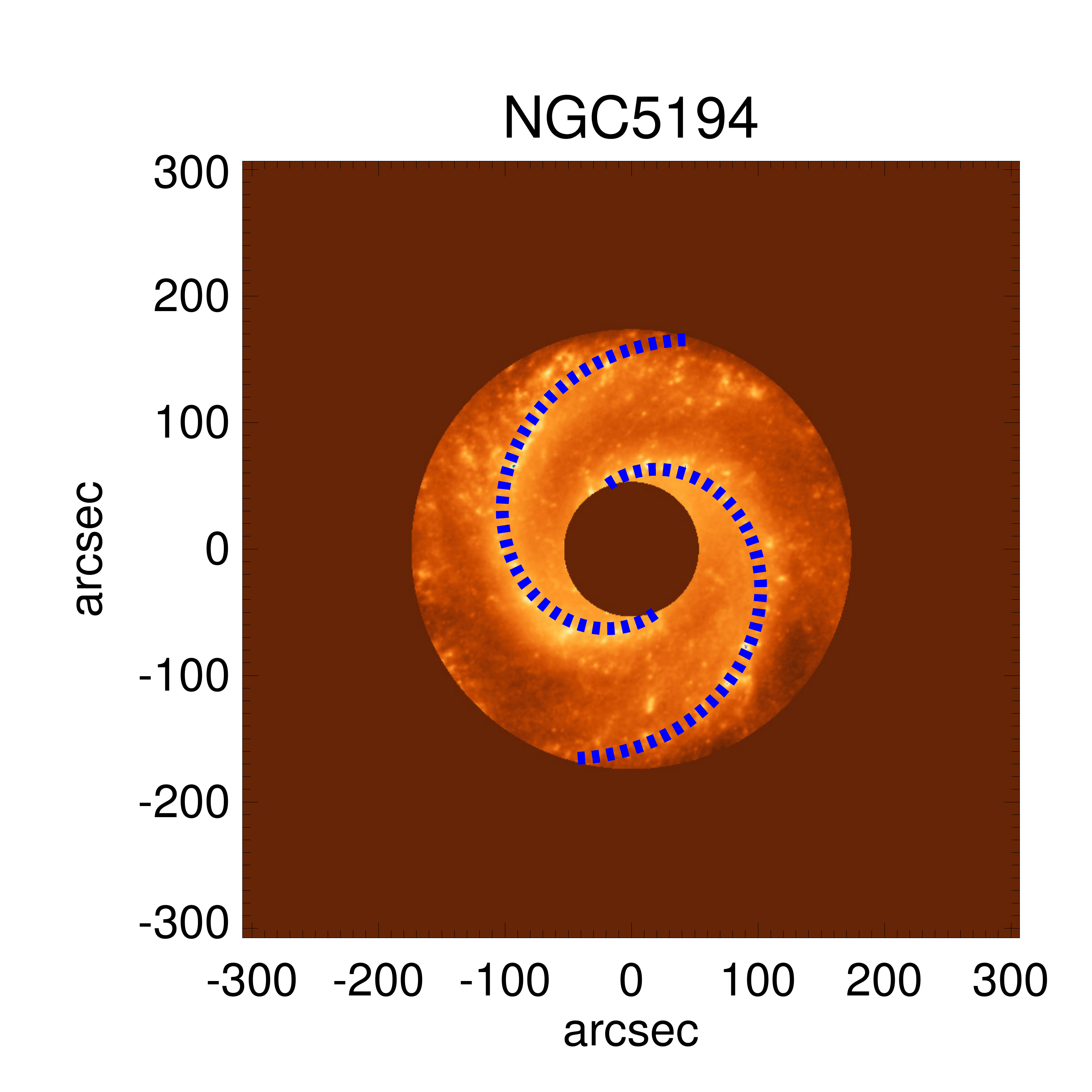}
\includegraphics[width=0.49\textwidth]{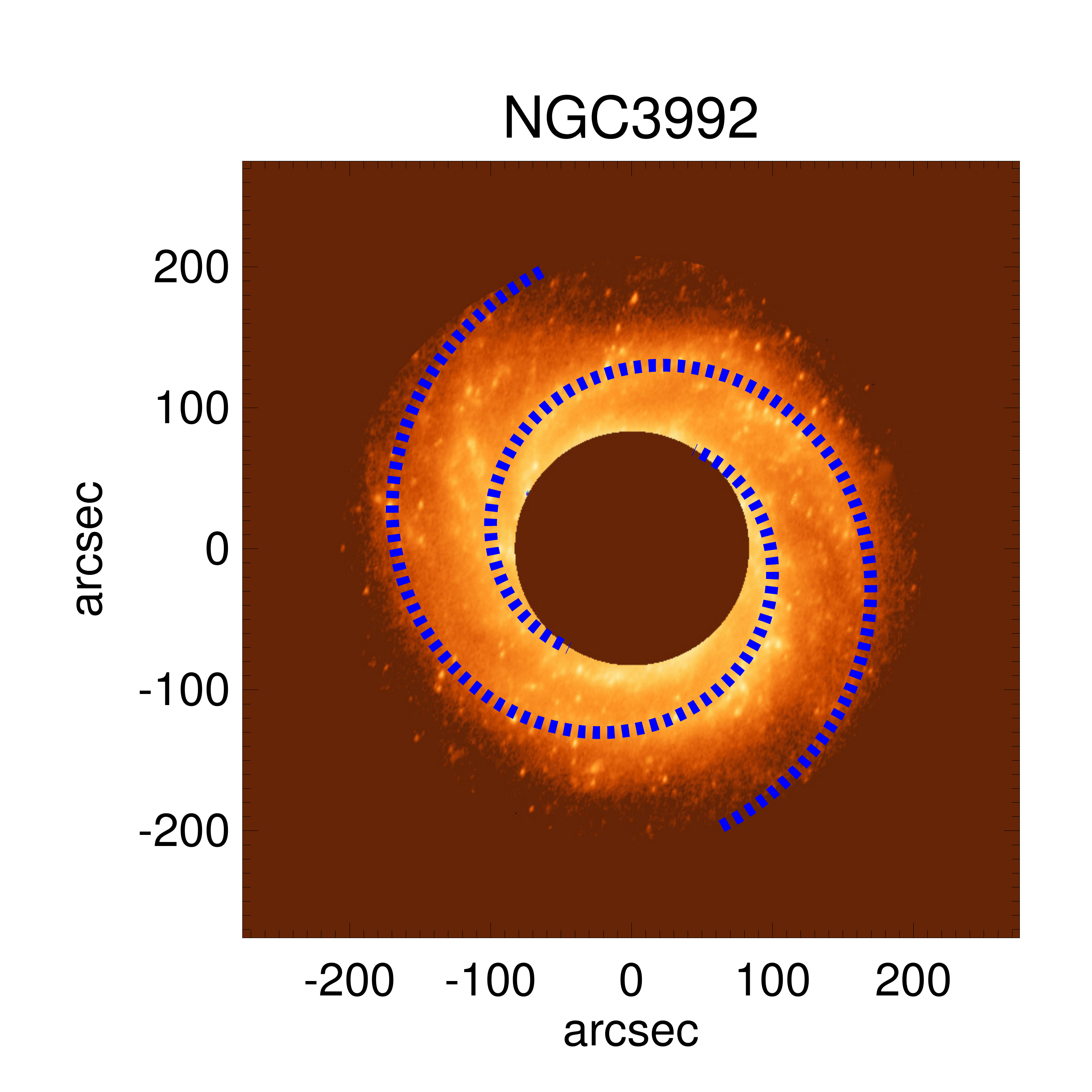}\\[-5ex]
\includegraphics[width=0.49\textwidth]{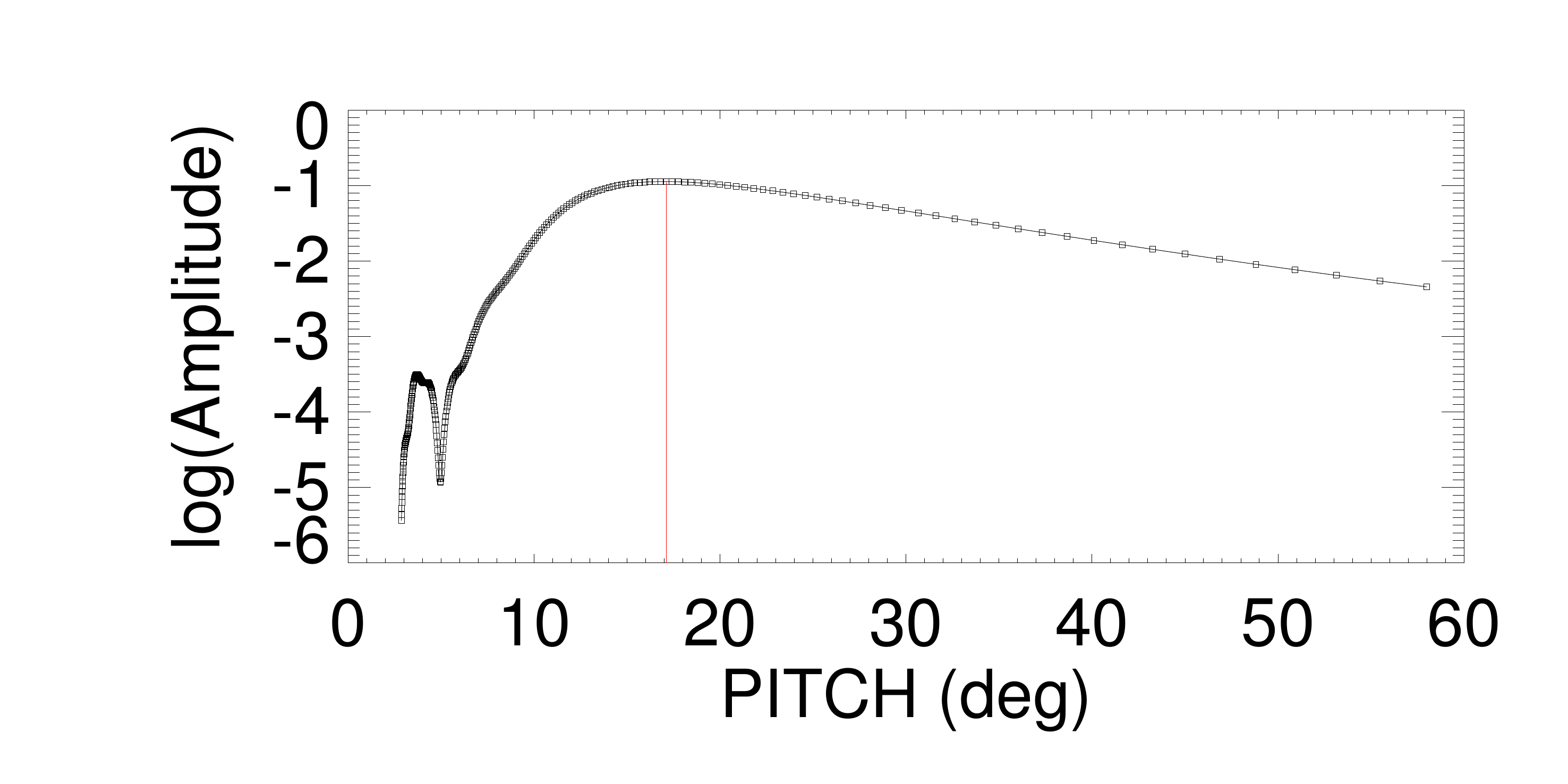}
\includegraphics[width=0.49\textwidth]{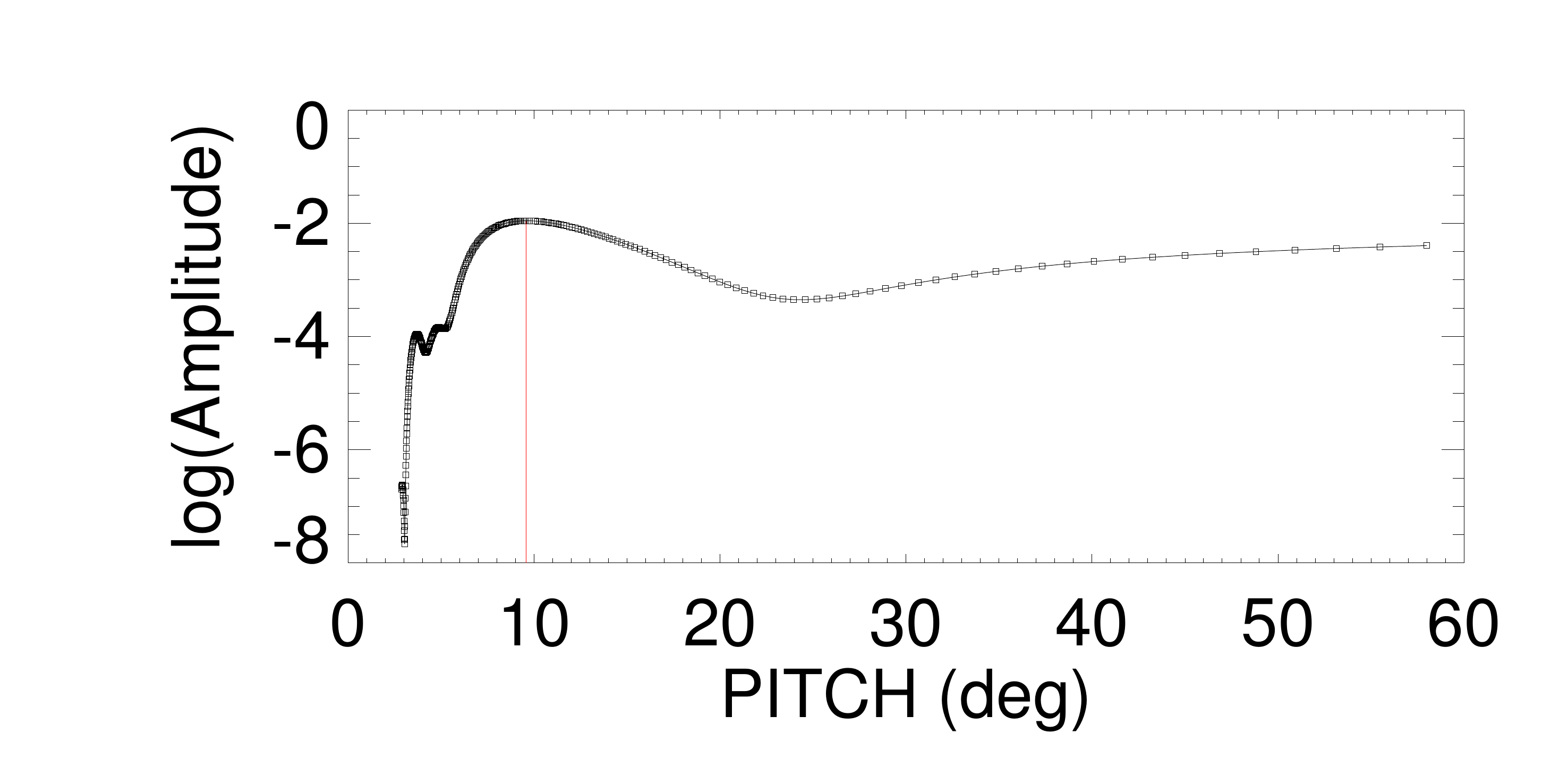}
\caption{
\emph{Upper panels:} Logarithmic fit to the spiral arms in the de-projected 3.6 $\mu$m image of 
NGC$\,$5194 (\emph{left}) (only central parts considered) (18-23 $\mu_{3.6\mu \rm m}(AB)$ magnitude scale) and 
NGC$\,$3992 (\emph{right}) (21-25 $\mu_{3.6\mu \rm m}(AB)$ magnitude scale), using 2-D Fourier transform spectral analysis (see text).
\emph{Lower panels:} 
Logarithm of the $m=2$ Fourier amplitude as a function of the pitch angle for the two galaxies shown above. 
The vertical line denotes the maximum value, that determines the global pitch angle assigned to the galaxy.
}
\label{spiral_fit_fourier}
\end{figure*} 
%
%
%
\subsection{Morphological classification}
%
%
We use the morphological classification of the galaxies in the S$^4$G sample made by \citet[][]{2015ApJS..217...32B}, 
that includes families ($\rm SB$, $\rm SA\underline{B}$, $\rm SAB$, $\rm S\underline{A}B$, $\rm SA$) and the revised Hubble stages ($T$). 
We also use arm classes from \citet[][]{2015ApJS..217...32B} and \citet[][]{2011ApJ...737...32E}. 
%
%
\subsection{Pitch angles of spiral arms}
%
%
\citet[][]{2015A&A...582A..86H} measured the pitch angles of spiral arm segments in S$^4$G galaxies using 3.6~$\mu$m photometry, 
which are analysed in this paper. Only measurements with "ok" quality flags ($=1-2$) are used.

\citet[][]{2015A&A...582A..86H} identified the different spiral segments in unsharp-masked images, 
created by dividing the original images by their smoothed versions. In unsharp-masked images one can 
easily identify subtle galactic structures that appear against a bright and diffuse background. 
These images were displayed in different scales to make the spiral arms stand out. 
They visually marked points tracing the spiral segments (and thus there are uncertainties 
in the parametrisation associated to human errors). 
Then, they performed a linear fit in the disc plane using polar coordinates ${\rm log}(r)$ versus $\theta$, 
where logarithmic arms appear as straight lines. 
Thus, several measurements of the pitch angle can appear for a single galaxy, 
corresponding to logarithmic segments at different radial distances. 
In Fig.~\ref{Figpitch} we illustrate these measurements for M$\,$51 and NGC$\,$3992. 
For further details, the reader is referred to \citet[][]{2015A&A...582A..86H}.
%
%
\subsection{Sample of not-highly inclined spiral galaxies}
%
%
For this study, we use a sample of 391 galaxies with inclinations lower than 65$^{\circ}$ 
\citep[according to][]{2015ApJS..219....4S}, 
with reliable measurements of pitch angles of individual spiral segments\footnote{
Only six galaxies in our sample (1.5\%) have a single measurement of the pitch angle, 
74 galaxies (18.9\%) have two measured logarithmic segments, 
86 (22.0\%) have three, 
105 (26.9\%) have four, 
58 (14.8\%) have five, 
and 62 galaxies (15.9\%) have more than five (with two cases having a maximum of nine segments).
}\citep[from][]{2015A&A...582A..86H}, 
and classified spiral types \citep[from][]{2015ApJS..217...32B}. 
Of these, 269 galaxies are barred 
(i.e. classified as $\rm SB$, $\rm SA\underline{B}$, $\rm SAB$, or $\rm S\underline{A}B$), 
76 are grand-design, 157 are multi-armed, and 158 host flocculent spirals.
%
%
\section{Global pitch angles of spirals}\label{pitch_met}
%
%
In order to parameterise the winding of the spirals, 
so that it can be studied as a function of galaxy parameters, 
we need to determine the global pitch angle of the galaxies ($\phi$ hereafter). 
We will assume that the spiral segments are all trailing, and will therefore treat $\phi$ always as a positive quantity. 
Also, we do not care whether the apparent spiral opening angles in sky are clockwise or counter-clockwise 
\citep[the winding direction of arms in spiral galaxies is distributed at random in the local Universe, see e.g.][]{2017MNRAS.466.3928H}.
%
%
\subsection{Mean pitch angle of individual logarithmic segments}\label{fourier_visual}
%
%
We first calculate the mean and median of the absolute value of the pitch angle measurements of logarithmic segments 
($|\phi|_{\rm mean}$ and $|\phi|_{\rm median}$, respectively). 
Uncertainties in the determination of $|\phi|$ are estimated 
by computing the standard deviation of the mean of the measurements of different segments. 
We compare $|\phi|_{\rm mean}$ and $|\phi|_{\rm median}$ in the top panel of Fig.~\ref{mean_vs_median}: 
the mean absolute difference is $3.49^{\circ} \pm 0.17^{\circ}$.

We also calculate the mean of the absolute value of the two innermost spiral segments 
(or one for the 6 galaxies with a single measurement), $|\phi|_{\rm inner}$, 
to trace the pitch angle in the central parts of galaxies and in the vicinity of the bar. 
The mean absolute difference between $|\phi|_{\rm mean}$ and $|\phi|_{\rm inner}$ is 
$4.70^{\circ} \pm 0.24^{\circ}$ (central panel of Fig.~\ref{mean_vs_median}).

In addition, we propose a measurement of the global pitch angle of disc galaxies by weighting the average 
by the relative arc length of the logarithmic segments, $|\phi|_{\rm weighted}$.  
The equation for a logarithmic spiral in polar coordinates ($r, \theta$) can be 
written as follows \citep[e.g.][]{1964ApJ...140..646L,1987ApJ...314....3E}:
\begin{equation}
r(\theta)=R \cdot e^{\theta \, {\rm tan} \phi},
\end{equation}
where $\theta=0$ at $r=R\in \mathds{R}$ and $\phi$ refers to the pitch angle. 
For a given galaxy, the arc length ($s$) of the logarithmic spiral segment $i$ 
extending between $(r_{i},\theta_{i})$ and $({r_{i}}^{\prime},{\theta_{i}}^{\prime})$ 
with pitch angle $\phi_{i}$ can be approximated as \citep[e.g.][]{1963Zwikker}:
\begin{equation}\label{eq_segm}
s_{i}=\int_{\theta_{i}}^{{\theta_{i}}^{\prime}} \sqrt{r^2+(dr/d\theta)^{2}} \cdot d \theta \approx |{r_{i}}^{\prime}-r_{i}| \cdot \dfrac{\sqrt{1+{\rm tan} |\phi_{i}|^{2}}}{{\rm tan} |\phi_{i}|},
\end{equation}
and then
\begin{equation}\label{eq_segm_2}
|\phi|_{\rm weighted}=\dfrac{1}{S} \sum_{i=1}^{\rm n} |\phi_{i}| \cdot s_{i},
\end{equation}
where $n$ is the total number of measured spiral arm segments, and $S=\sum_{i=1}^{\rm n}s_{i}$. 
This is in practice very close to weighting with the relative radial extent of 
the spirals segments over which measurements were performed. The mean absolute difference between $|\phi|_{\rm mean}$ and 
$|\phi|_{\rm weighted}$ is $2.9^{\circ} \pm 0.14^{\circ}$ (see bottom panel of Fig.~\ref{mean_vs_median}).

The global pitch angles of each of the galaxies in our sample\footnote{
Preliminary analysis of these measurements (and the properties of barred and non-barred galaxies) 
using unsupervised machine learning techniques (self-organising maps) has been done by Díaz-García et al. (2019).} 
are listed in Table~\ref{tabulate_mean_pitches}. 
Hereafter, we will use all proxies of $|\phi|$ for the analysis in this paper.
%
%
\subsection{Pitch angles from 2-D Fourier analysis}\label{fourier_pitch}
%
%
In order to test the sensitivity of the pitch angle determination to the employed methodology, 
we also fit the spiral structure using Fourier transform spectral analyses of the de-projected $3.6$~$\mu$m images. 
We apply a similar method as in \citet[][]{1992A&AS...93..469P} and \citet[][]{2000A&A...359..932P}, 
who had in turn followed the approach by \citet[][]{1975dgs..conf..103K} and \citet[][]{1982A&A...111...28C} 
\citep[for a more sophisticated determination, see][]{2014AJ....148..133P}.

Images are decomposed into a flux-weighted superposition of $m$-armed 
logarithmic spirals of different pitch angles \citep[see e.g.][]{2000A&A...359..932P,2001ChJAA...1..395M,2012ApJS..199...33D}, 
whose amplitudes are calculated as: 
\begin{equation}\label{amplitude_m_p}
A(p,m)=\frac{1}{D} \int_{-\pi}^{+\pi} \int_{R_{\rm inner}}^{R_{\rm outer}} I(u,\theta) e^{-i (m \theta+p u)} du d\theta,
\end{equation}
where $u={\rm ln}(r)$, $I(u,\theta)$ refers to the intensity distribution in polar coordinates, 
the variable $p \in \mathds{R}$ defines the pitch angle of the spiral as ${\rm arctan}(-m/p)$, 
$D=\int_{R_{\rm inner}}^{R_{\rm outer}}\int_{-\pi}^{+\pi} I(r,\theta) dr d\theta$ is the normalisation factor, 
and $R_{\rm inner}$ and $R_{\rm outer}$ are the limits in the radial range for the fit.

Choosing $R_{\rm inner}$ and $R_{\rm outer}$ is the greatest source of human error for the fit \citep[][]{2012ApJS..199...33D}. 
Where available, we use the minimum and maximum radii where \citet[][]{2015A&A...582A..86H} fit the logarithmic segments. 
By doing so, the central bulge is excluded from the fit and the outer radius reaches the limit of the arms in the images. 
Whenever clear outer arm breaks are seen (e.g. M$\,$51), we limit the fit to the inner segments. 
When the galaxies are barred, the masked region has a radius $\gtrsim r_{\rm bar}$ 
(at times, we mask regions well beyond the bar radius to avoid short spurious spiral segments that can contaminate the fit). 
Occasionally, when the spiral pattern presents multiple breaks, we limit the fit to the range $[r_{\rm bar}, 2r_{\rm bar}]$. 
By visual inspection of the images, we also choose the range of pitch angles for the fit (positive or negative depending on winding direction).

We use the limits $p\in[-50,50]$ and step-size ($0.25$) advocated by \citet[][]{2000A&A...359..932P}. 
Finally, for $m=2$, we select the value of $p$ with the highest amplitude, which defines the final pitch angle of the spiral arms, 
called as $|\phi|_{\rm Fourier}$ (for NGC$\,5054$ we used $m=3$ to fit the three arms). 
Examples of the resulting fits for M$\,$51 and NGC$\,$3992 are shown in Fig.~\ref{spiral_fit_fourier}.

We could reliably identify the maximum of the pitch angle $p$ associated to the spiral pattern ($|\phi|_{\rm Fourier}$) 
for a subsample of 32 of the galaxies in our master sample ($\sim 10\%$), 
of which most (25) are barred (i.e. they are not SA) according to \citet[][]{2015ApJS..217...32B}. 
In Table~\ref{tabulate_Fourier_pitches} we list the measurements of the pitch angle using the 2-D Fourier method. 
We also indicate the radial range where the fit was done. 
A rough estimate of the uncertainties in $|\phi|_{\rm Fourier}$ is calculated by 
reducing by an arbitrary $20\,\%$ the radial range for the fit: 
an offset of $(R_{\rm outer}-R_{\rm inner})/10$ is added (subtracted) to $R_{\rm inner}$ (from $R_{\rm outer}$).
%
%
\begin{figure}
\centering
\includegraphics[width=0.5\textwidth]{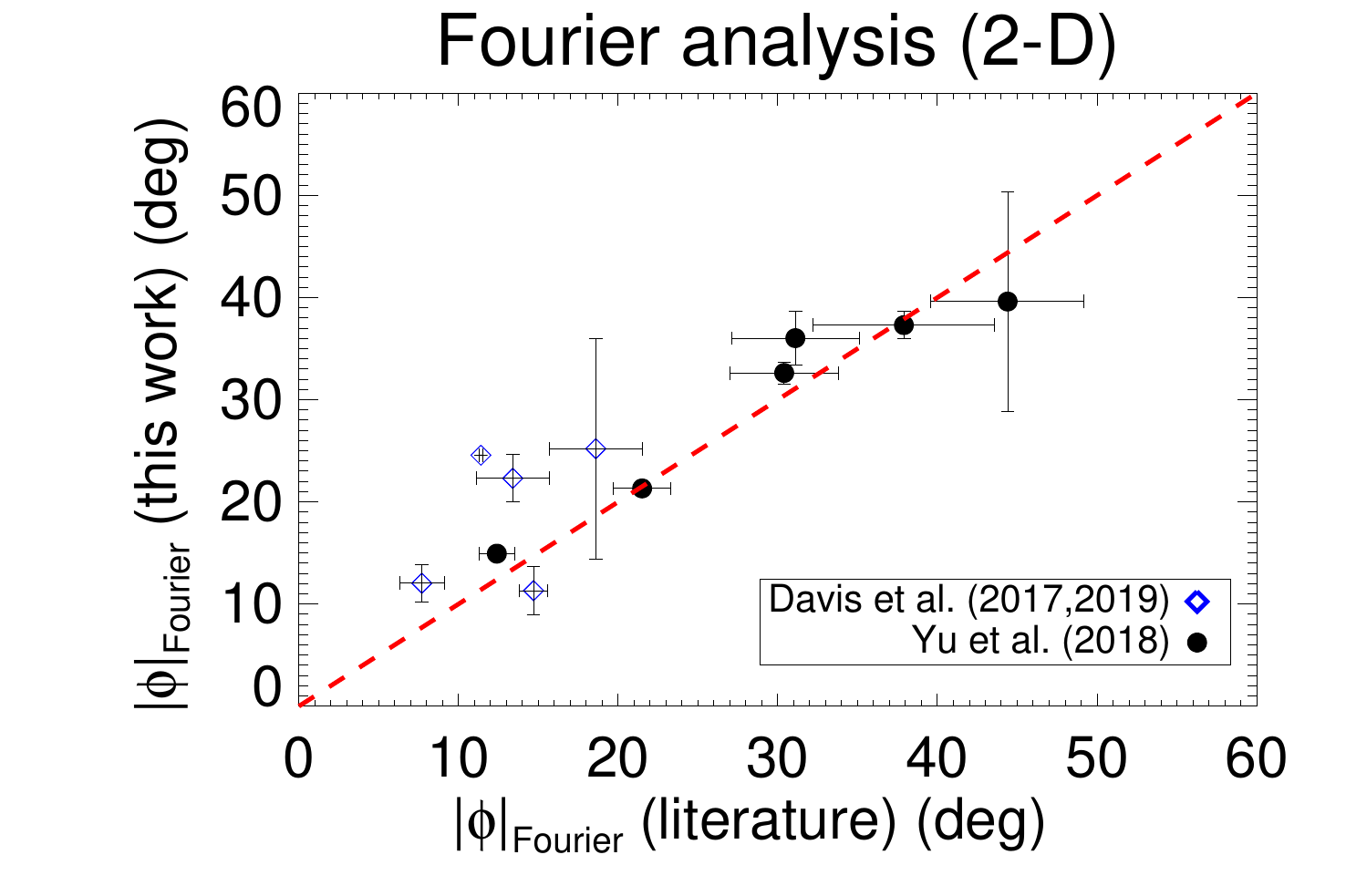}
\caption{
For a subsample of 11 S$^4$G galaxies, 
comparison between the pitch angles obtained in this work and in the literature (see legend) applying Fourier analysis (see text). 
The $y=x$ straight line is shown in red. 
}
\label{fourier_comparison_literature}
\end{figure}
%
%
\begin{table}
\centering
\small
\begin{tabular}{| c| c| c| c| c|}
\hline
\multicolumn{1}{|c}{\centering Galaxy} &
\multicolumn{1}{|c}{\centering $R_{\rm inner}$ } & \multicolumn{1}{|c}{\centering $R_{\rm outer}$} &
\multicolumn{1}{|c}{\centering $|\phi|_{\rm Fourier}$} & \multicolumn{1}{|c|}{\centering Error |$\phi|_{\rm Fourier}$} \tabularnewline
\multicolumn{1}{|c}{\centering } &
\multicolumn{1}{|c}{\centering ($\arcsec$)} & \multicolumn{1}{|c}{\centering ($\arcsec$)} &
\multicolumn{1}{|c}{\centering ($^{\circ})$} & \multicolumn{1}{|c|}{\centering ($^{\circ})$} \tabularnewline
\hline
\hline
ESO027-001& 15.9& 90.0& 18.1&$\pm$  2.4 \\
IC0167& 14.8& 80.2& 48.8&$\pm$  2.1 \\
IC0769& 12.7& 68.2& 14.7&$\pm$  0.4 \\
IC2627&  9.0& 82.6& 36.0&$\pm$  2.6 \\
NGC2543& 29.2& 71.0& 13.8&$\pm$ 16.8 \\
NGC2710& 10.1& 55.9& 33.7&0.0 \\
NGC2712& 26.8& 53.2&  9.2&$\pm$  1.5 \\
NGC2750& 14.7& 72.5& 17.1&$\pm$  1.4 \\
NGC3031&260.4&654.0& 22.3&$\pm$  2.3 \\
NGC3227& 54.8&127.5& 12.0&$\pm$  1.9 \\
NGC3310& 13.2& 69.1& 41.6&0.0 \\
NGC3507& 23.0& 72.8& 14.7&$\pm$  0.4 \\
NGC3513& 17.8& 71.2& 21.3&0.0 \\
NGC3583& 24.1& 70.2& 18.8&$\pm$  3.1 \\
NGC3627& 61.8&215.4& 25.2&$\pm$ 10.8 \\
NGC3893& 10.6&112.5& 20.0&0.0 \\
NGC3953& 48.5&173.6&  8.5&$\pm$  0.2 \\
NGC3992& 82.3&205.1&  9.5&0.0 \\
NGC4303& 38.1&175.8& 11.3&$\pm$  2.4 \\
NGC4430& 27.9& 73.1& 17.4&$\pm$  0.3 \\
NGC4902& 27.1& 74.1& 14.9&$\pm$  0.2 \\
NGC5054& 25.1&152.2& 39.6&$\pm$ 10.7 \\
NGC5194& 53.2&174.0& 17.1&$\pm$  0.6 \\
NGC5247& 21.9&152.1& 37.3&$\pm$  1.4 \\
NGC5364& 17.0&186.3&  9.7&$\pm$  0.1 \\
NGC5430& 33.0& 59.9& 12.9&$\pm$  0.2 \\
NGC5668& 12.8& 91.6& 29.7&$\pm$  3.9 \\
NGC5774& 17.1& 88.2& 14.7&0.0 \\
NGC5921& 39.1&112.6& 26.6&$\pm$  4.8 \\
NGC5970& 17.2& 72.4& 23.4&$\pm$  3.4 \\
NGC7412& 20.4&103.6& 32.6&$\pm$  1.1 \\
NGC7741& 48.7&113.5& 18.1&$\pm$  5.4 \\
\hline
\end{tabular}
\caption{
Pitch angle of a subsample of S$^4$G galaxies determined based on 2-D Fourier analysis.
We also indicate the uncertainty in the measurement and the radial boundaries (in the disk plane) used for the fit.
}
\label{tabulate_Fourier_pitches}
\end{table}

%
%
\begin{figure}
\centering
\includegraphics[width=0.47\textwidth]{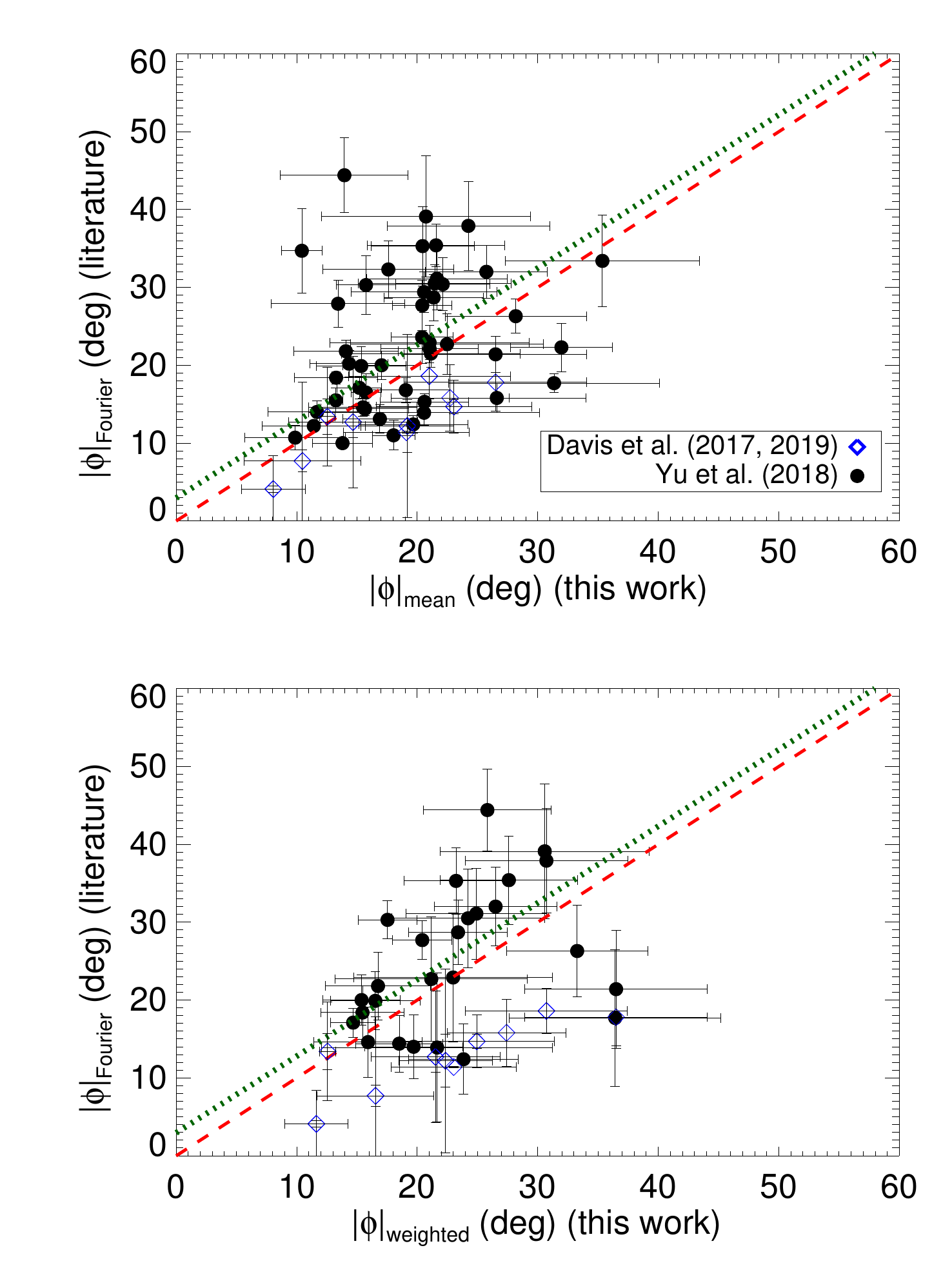}
\caption{
Comparison between mean and weighted mean pitch angles calculated in this work, 
and the pitch angle obtained by \citet{2018ApJ...862...13Y} (black) and 
\citet{2017MNRAS.471.2187D} (blue diamonds) via Fourier analysis (see the text). 
The dashed red line corresponds to the straight line $y=x$, 
while the dotted green line shows the linear fit to the data cloud.
}
\label{mean_vs_median_literature}
\end{figure}
%
%
\begin{figure}
\centering
\includegraphics[width=0.45\textwidth]{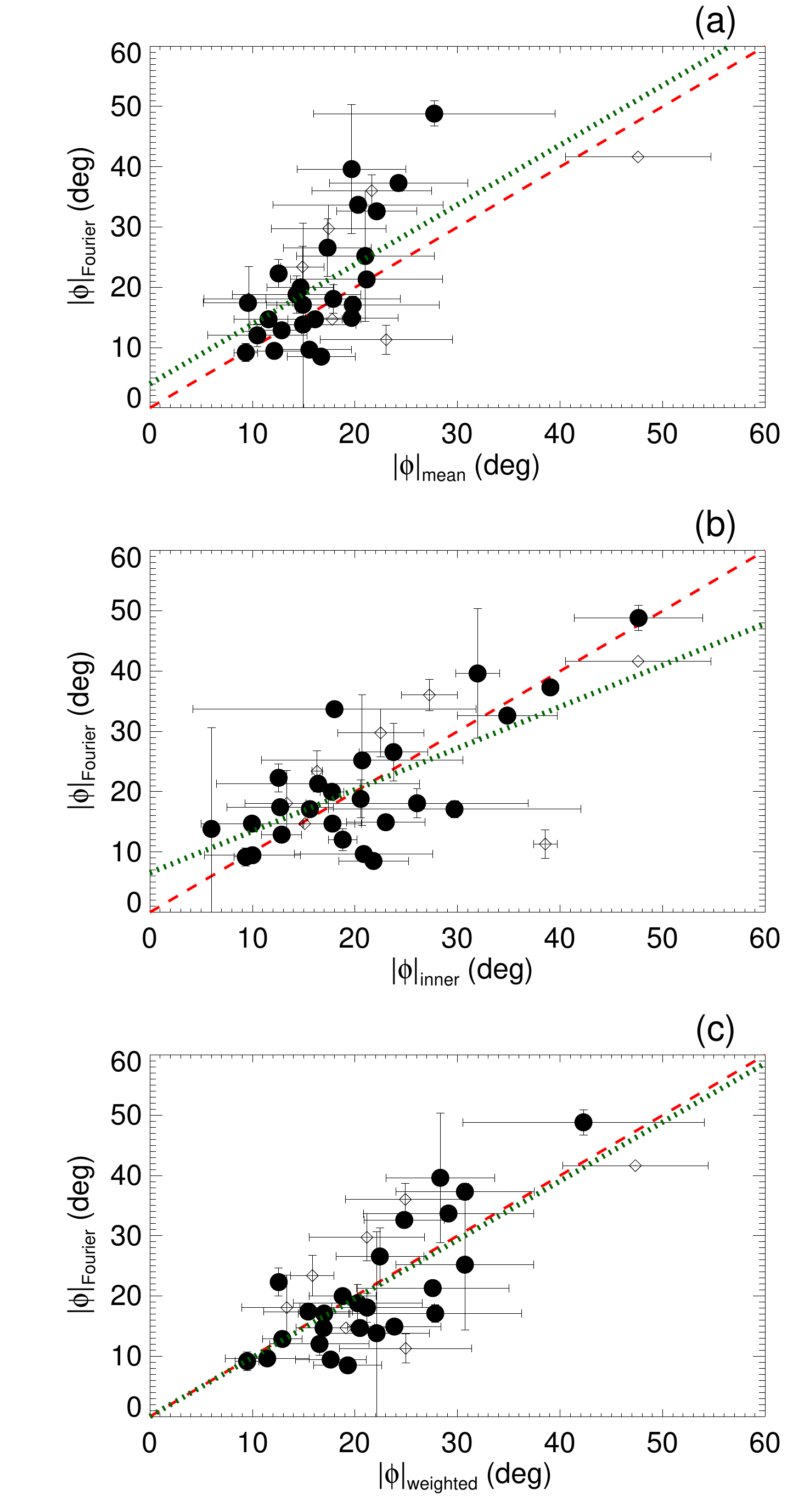}
\caption{
For a subsample of S$^4$G galaxies, comparison between the pitch angles determined with Fourier methods (from this work, see text) 
and a) mean pitch angle, b) inner mean pitch angles, and c) mean pitch angle weighted by the arc length of the segments. 
We show measurements with very reliable ($=1$, filled circle) and acceptable ($=2$, empty diamonds) flags in \citet[][]{2015A&A...582A..86H}. 
As in Fig.~\ref{mean_vs_median_literature}, 
the red dashed line corresponds to the straight line $y=x$, while the green dotted line shows the linear fit to the cloud of points.
}
\label{Fourier_visual}
\end{figure}
%
%
\begin{figure}
\centering
\includegraphics[width=0.49\textwidth]{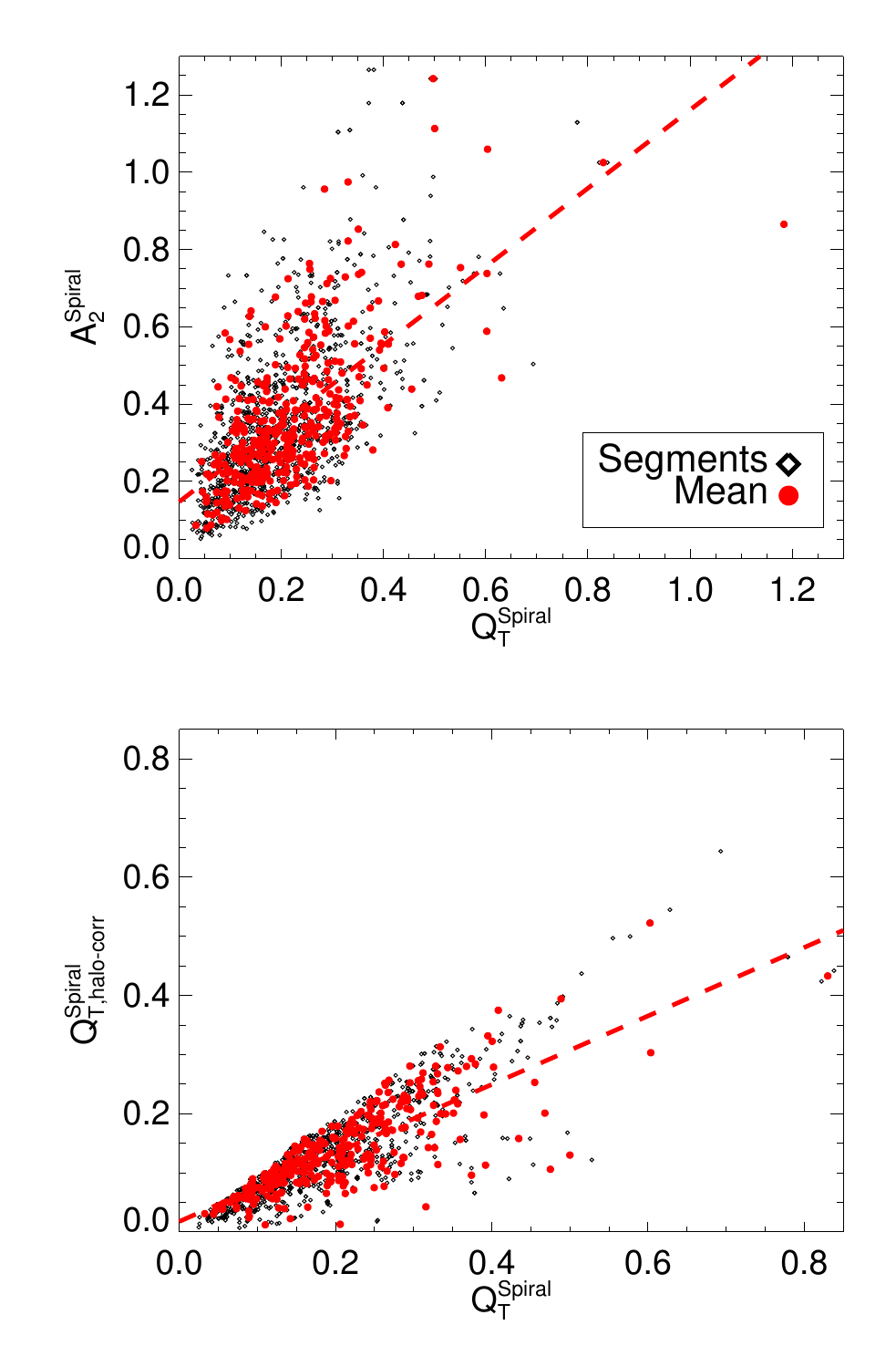}
\caption{
Comparison between the proxies of the amplitude of the spirals used in this work (Eqs.~\ref{qt_sp_mean}, \ref{qt_halo_sp_mean}, and \ref{a2_sp_mean}). 
We show the mean value per galaxy, and measurements over individual segments. 
The Spearman's rank correlation coefficients (significances) are 0.62 ($4.14\cdot 10^{-43}$) 
and 0.82 (0.0) for the upper and lower panels, respectively. 
With a red dashed line we show the linear fit to the red data points.
}
\label{comparison_spiral_amp}
\end{figure}
%
%
\begin{figure}
\centering
\includegraphics[width=0.5\textwidth]{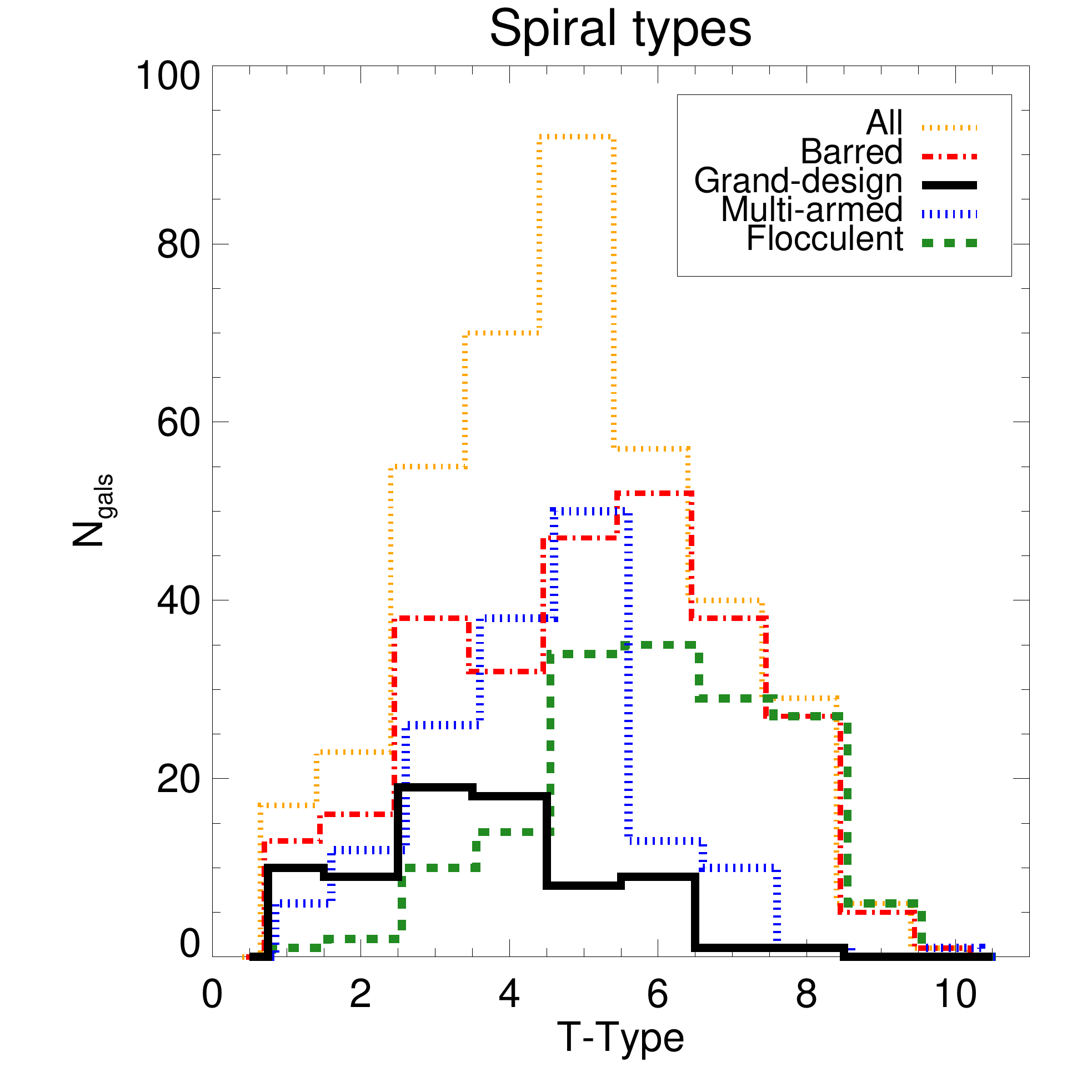}
\includegraphics[width=0.5\textwidth]{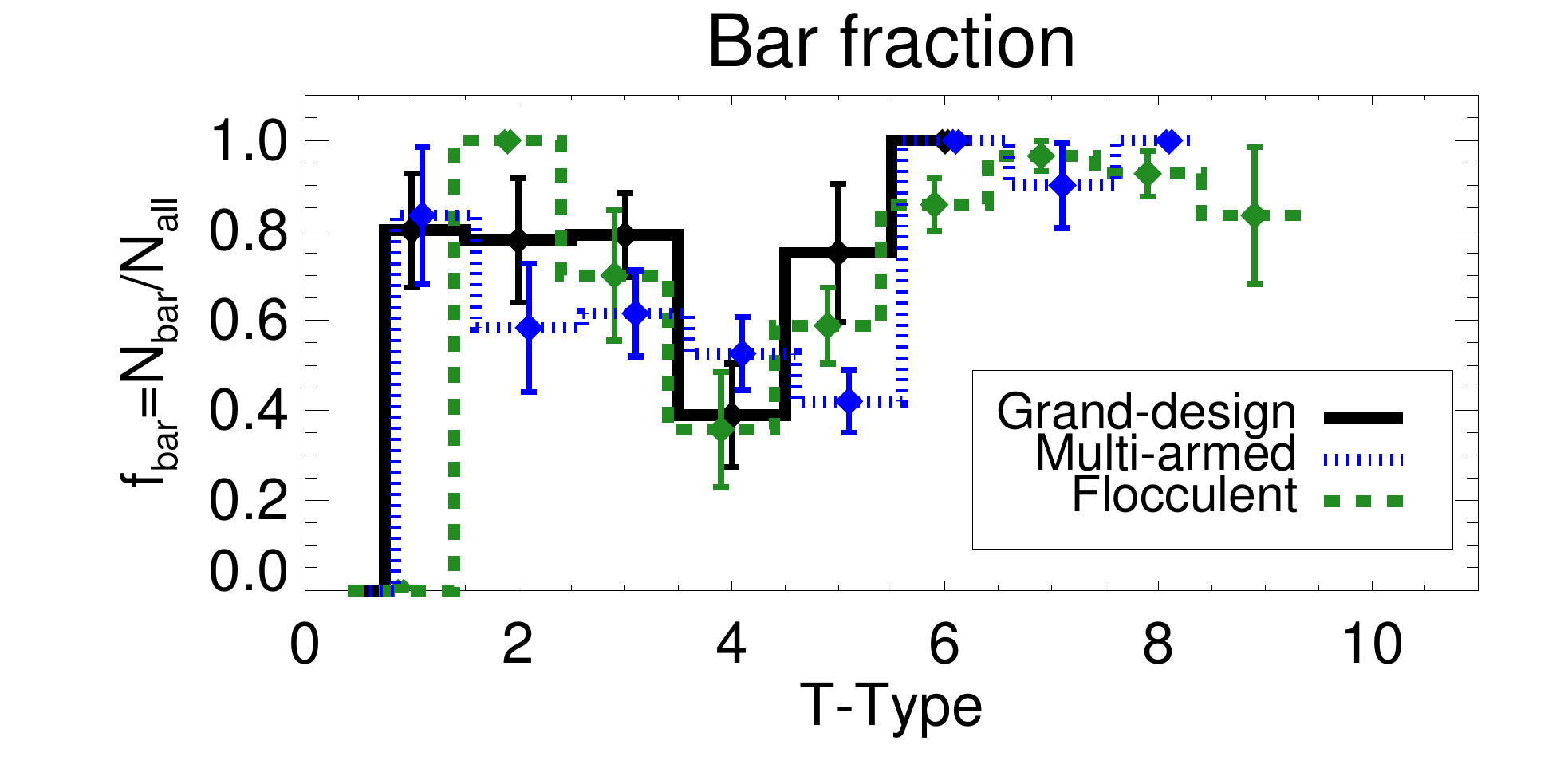}
\caption{
\emph{Upper panel:}
Histograms of the distributions of the galaxies in our sample (showing the barred galaxies separately) and of the different arm classes 
(grand-design, multi-armed, and flocculent) as a function of revised Hubble stage. 
Small offsets ($\le 0.1$) have been added in the $x$-axis for the sake of avoiding line overlapping.
\emph{Lower panel:} 
Bar fraction for the different spiral types. With vertical lines we represent binomial counting errors. 
}
\label{histo_spiral_types_ttype}
\end{figure}
%
%
\begin{figure*}
\centering
\begin{tabular}{c c c}
\includegraphics[width=0.33\textwidth]{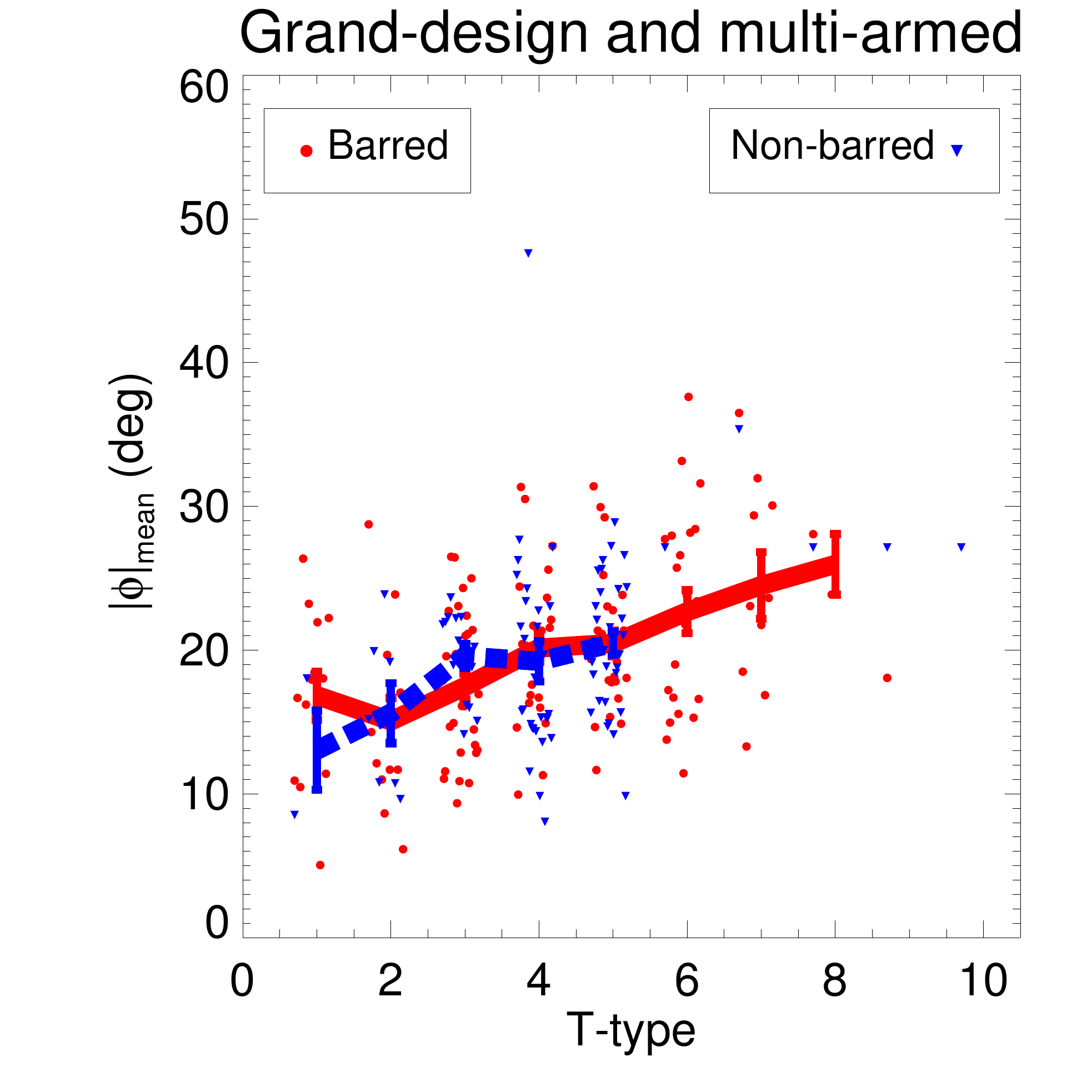}
\includegraphics[width=0.33\textwidth]{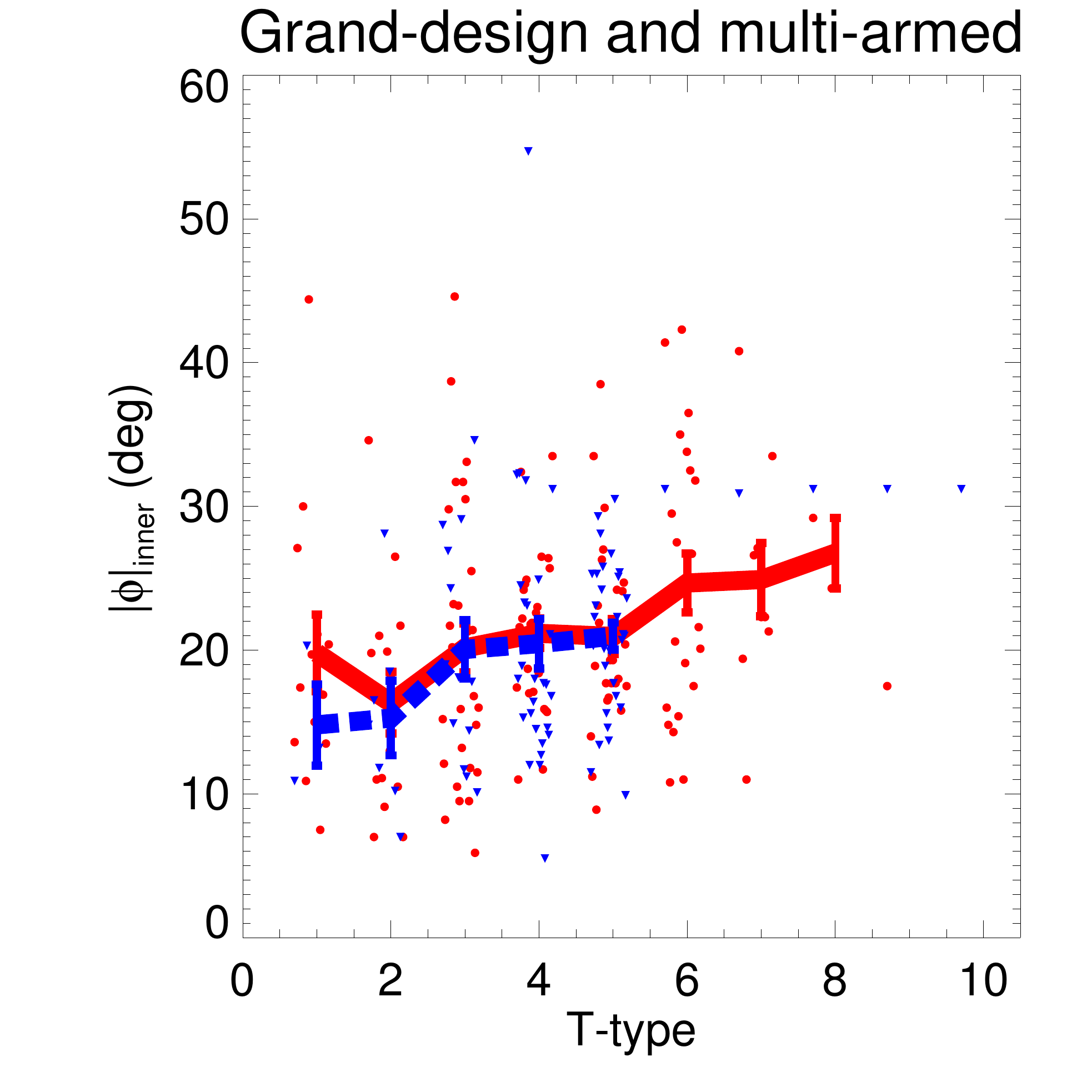}
\includegraphics[width=0.33\textwidth]{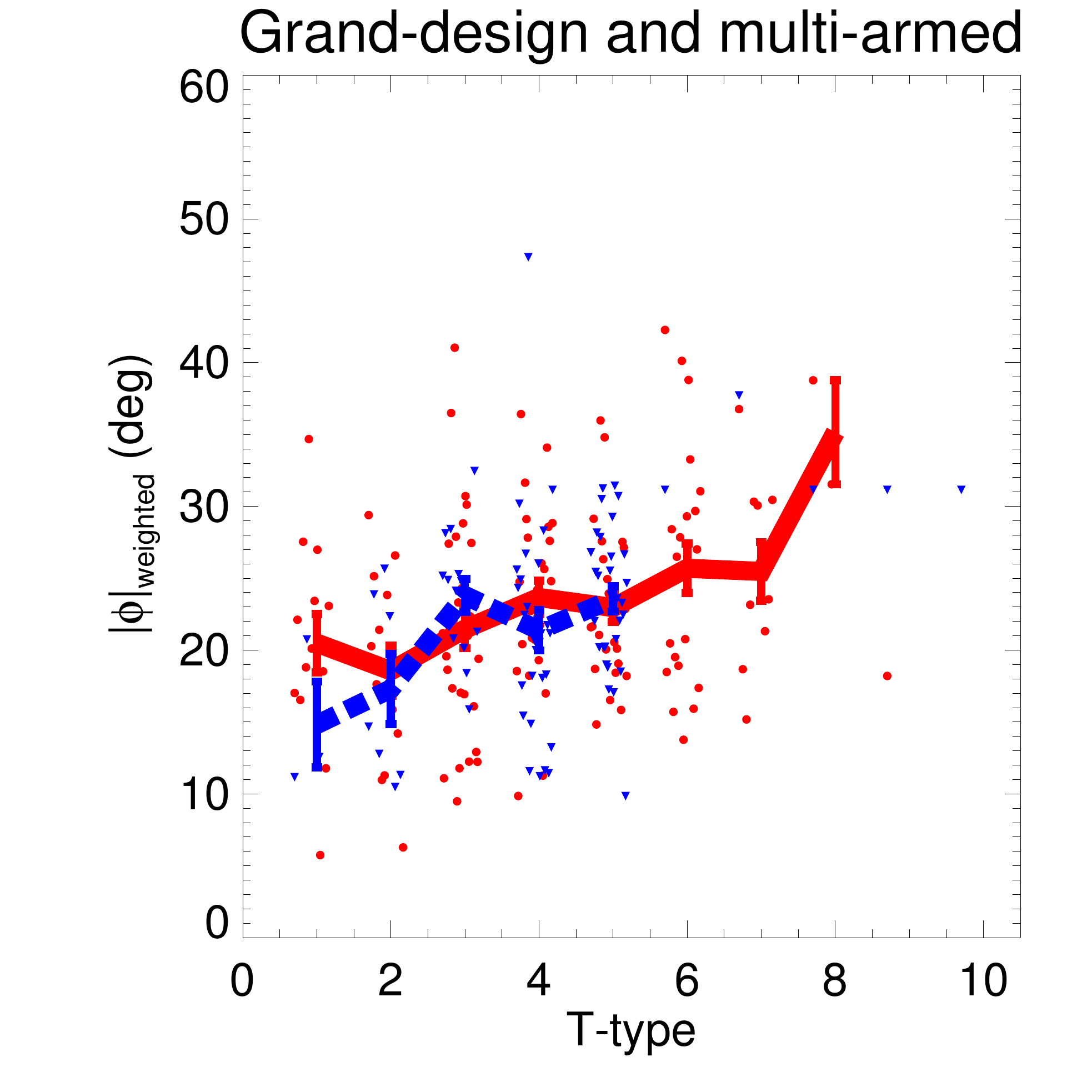}
\end{tabular}
\caption{
Mean pitch angle (\emph{left panel}), mean of the two innermost spiral segments (\emph{middle panel}), 
and weighted mean pitch angle (\emph{right panel}) 
as a function of the integer value of the revised numerical Hubble stage, for all the grand-design and multi-armed spirals in our sample. 
The running mean and standard deviation of the mean are shown for the barred (red) and non-barred (blue) galaxies separately, 
whenever more than one data point lies within the bin (i.e. when $T\le 8$ for barred galaxies and $T\le 5$ for non-barred ones). 
We have added small random offsets ($\lesssim 0.3$) to the $T$ values in the $x$-axis (integers) to avoid point overlapping.
}
\label{Fig_pitch_angles}
\end{figure*}
%
%
\begin{figure*}
\centering
\begin{tabular}{c c c}
\includegraphics[width=0.33\textwidth]{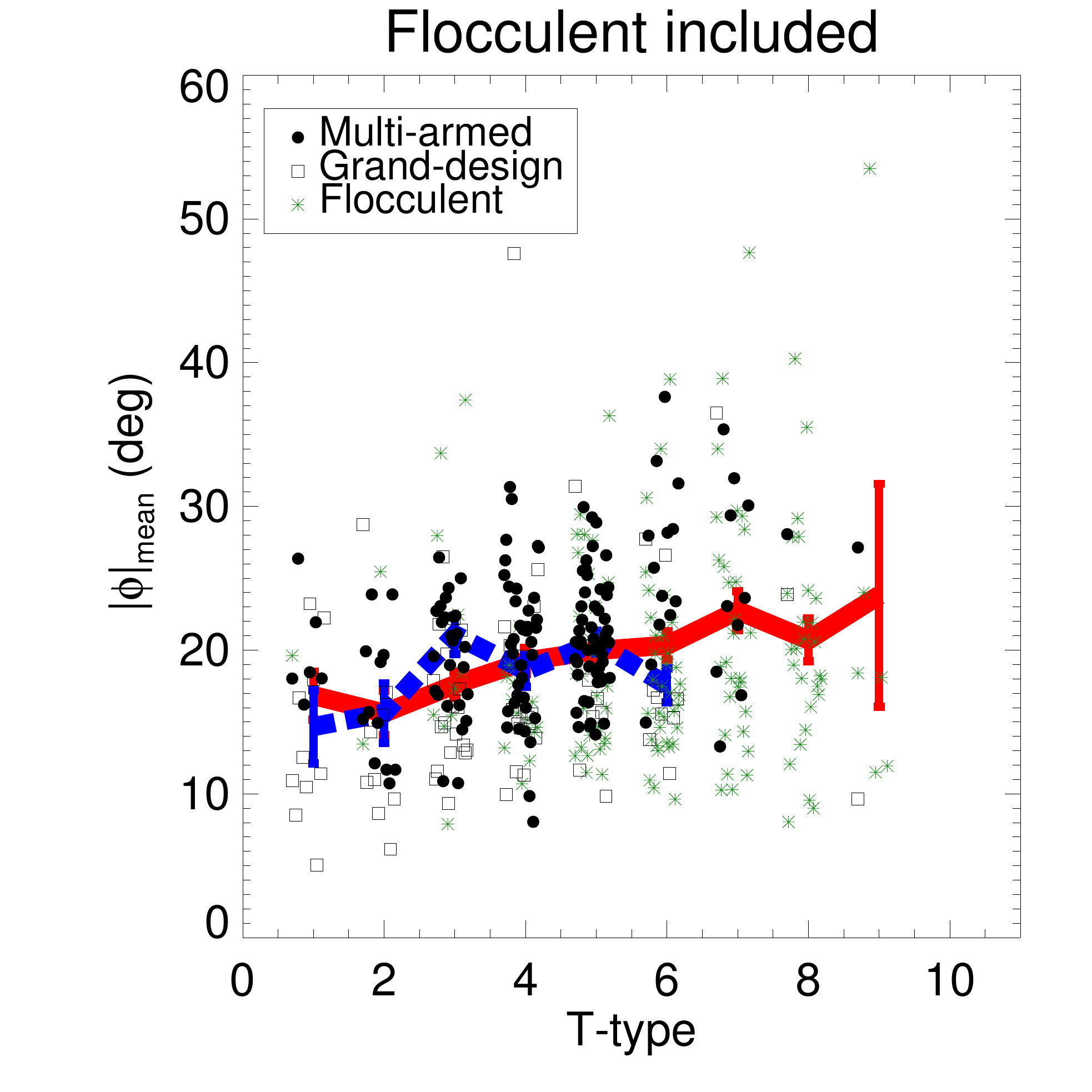}
\includegraphics[width=0.33\textwidth]{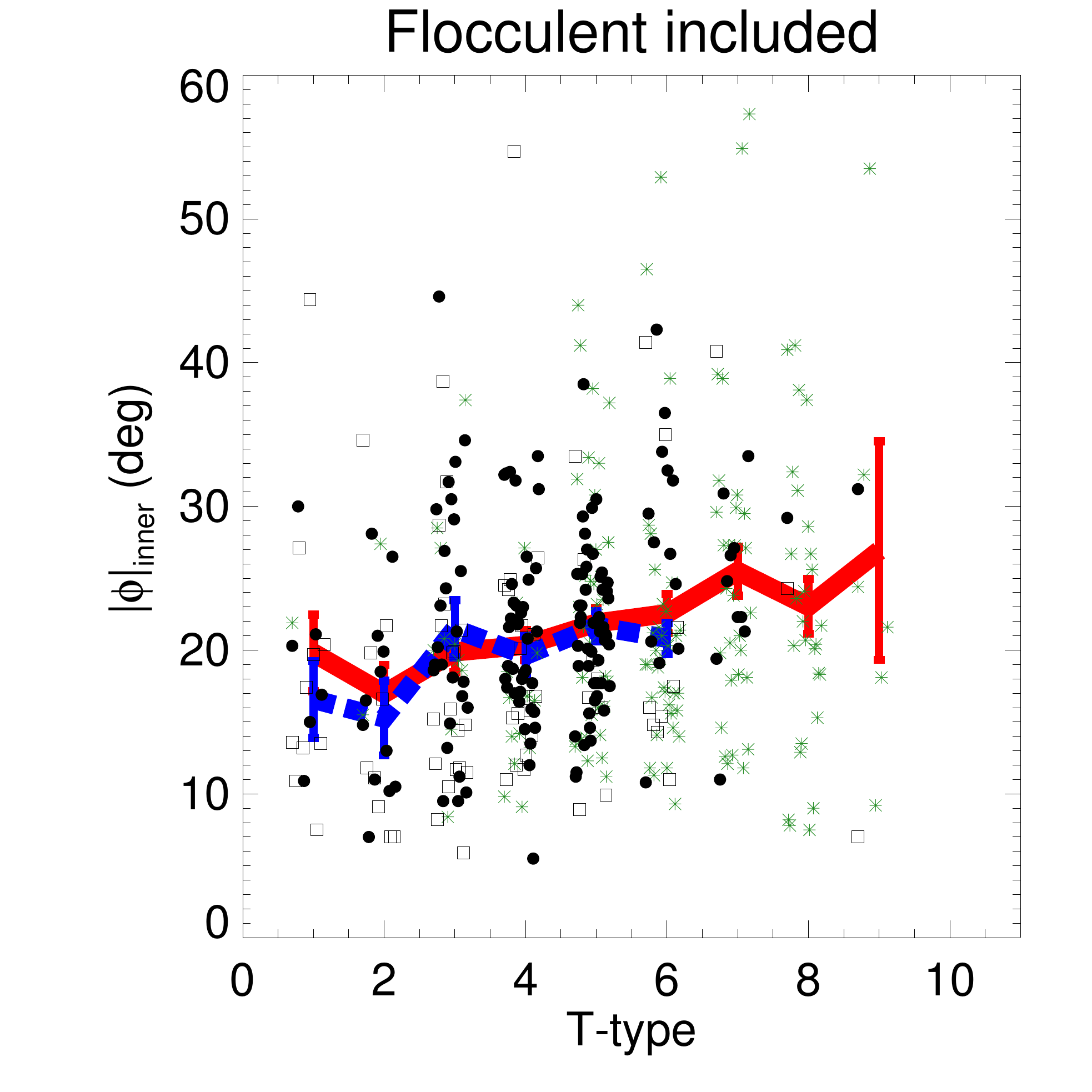}
\includegraphics[width=0.33\textwidth]{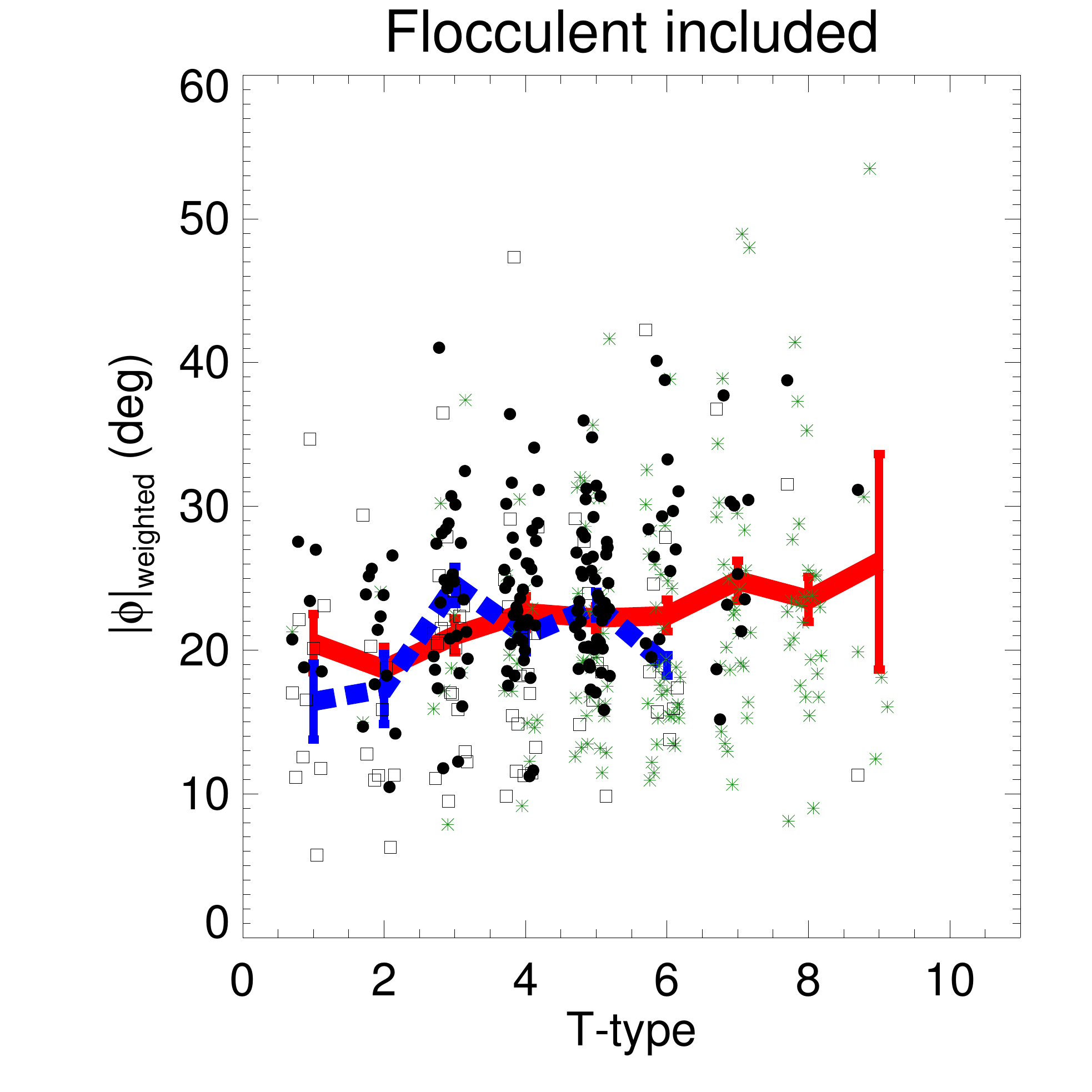}
\end{tabular}
\caption{
As in Fig.~\ref{Fig_pitch_angles}, but including flocculent spirals (green). 
The running mean and standard deviation of the mean are shown for the galaxies that are either barred (red) 
or non-barred (blue, covering only the $T-$type range $1\le T \le 6$ where they appear).
}
\label{Fig_pitch_angles_fgm}
\end{figure*}
%
%
\begin{figure}
\centering
\includegraphics[width=0.5\textwidth]{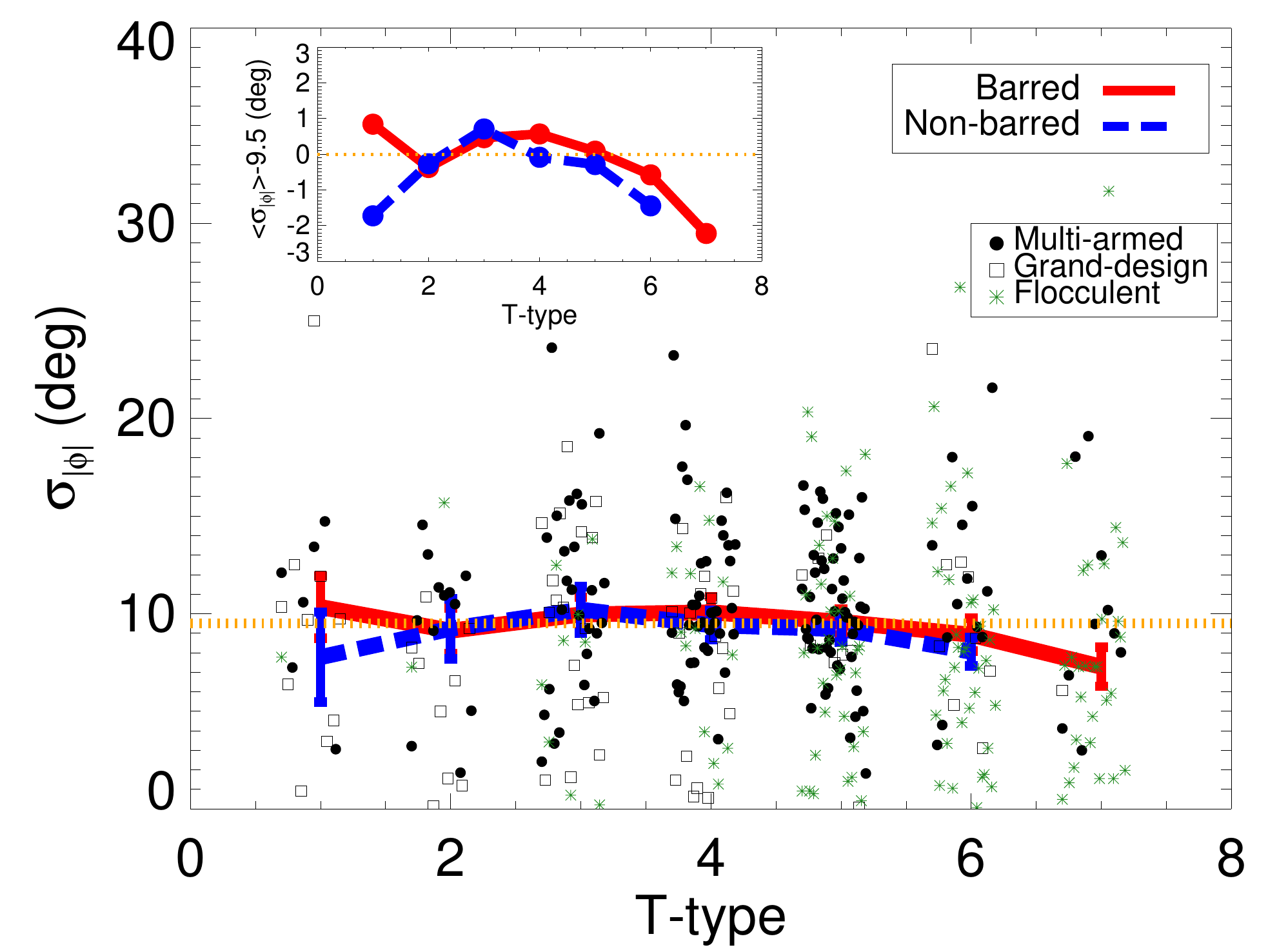}
\caption{
As in Fig.~\ref{Fig_pitch_angles_fgm}, but for the standard deviation of $|\phi|$ in each of the galaxies. 
The lines indicate the running median per $T-$type for barred and non-barred galaxies, 
and the vertical lines correspond to the standard deviation of the mean. 
The horizontal orange dotted line traces the mean $\sigma_{|\phi|}$ ($\approx 9.5^{\circ}$). 
Mean residual differences with respect to this value for each $T$ bin (typically $\lesssim 2^{\circ}$) 
are shown in the inner panel on the upper left corner.
}
\label{sigma_pitch_angles_fgm}
\end{figure}
%
%
\begin{figure}
\centering
\begin{tabular}{c c}
\includegraphics[width=0.5\textwidth]{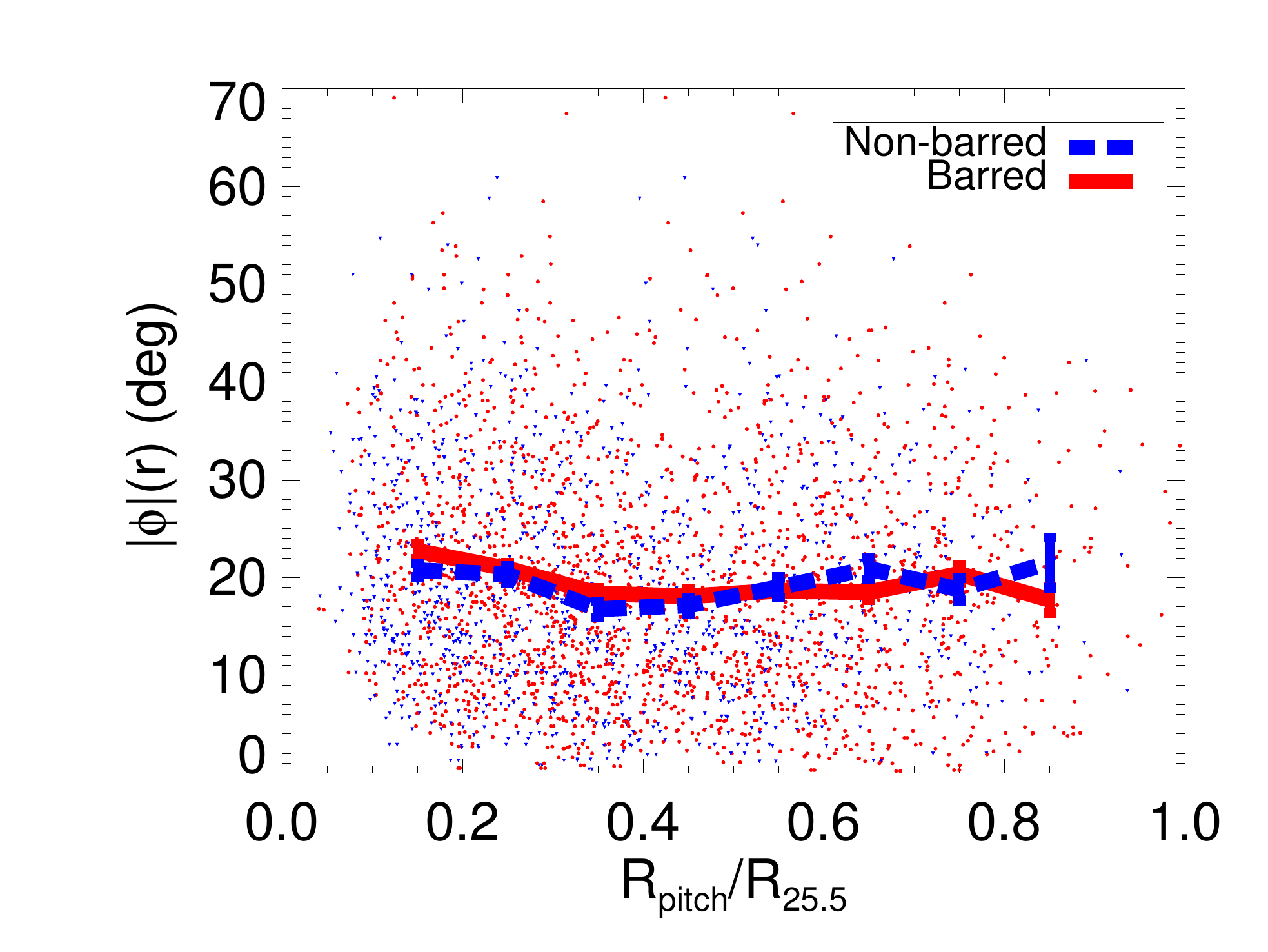}\\[-4ex]
\includegraphics[width=0.5\textwidth]{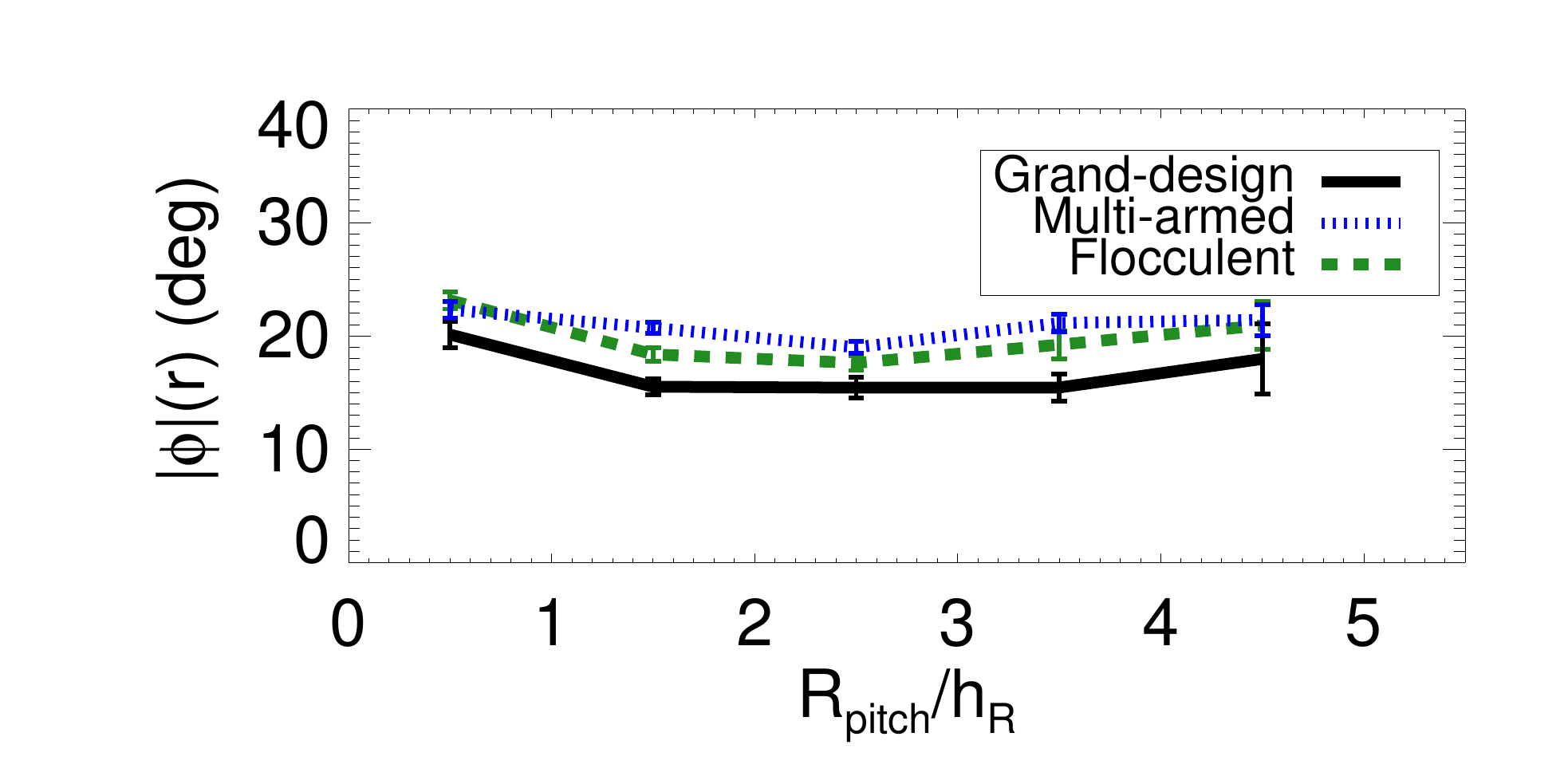}
\end{tabular}
\caption{
\emph{Upper panel:} 
Radial variation of the pitch angle based on the individual measurements of logarithmic segments by \citet[][]{2015A&A...582A..86H}. 
In the $x$-axis we plot the boundaries of the radial interval where the pitch was measured, 
normalised by the 25.5~mag~arcsec$^{-2}$ isophotal radius. 
The lines correspond to the running mean (bin size of 0.1), 
and the error bars are the standards deviation of the mean (the values appear listed in Table~\ref{pitch_vs_radius_R255}). 
Barred and non-barred galaxies are plotted separately. 
\emph{Lower panel:} 
Same as above but using the disc scale-length for the normalization, 
and showing only the running mean (bin size of 1, see Table~\ref{pitch_vs_radius_HR}). 
The colours indicate flocculent, multi-armed, and grand-design galaxies, as indicated in the legend. 
}
\label{pitch_radial}
\end{figure}
%
%
\subsection{Comparison between methods and with values in the literature}\label{comp_vis_lit}
%
%
Recent work by \citet[][]{2018ApJ...862...13Y} characterised the pitch angle of the spiral arms in the Carnegie-Irvine Galaxy Survey 
\citep[][]{2011ApJS..197...21H} using Fourier methods\footnote{
We note that the mean difference between the pitch angles obtained at 3.6~$\mu$m and from $I$-band images by \citet{2018ApJ...862...13Y} is very small: 
$\approx 0.1^{\circ}$ for the 2-D method and $\approx 0.3^{\circ}$ for the 1-D method \citep[][]{2018ApJ...869...29Y}.
}: there is an overlap of 49 galaxies between their survey and our sample of 391 galaxies. 
In addition, there are seven S$^4$G galaxies from our sample with measurements available from \citet[][]{2017MNRAS.471.2187D}\footnote{ 
These measurements were performed using using the software \emph{Spirality} \citep[][]{2015arXiv151106365S,2015ascl.soft12015S}, 
2DFFT software \citep[][]{2012ApJS..199...33D,2016ascl.soft08015D}, 
and compared to computer vision software \citep[][]{2014ApJ...790...87D}.}. 
Of those 56 galaxies with pitch angle measurements in the literature, 
we could reliably apply the Fourier method to 11 (Sect.~\ref{fourier_pitch}): 
in Fig.~\ref{fourier_comparison_literature} we show a comparison of the obtained values, showing a good agreement.
 
In general, there is also good agreement between the average pitch angle obtained in this work 
- based on the measurements of individual logarithmic segments - 
and the values of $|\phi|_{\rm Fourier}$ from the literature (Fig.~\ref{mean_vs_median_literature}): 
we obtain a mean absolute difference of $3.9^{\circ} \pm 1.0^{\circ}$ and $8.4^{\circ} \pm 1.4^{\circ}$ for 
$|\phi|_{\rm mean}$ and $|\phi|_{\rm weighted}$, respectively. 
The scatter and outliers in the comparison are, most likely, due to the employed methodology to estimate the pitch angle. 

Differences in $|\phi|$ for the two methods applied in this paper 
(Fourier analysis and average of the pitch angles of individual logarithmic segments) 
over 3.6~$\mu$m images for a subsample of 32 galaxies are shown 
in Fig.~\ref{Fourier_visual}. We compare $|\phi|_{\rm Fourier}$ with 
$|\phi|_{\rm mean}$, $|\phi|_{\rm inner}$, and $|\phi|_{\rm weighted}$, 
obtaining mean absolute differences of $6.8^{\circ} \pm 1.0^{\circ}$, $10.0^{\circ} \pm 2.2^{\circ}$, and $5.9^{\circ} \pm 0.7^{\circ}$, 
respectively. Discrepancies are associated to decoupled spiral modes that are not fitted with the Fourier method.

Obtaining a single value that characterises the winding of the spirals is not straightforward, 
and it is sensitive to the utilised methodology and its drawbacks \citep[see also discussion in e.g.][]{1992ApJS...79...37E}. 
Given the uncertainties inherent to the various methodologies, the analysis of the possible dependence of pitch angle 
on bar strength (Sect.~\ref{pitch_bar_strength}) is done based on visual measurements of $|\phi|$ as well as with 
the Fourier transform (using values from this work and from the literature).
%
%
\section{Amplitude of non-axisymmetries}\label{non_axis}
%
%
\subsection{Bar strength and size}\label{bar_strength_methods}
%
%
We use deprojected visual measurements of bar sizes ($r_{\rm bar}$) and 
ellipticities ($\epsilon$) from \citet[][]{2015A&A...582A..86H}. 
We use normalised Fourier amplitudes $A_{m}=I_{m}/I_{0}$, where $I_{0}$ indicates the $m=0$ surface density component, 
calculated by \citet[][]{2016A&A...587A.160D} by applying the NIR-QB code \citep[][]{1999AJ....117..792S,2002MNRAS.337.1118L} 
to $3.6\,\mu$m S$^4$G images. 
They also derived radial profiles of tangential forces normalised to the mean radial force field \citep[][]{1981A&A....96..164C}: 
\begin{equation}\label{torquerad}
Q_{\rm T}(r)=\frac{{\rm max}\left( |F_{\rm T}(r,\phi)| \right)}{\langle |F_{\rm R}(r,\phi)|\rangle}.
\end{equation}
The maximum of $A_2$ and $Q_{\rm T}$ at the bar region are used as proxies of the bar strength 
and named $A_2^{\rm max}$ and $Q_{\rm b}$, respectively \citep[e.g.][]{2001ApJ...550..243B,2002MNRAS.331..880L,2004ApJ...607..103L}.

In addition, \citet[][]{2016A&A...587A.160D} obtained a first-order model of the dark matter halo rotation curve and 
implemented a correction on $Q_{\rm T}$ for the contribution of the halo to the radial forces 
($Q_{\rm T}^{\rm halo-corr}$), following \citet[][]{2004AJ....127..279B}.

Finally, \citet[][]{2016A&A...587A.160D} eliminated the spiral contribution to the bar local force by setting to zero the 
Fourier density amplitudes beyond the bar radius for $m > 0$, as done in \citet[][]{2010ApJ...715L..56S}. 
We use the maximum at the bar region ($Q_{\rm b}^{\rm bar-only}$) as an intrinsic measure on the bar-only force.
%
%
\subsection{Spiral strength}
%
%
Here we define a method to calculate the strength of spiral arms in the same regions $[r_{i},{r_{i}}^{\prime}]$ 
where the pitch angles of logarithmic segments were measured:
\begin{equation}\label{qt_sp}
Q_{\rm T}^{{\rm Segment} \,i}=\max_{r_{i} \leq r \leq {r_{i}}^{\prime}} Q_{\rm T}(r),
\end{equation}
\begin{equation}\label{qt_halo_sp}
Q_{\rm T, halo-corr}^{{\rm Segment} \,i}=\max_{r_{i} \leq r \leq {r_{i}}^{\prime}} Q_{\rm T}^{\rm halo-corr}(r),
\end{equation}
\begin{equation}\label{a2_sp}
A_{2}^{{\rm Segment} \,i}=\max_{r_{i} \leq r \leq {r_{i}}^{\prime}} A_{2}(r).
\end{equation}
%
%
For each galaxy, we then calculate the mean of the amplitude of the spiral segments: 
\begin{equation}\label{qt_sp_mean}
Q_{\rm T}^{\rm Spiral}=<Q_{\rm T}^{{\rm Segment} \,i}>_{i\in[1,\rm n]}, 
\end{equation}
\begin{equation}\label{qt_halo_sp_mean}
Q_{\rm T,halo-corr}^{\rm Spiral}=<Q_{\rm T, halo-corr}^{{\rm Segment} \,i}>_{i\in[1,\rm n]},
\end{equation}
\begin{equation}\label{a2_sp_mean}
A_{2}^{\rm Spiral}=<A_{2}^{{\rm Segment} \,i}>_{i\in[1,\rm n]},
\end{equation}
where $n$ refers to the number of logarithmic segments fitted by \citet[][]{2015A&A...582A..86H}. 
In Appendix~\ref{Bar_strength_spiral_strength} we also make use of 
the mean amplitude of the two innermost and two outermost spiral segments.

There is a clear correlation between all the proxies of the spiral strength (Fig.~\ref{comparison_spiral_amp})\footnote{
\citet[][]{2016A&A...587A.160D} reported a bimodality in the comparison between $Q_{\rm b}$ and $A_{2}^{\rm max}$, which 
might be due to the sensitivity of $Q_{\rm b}$ to the enhanced radial forces 
by galactic bulges \citep[][]{2001A&A...375..761B,2002MNRAS.337.1118L,2016A&A...587A.160D}. 
As a matter of fact, no bimodality is present when $A_{2}^{\rm Spiral}$ is plotted versus $Q_{\rm T}^{\rm Spiral}$, 
which are measured at a radii where the bulge dilution should be much smaller.
}. 
At the radii at which these segments were measured 
the contribution of the dark matter halo to the overall radial force field might be significant, 
and thus it may be important to perform the halo correction: 
this explains the scatter in the comparison of $Q_{\rm T}^{\rm Spiral, halo-corr}$ versus $Q_{\rm T}$ 
(lower panel of Fig.~\ref{comparison_spiral_amp}). 
%
%
\begin{table*}
\centering
\small
\begin{tabular}{| c| c| c| c| c| c| c|}
\hline
\multicolumn{1}{|c}{\centering } & \multicolumn{1}{|c}{\centering Sample} & \multicolumn{1}{|c}{\centering $T\in [0,2)$} &
\multicolumn{1}{|c}{\centering $T\in [2,4)$} & \multicolumn{1}{|c}{\centering $T\in [4,6)$} & \multicolumn{1}{|c|}{\centering $T\in [6,8)$} & \multicolumn{1}{|c|}{\centering $T\in [8,10]$} \tabularnewline
\hline
\hline
 
$|\phi|_{\rm mean}$&All& 15.9 $\pm$   1.3 (19)& 17.8 $\pm$   0.7 (80)& 19.7 $\pm$   0.4 (159)& 21.2 $\pm$   0.8 (97)& 21.3 $\pm$   1.5 (36) \\
\hline
 
 &Barred& 16.7 $\pm$   1.6 (14)& 17.2 $\pm$   0.8 (55)& 19.7 $\pm$   0.6 (77)& 21.4 $\pm$   0.8 (90)& 21.4 $\pm$   1.6 (33) \\
\hline
 
 &Non-barred& 13.7 $\pm$   2.2 (5)& 19.1 $\pm$   1.2 (25)& 19.8 $\pm$   0.7 (82)& 19.0 $\pm$   3.0 (7)& 19.7 $\pm$   0.7 (3) \\
\hline
 
 &Grand-design& 13.6 $\pm$   1.6 (12)& 15.3 $\pm$   1.0 (27)& 18.1 $\pm$   1.5 (26)& 19.7 $\pm$   2.5 (10)&- \\
\hline
 
 &Multi-armed& 19.8 $\pm$   1.5 (6)& 18.5 $\pm$   0.7 (41)& 20.8 $\pm$   0.5 (85)& 25.3 $\pm$   1.3 (23)& 28.4 $\pm$   0.3 (2) \\
\hline
 
 &Flocculent&-& 20.8 $\pm$   2.5 (12)& 18.7 $\pm$   0.8 (48)& 19.9 $\pm$   1.0 (64)& 20.8 $\pm$   1.6 (33) \\
\hline
\hline
 
$|\phi|_{\rm inner}$&All& 21.3 $\pm$   2.3 (19)& 19.7 $\pm$   1.0 (80)& 21.5 $\pm$   0.7 (159)& 23.5 $\pm$   1.0 (97)& 23.2 $\pm$   1.7 (36) \\
\hline
 
 &Barred& 22.5 $\pm$   3.0 (14)& 18.6 $\pm$   1.3 (55)& 21.7 $\pm$   1.1 (77)& 23.9 $\pm$   1.1 (90)& 23.0 $\pm$   1.8 (33) \\
\hline
 
 &Non-barred& 17.7 $\pm$   3.2 (5)& 22.0 $\pm$   1.6 (25)& 21.3 $\pm$   1.0 (82)& 19.2 $\pm$   2.3 (7)& 24.6 $\pm$   3.3 (3) \\
\hline
 
 &Grand-design& 18.3 $\pm$   3.3 (12)& 16.0 $\pm$   1.7 (27)& 20.8 $\pm$   1.9 (26)& 21.9 $\pm$   3.7 (10)&- \\
\hline
 
 &Multi-armed& 26.8 $\pm$   2.5 (6)& 21.1 $\pm$   1.4 (41)& 22.2 $\pm$   1.0 (85)& 27.1 $\pm$   1.8 (23)& 38.2 $\pm$   1.8 (2) \\
\hline
 
 &Flocculent&-& 23.0 $\pm$   3.0 (12)& 20.7 $\pm$   1.2 (48)& 22.5 $\pm$   1.2 (64)& 22.4 $\pm$   1.7 (33) \\
\hline
\hline
 
$|\phi|_{\rm weighted}$&All& 19.2 $\pm$   1.5 (19)& 21.0 $\pm$   0.8 (80)& 22.3 $\pm$   0.5 (159)& 23.3 $\pm$   0.8 (97)& 24.1 $\pm$   1.5 (36) \\
\hline
 
 &Barred& 20.5 $\pm$   1.9 (14)& 20.6 $\pm$   1.0 (55)& 22.3 $\pm$   0.7 (77)& 23.4 $\pm$   0.9 (90)& 24.2 $\pm$   1.7 (33) \\
\hline
 
 &Non-barred& 15.8 $\pm$   2.2 (5)& 22.0 $\pm$   1.3 (25)& 22.3 $\pm$   0.7 (82)& 21.0 $\pm$   2.9 (7)& 22.8 $\pm$   1.5 (3) \\
\hline
 
 &Grand-design& 17.4 $\pm$   2.2 (12)& 18.5 $\pm$   1.3 (27)& 19.9 $\pm$   1.6 (26)& 23.2 $\pm$   3.1 (10)&- \\
\hline
 
 &Multi-armed& 22.7 $\pm$   1.6 (6)& 22.6 $\pm$   1.0 (41)& 23.9 $\pm$   0.5 (85)& 27.2 $\pm$   1.4 (23)& 35.9 $\pm$   2.8 (2) \\
\hline
 
 &Flocculent&-& 21.3 $\pm$   2.3 (12)& 20.8 $\pm$   1.0 (48)& 21.8 $\pm$   1.0 (64)& 23.2 $\pm$   1.6 (33) \\
\hline
\hline
\end{tabular}
\caption{
Average pitch angles (in degrees) as a function of Hubble stage.
It is presented the mean, the standard deviation of the mean, and the number of galaxies within the different $T$-type bins.
We study separately barred and non-barred galaxies, and also grand-design, multi-armed, and flocculent spirals.
}
\label{pitch_vs_ttype}
\end{table*}

\begin{table}
\centering
\small
\begin{tabular}{| c| c| c|}
\hline
\multicolumn{1}{|c}{\centering $R_{\rm pitch}/R_{25.5}$} & \multicolumn{1}{|c|}{\centering $<|\phi|>(^{\circ})$ (Barred)} & \multicolumn{1}{|c|}{\centering $<|\phi|>(^{\circ})$ (Non-barred)} \tabularnewline
\hline
\hline
0.1-0.2& 22.72 $\pm$  0.80& 20.75 $\pm$  0.70 \\
0.2-0.3& 20.91 $\pm$  0.61& 20.30 $\pm$  0.89 \\
0.3-0.4& 18.37 $\pm$  0.58& 16.76 $\pm$  0.82 \\
0.4-0.5& 18.08 $\pm$  0.79& 17.15 $\pm$  0.84 \\
0.5-0.6& 18.65 $\pm$  0.71& 19.01 $\pm$  1.03 \\
0.6-0.7& 18.50 $\pm$  0.80& 20.91 $\pm$  1.12 \\
0.7-0.8& 20.37 $\pm$  0.90& 18.76 $\pm$  1.16 \\
0.8-0.9& 17.73 $\pm$  1.36& 21.53 $\pm$  2.60 \\
\hline
\end{tabular}
\caption{
Average pitch angles as a function of radius.
It is presented the mean and the standard deviation of the mean in different radial ranges, normalized by the isophotal radii at the surface brightness 25.5~mag~arcsec$^{-2}$ in the $3.6\,\mu$m images ($R_{25.5}$).
We study separately barred and non-barred galaxies.
}
\label{pitch_vs_radius_R255}
\end{table}

\begin{table*}
\centering
\small
\begin{tabular}{| c| c| c| c|}
\hline
\multicolumn{1}{|c}{\centering $R_{\rm pitch}/h_{\rm R}$} & \multicolumn{1}{|c|}{\centering $<|\phi|>(^{\circ})$ (Grand-design)}  & \multicolumn{1}{|c|}{\centering $<|\phi|>(^{\circ})$ (Multi-armed)} & \multicolumn{1}{|c|}{\centering $<|\phi|>(^{\circ})$ (Flocculent)} \tabularnewline
\hline
\hline
0-1& 20.12 $\pm$  1.14& 22.32 $\pm$  0.72& 23.14 $\pm$  0.76 \\
1-2& 15.52 $\pm$  0.73& 20.72 $\pm$  0.50& 18.37 $\pm$  0.61 \\
2-3& 15.43 $\pm$  0.93& 19.00 $\pm$  0.54& 17.61 $\pm$  0.66 \\
3-4& 15.45 $\pm$  1.22& 21.14 $\pm$  0.77& 19.23 $\pm$  1.24 \\
4-5& 17.97 $\pm$  3.10& 21.42 $\pm$  1.38& 20.93 $\pm$  2.10 \\
\hline
\end{tabular}
\caption{
As in Table.~\ref{pitch_vs_radius_R255}, but defining the radial ranges with respect to the disk scale-length ($h_{\rm R}$).
}
\label{pitch_vs_radius_HR}
\end{table*}

%
%
%
\section{Spiral arms across the Hubble sequence}\label{pitches_hubble}
%
%
\subsection{Frequency of spiral types versus $T-$type}\label{spiral_types_ttype}
%
%
The distribution of the different arm classes as a function of the Hubble stage in our sample is shown in Fig.~\ref{histo_spiral_types_ttype} 
(upper panel). 
Flocculent arms are mainly found among late-type spirals (peak for Scd, i.e. $T=6$), 
multi-armed spirals mainly have intermediate $T-$types (peak for Sc, i.e. $T=5$), 
while grand-design galaxies are typically early-types (peak for Sb, i.e. $T=3$). 
All types of spirals are found for $3 \lesssim T \lesssim 6$.

In the lower panel of Fig.~\ref{histo_spiral_types_ttype} we show the bar fraction ($f_{\rm bar}$) for the different spiral types. 
$f_{\rm bar}$ is known to have a bimodal distribution versus $T$ in the S$^4$G, 
with a minimum (40-45 $\%$) for $T\approx 4$ (Sbc galaxies) \citep[][]{2015ApJS..217...32B,2016A&A...587A.160D}. 
This trend is maintained when different classes of spirals are studied separately.

The global bar fraction in our sample of spiral galaxies is $69 \pm 2.3\%$\footnote{We calculate 
binomial counting errors: $\Delta f_{\rm bar}=\sqrt{f_{\rm bar}\cdot (1-f_{\rm bar})/N_{\rm gals}}$, 
where $f_{\rm bar}$ refers to the bar fraction and $N_{\rm gals}$ to the total number of galaxies.}. 
It is slightly lower for multi-armed spirals ($59 \pm 3.9\%$) and 
gets higher for grand-design ($71 \pm 5.2\%$) and flocculent galaxies ($77 \pm 3.3\%$). 
Among the 133 galaxies with $T>5$, which are known to be faint ($M_{\ast}\lesssim 10^{10}M_{\odot}$) 
\citep[see e.g. Fig.~1 in][]{2016A&A...596A..25L}, the fraction of barred galaxies is very high for all kinds of spirals, 
namely $100 \%$ for grand-design, $96.0 \pm 3.9 \%$ for multi-armed, and $90.7 \pm 2.9 \%$ for flocculent spirals.
%
%
\subsection{Pitch angles versus $T-$type}\label{ttype_pitch}
%
%
For grand-design and multi-armed spirals, 
the global pitch angle increases with increasing $T$ (Fig.~\ref{Fig_pitch_angles}), as expected, but a large scatter is found. 
The pitch angle is independent of $T$ for flocculent spirals (Fig.~\ref{Fig_pitch_angles_fgm}), 
slightly flattening out the global distribution (see also Table~\ref{pitch_vs_ttype}). 
We do not find statistical and significant differences in the distribution of the pitch angle for barred and non-barred galaxies, 
regardless of spiral type or method used to estimate the average $|\phi|$.

In Fig.~\ref{sigma_pitch_angles_fgm} we show the internal scatter of $|\phi|$ 
versus $T$ for all types of spirals in our sample. 
We note that $\sigma_{|\phi|}$ is calculated, for each galaxy, from the pitch angle of the segments, 
whenever the host has more than one reliable measurement of $|\phi|$ ($98.5 \%$ of the galaxies in our sample). 
The distribution is fairly flat. The mean $\sigma_{|\phi|}$ is $9.5^{\circ} \pm 0.3^{\circ}$, 
regardless of the presence of a bar, with a mild decrease in the end of the Hubble sequence.

\section{Pitch angle as a function of radius}\label{pitches_rad}

Pitch angles in individual spiral galaxies are not necessarily constant 
as a function of galactocentric radius \citep[e.g.][]{1981AJ.....86.1847K} (see also Fig.~\ref{Figpitch}), 
and spiral arms are often asymmetric \citep[e.g.][]{2011ApJ...737...32E}. 
In Fig.~\ref{pitch_radial} we study the radial variation of the pitch angle for all the galaxies in our sample, 
showing all the individual measurements on logarithmic spiral segments 
(see also Tables~\ref{pitch_vs_radius_R255}~and~\ref{pitch_vs_radius_HR}). 
Pitch angles that were measured within the radial interval 
$[r_{i},{r_{i}}^{\prime}]$ are represented as a function of the boundaries $r_{i}$ and ${r_{i}}^{\prime}$, 
normalised by the disc size (traced by $R_{25.5}$ and $h_{\rm R}$). 

On average, the distribution of pitch angles barely change with radius (Fig.~\ref{pitch_radial}), regardless of spiral class (lower panel). 
The statistical trend is the same for barred and non-barred galaxies (upper panel). 
%
%
\section{Spiral properties as a function of bar strength}\label{spiral_bar}
%
%
%
\subsection{Pitch angle versus bar strength}\label{pitch_bar_strength}
\begin{figure}
\centering
\begin{tabular}{c c c}
\includegraphics[width=0.4\textwidth]{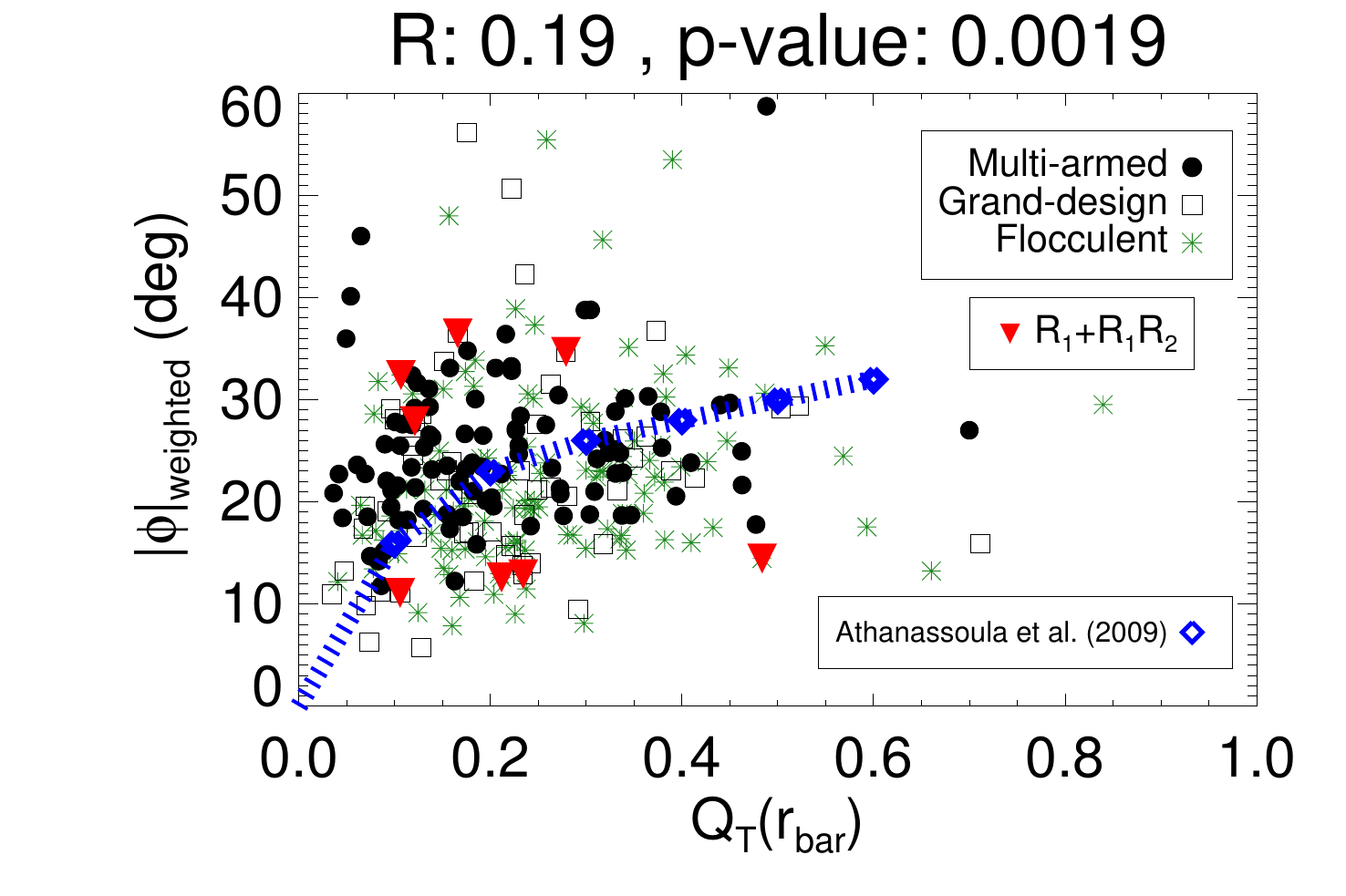}\\
\includegraphics[width=0.4\textwidth]{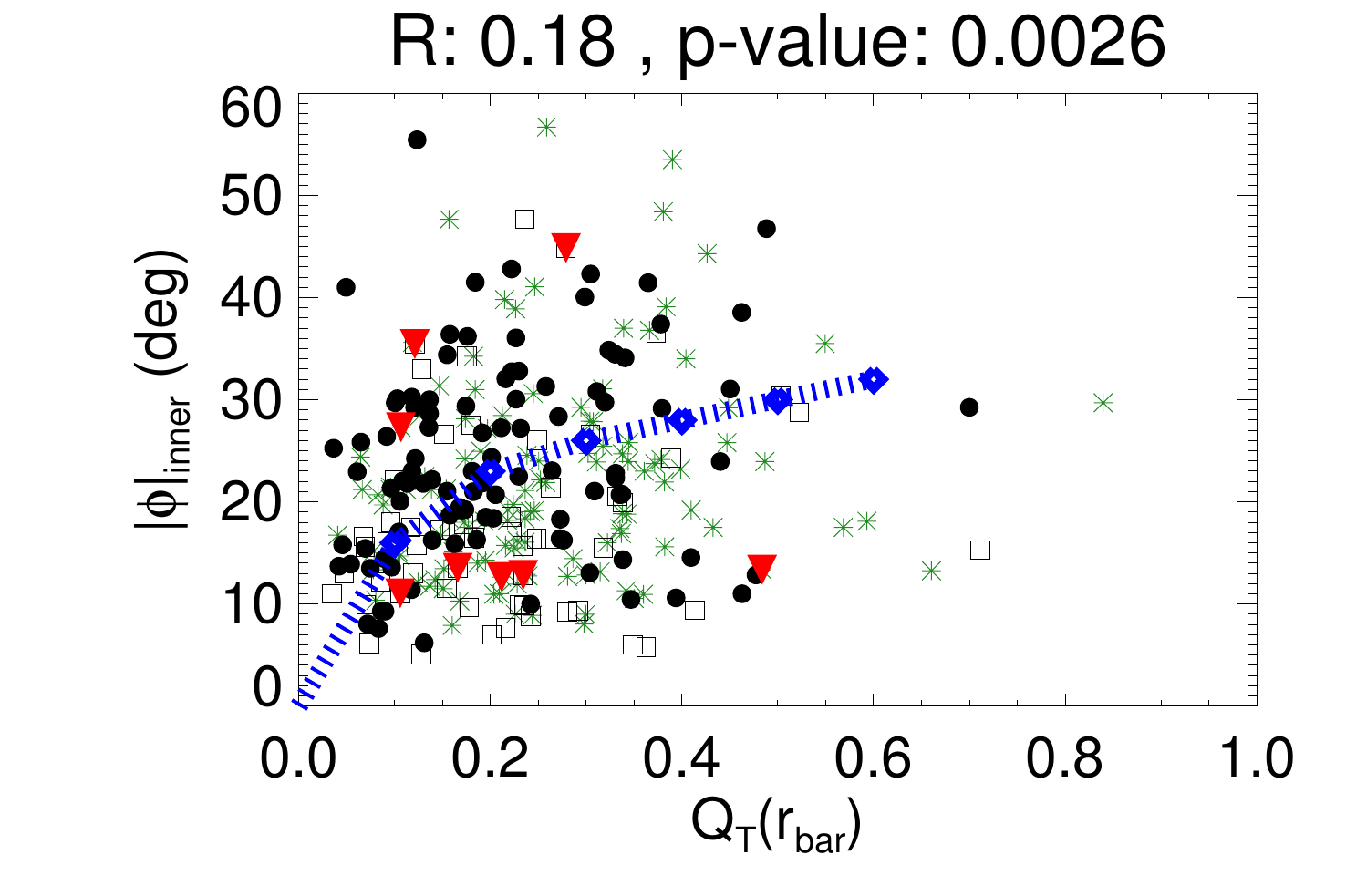}\\
\includegraphics[width=0.4\textwidth]{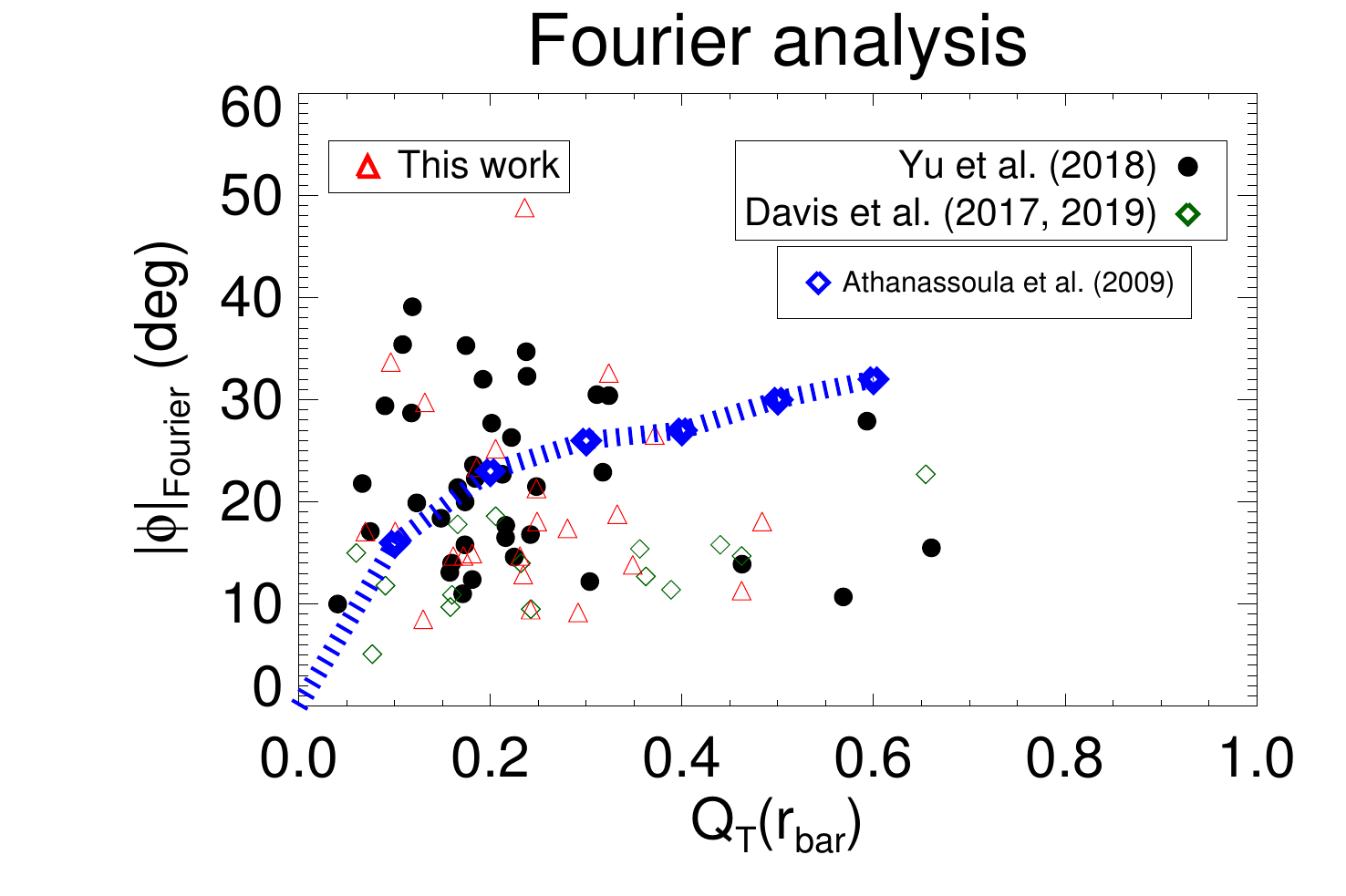}
\end{tabular}
\caption{
Weighted mean pitch angle (\emph{upper panel}) and mean of the innermost spiral segments (\emph{middle panel}) versus 
tangential-to-radial forces evaluated at the bar radius. 
The blue diamonds and line outline the trend in the simulations by \citet[][]{2009MNRAS.400.1706A} for $Q_{\rm T}(L_{1})$. 
With red triangles we show the few galaxies in our sample hosting outer rings of type $R_{1}$ or $R_{1}R_{2}$. 
On the \emph{lower panel} we show the same plot but using pitch angles obtained via 2-D Fourier analysis from this work and from the literature, 
using the measurements from \citet{2018ApJ...862...13Y} (black points), 
and from \citet{2017MNRAS.471.2187D, 2019ApJ...873...85D} (green diamonds).
}
\label{fig_bar_pitch}
\end{figure}

%
%
In Fig.~\ref{fig_bar_pitch} we probe a possible dependence of winding of the arms on the bar-induced gravitational torque, 
which is expected from the manifold theory (see Sect.~\ref{manifold}). 
We compare $|\phi|_{\rm weighted}$ and $|\phi|_{\rm inner}$ to the tangential-to-radial forcing evaluated 
at the bar radius: $Q_{\rm T}(r_{\rm bar})$ (upper and central panels). 
We also use measurements of $|\phi|_{\rm Fourier}$ (lower panel), from this work and from the literature, 
to make sure that the results are not biased by the employed methodology. Overlaid in the upper panels of Fig.~\ref{fig_bar_pitch} 
is the expected correlation (roughly traced from their Fig.~5) from the numerical models by \citet[][]{2009MNRAS.400.1706A}, 
but we note that they evaluated tangential-to-radial forces exactly at the $L_{1}$ point 
\citep[typically an unstable saddle point,][]{1987gady.book.....B}. $L_{1}$ and $L_{2}$ are hard to determine accurately in our images, 
since the kinematic data for our sample of galaxies is scarce and their dynamical modelling is beyond the scope of this paper. 

The average values of $|\phi|$ for a given $Q_{\rm T}(r_{\rm bar})$-bin obtained in this work are consistent with those in the simulations 
by \citet[][]{2009MNRAS.400.1706A}. 
However, we hardly find any dependence (statistical tests\footnote{The Spearman's rank correlation assesses 
the existence of a monotonic relation between two variables. 
$\rho=(-1)+1$ implies perfect (anti)correlation. Small p-values ($<0.01$) indicate significant correlation.} confirm a very weak correlation: 
$\rho \approx 0.18-0.2$, $p<0.01$ for $|\phi|_{\rm weighted}$, $|\phi|_{\rm mean}$, and $|\phi|_{\rm inner}$). 
For a better comparison with the aforementioned models, 
we analyse separately the eight galaxies in our sample hosting $R_{1}$ or $R_{1}R_{2}$ outer rings. 
Their morphology is reproduced with manifold calculations \citep[see Fig. 2 in][]{2009MNRAS.400.1706A} 
and their $L_{1}$ point location is most likely close to $r_{\rm bar}$. 
In spite of this, we do not find a clear trend between pitch angle and bar torque\footnote{
The low number of rings of type $R_{1}$, $R_{2}$, or $R_{1}R_{2}$ is most likely due to the bias of the S$^4$G towards late-type galaxies. 
We note that we do not sample galaxies hosting rings of type $R_{2}$ exclusively.}. 
$|\phi|_{\rm Fourier}$ is not found to correlate with bar strength either. 
We confirmed the lack of a strong correlation between bar strength and pitch angles when the maximum bar torque - $Q_{\rm b}$ - or 
the halo-corrected forces evaluated at the bar end - $Q_{\rm T}^{\rm halo-corr}(r_{\rm bar})$ - 
are used instead of $Q_{\rm T}(r_{\rm bar})$: 
we obtain correlation coefficients (significances) of 0.18 (0.006) and 0.17 (0.01), respectively, when using $|\phi|_{\rm mean}$ 
(and roughly the same values for $|\phi|_{\rm weighted}$ and $|\phi|_{\rm inner}$).
%
%
%
%
\begin{figure*}[h]
\begin{center}
\includegraphics[width=0.99\textwidth]{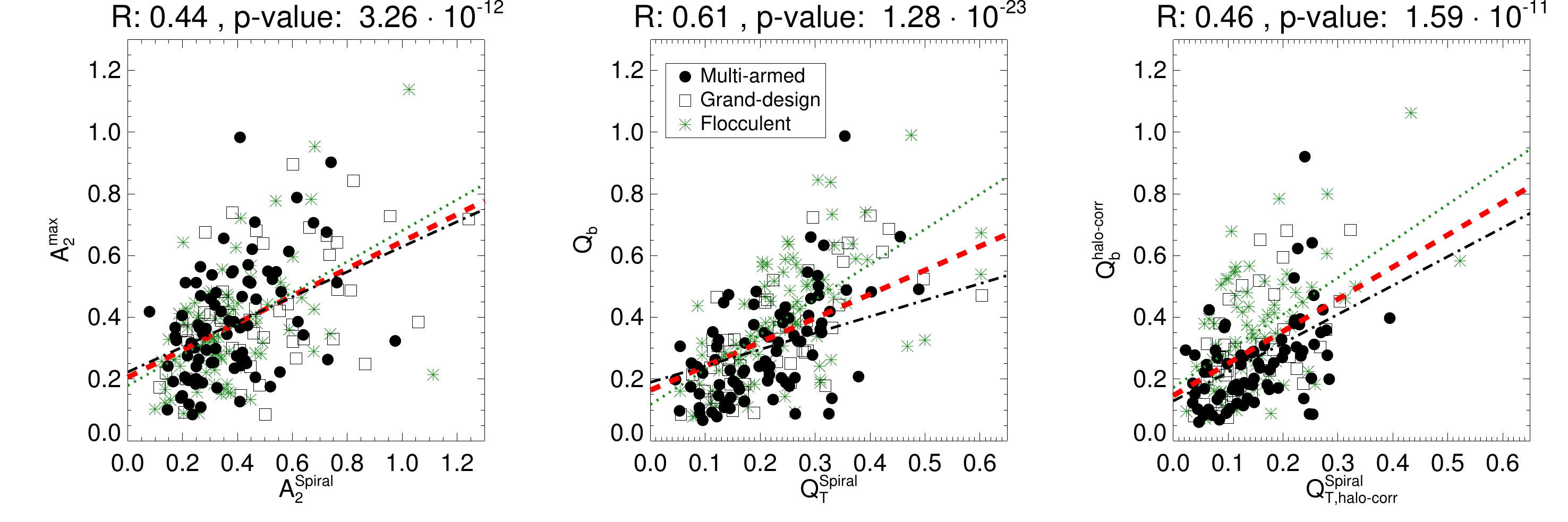}
\caption{
Bar strength \citep[from][]{2016A&A...587A.160D} as a function of spiral strength, 
measured from $m=2$ Fourier amplitudes (Eq.~\ref{a2_sp_mean}) (\emph{left}), 
from tangential-to-radial forces (Eq.~\ref{qt_sp_mean}) (\emph{middle}), 
also including the correction for the halo dilution (Eq.~\ref{qt_halo_sp_mean}) (\emph{right}). 
Different types of spirals are plotted with different symbols, as indicated in the legend. 
The red dashed line shows the linear fit to the cloud of points 
(in green and black we show the fit for flocculent and grand-design+multi-armed, respectively).
}
\label{spiral_bar_strength}
\end{center}
\end{figure*}
%
%
\begin{figure}[h]
\begin{center}
\includegraphics[width=0.35\textwidth]{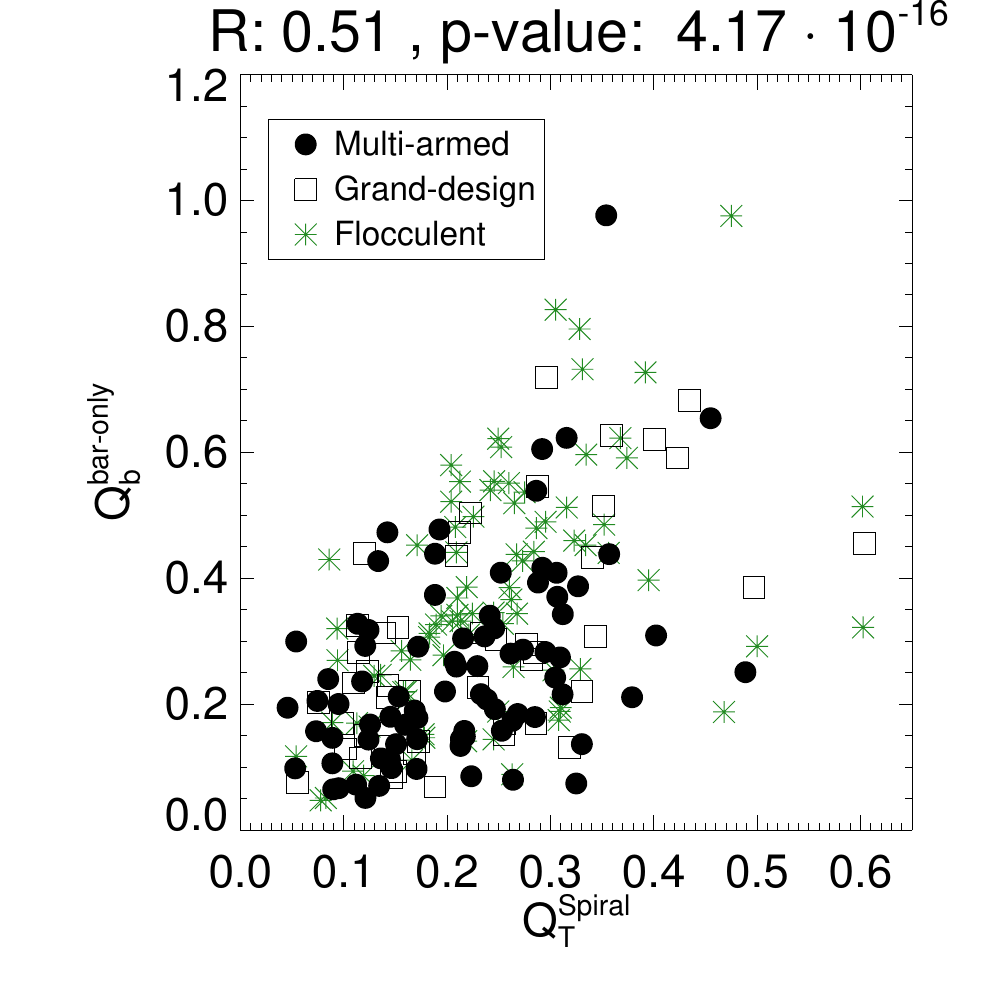}
\caption{
Bar-only force versus spiral strength (Eq.~\ref{qt_sp_mean}). 
The colour palette and symbols are the same as in Fig~\ref{spiral_bar_strength}.
}
\label{pitch_vs_spiral_strength_qb_baronly}
\end{center}
\end{figure}
%
%
\begin{figure*}[h]
\begin{center}
\includegraphics[width=0.99\textwidth]{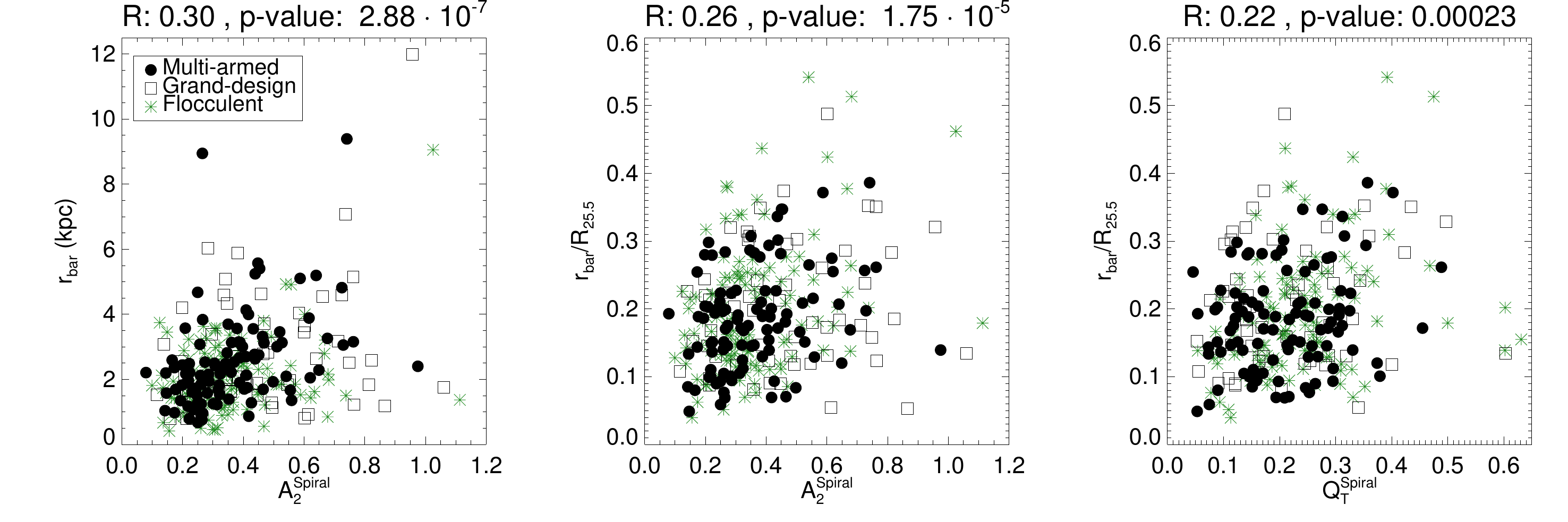}
\caption{
Bar size, in physical units (\emph{left}) and normalised to the disc size (\emph{central and right panels}), 
as a function of the spiral strength, measured from the $m=2$ Fourier amplitude and from the tangential-to-radial forces. 
The colour palette and symbols are the same as in Fig~\ref{spiral_bar_strength}.
}
\label{spiral_bar_length}
\end{center}
\end{figure*}
%
%
\begin{figure}[h]
\begin{center}
\includegraphics[width=0.35\textwidth]{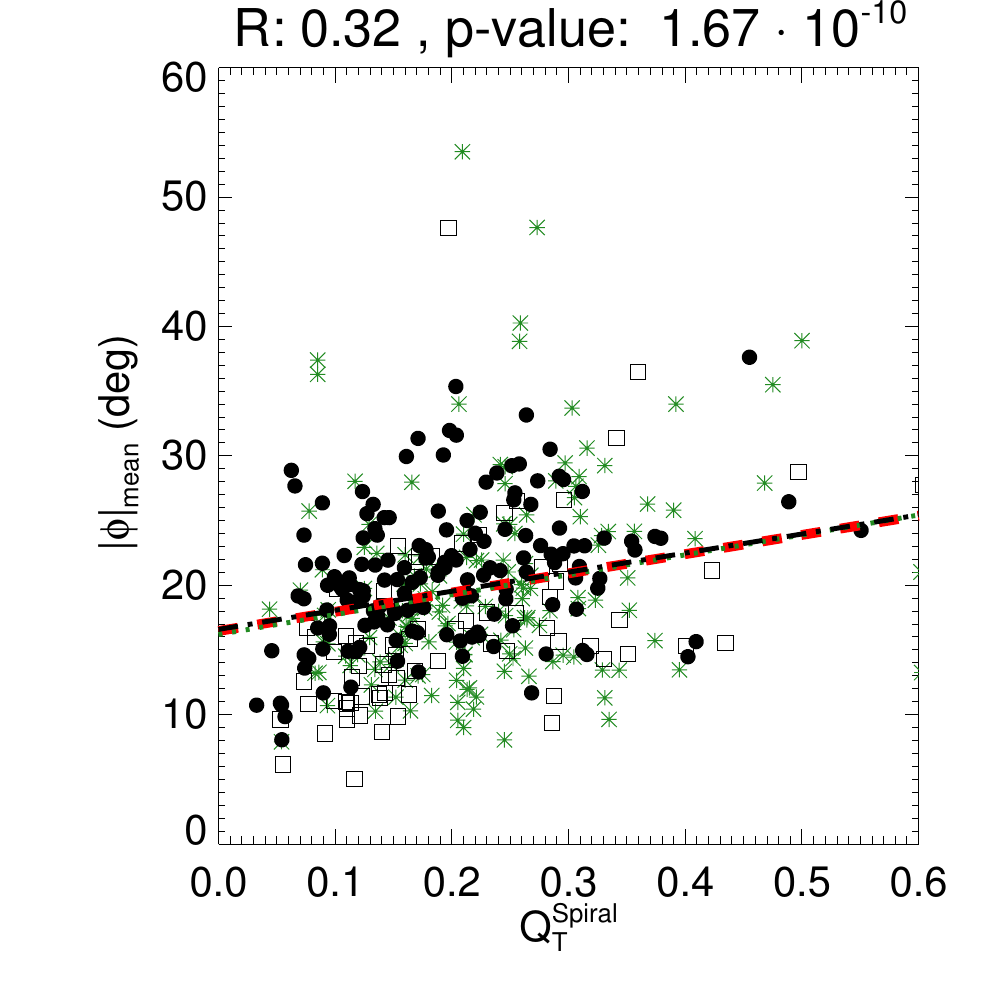}
\caption{
Mean pitch angle as a function of spiral strength, measured from tangential-to-radial forces (Eq.~\ref{qt_sp}). 
The colour palette and symbols are the same as in Fig~\ref{spiral_bar_strength}. 
The red dashed line represents the linear fit to the data cloud, 
while the green and black lines show the fit for flocculent and grand-design+multi-armed galaxies, respectively.
}
\label{pitch_vs_spiral_strength}
\end{center}
\end{figure}
%
%
\subsection{Spiral strength versus bar strength}\label{spiral_bar_strength_sect}
%
%
In order to shed more light on the coupling between bars and spirals, 
here we assess, with unprecedented sample size, the possible dependence between their strengths, 
which has been extensively debated in the literature to this day (see Sect.~\ref{bars_spirals_amplitude}). 
The amplitude of spirals in the S$^4$G is correlated with bar strength (Fig.~\ref{spiral_bar_strength}). 
This is clear for all proxies of the strength of spirals 
($A_{2}^{\rm Spiral}$, $Q_{\rm T}^{\rm Spiral}$, $Q_{\rm T,halo-corr}^{\rm Spiral}$), 
shown in the different panels (see Spearman's correlation coefficient and significance above the panels). 
In Fig.~\ref{pitch_vs_spiral_strength_qb_baronly} we show that the correlation is clear 
when the bar strength is estimated from $Q_{\rm b}^{\rm bar-only}$ (spiral contribution to bar forcing suppressed).
	
The correlation between bars and spirals holds true when different types of spirals are studied separately. 
Interestingly, when using $Q_{\rm T}$ and $A_{2}$, 
the trend for flocculent spirals is fairly similar as for grand-design and multi-armed. 
We note that the effect of the correction for the halo-dilution on the force profiles is slightly 
larger for $T\gtrsim 5$ \citep[][]{2016A&A...587A.160D}. 
Thus, the difference between $Q_{\rm T, halo-corr}^{\rm Spiral}$ and $Q_{\rm T}^{\rm Spiral}$ 
is more pronounced in flocculent spirals (typically hosted by late-type galaxies, see previous Sect.~\ref{spiral_types_ttype}) 
than in other types of spirals.

Likewise, we find that the strength of the spirals increases with increasing bar size (Fig.~\ref{spiral_bar_length}), 
regardless of whether $r_{\rm bar}$ is measured in physical units or relative to the disc size. 
In addition, the gravitational torques associated with the spirals (with and without halo-correction) 
increase with increasing pitch angle (Fig.~\ref{pitch_vs_spiral_strength}, 
see also Fig.~\ref{pitch_vs_spiral_strength_mean_all_segments}): 
this is somehow expected, as more tightly-wrapped spirals tend to be more axisymmetric than loosely-wound arms.
%
%
\section{Pitch angle and central mass concentration}\label{BH_mass_sec}
%
%
The winding of the spiral arms has been claimed to depend on the prominence of the bulge \citep[e.g.][]{1981AJ.....86.1847K}, 
the steepness of the rotation curve \citep[e.g.][]{2005MNRAS.359.1065S}, 
and on the mass of the central supermassive black holes ($M_{\rm BH}$) \citep[][]{2017MNRAS.471.2187D}. 
The small scatter in the latter scaling relation would make the pitch angle suitable for the prediction of $M_{\rm BH}$ even in bulge-less galaxies. 
Here we assess these scaling relations in the S$^4$G survey.
%
%
\subsection{Pitch angle versus $M_{\rm BH}$}
%
%
We estimate $M_{\rm BH}$ from the central velocity dispersion ($\sigma_{\ast}$) 
from HyperLEDA (available for 117 galaxies in our sample) and the calibration from \citet[][]{2009ApJ...698..198G}:
\begin{equation}
{\rm log_{10}}(M_{\rm BH})=8.12+4.24\cdot {\rm log_{10}}(\sigma_{\ast}/200).
\end{equation}
In order to have an estimate of the uncertainties in $M_{\rm BH}$, 
for a subsample of S$^4$G galaxies with disc inclinations lower than 65$^{\circ}$, 
we checked that the inferred black hole masses are consistent with other measurements in the literature 
\citep[e.g. compilations by][]{2013ApJ...776...50C,2013ApJ...764..151G,2019arXiv190304738S,2019ApJ...873...85D}, 
using direct measurements and also the $M_{\rm BH}-\sigma_{\ast}$ relation (Fig.~\ref{blackholescomparison}), 
but we find scatter and outliers (e.g. NGC$\,$5055) in the comparison that emphasise the uncertainties in our analysis.

We do not find a strong dependence between pitch angle and $M_{\rm BH}$ (Fig.~\ref{bh_mass}): 
correlation coefficients (significances) of $-0.24$ $(0.008)$, $-0.10$ $(0.27)$, 
and $-0.03$ $(0.78)$ for $|\phi|_{\rm mean}$, $|\phi|_{\rm inner}$, and $|\phi|_{\rm weighted}$, respectively. 
Only a weak correlation is detected with $|\phi|_{\rm mean}$, that becomes slightly stronger ($\rho=-0.34$ and $p=0.0008$) 
when only grand-design and multi-armed galaxies are considered (for the flocculent spirals alone, no correlation is seen whatsoever). 
Since the correlation is weak, the scatter in $M_{\rm BH}$ for a given $|\phi|$ (about two decades) is much too large for this relation to 
yield any useful predictions concerning black hole mass. 
This contradicts previous reports in the literature \citep[e.g.][]{2017MNRAS.471.2187D}: 
this could be explained by their use of smaller samples, 
but we note the lack of clear correlation can be due to the uncertainty in $|\phi|$ or to the use of indirect estimates of $M_{\rm BH}$.
%
%
\subsection{Pitch angle versus $d_{\rm R}v_{\ast}(0)$}
%
%
\citet[][]{2016A&A...587A.160D} calculated the inner gradient of the stellar 
component of the rotation curve ($d_{\rm R}v_{\ast}(0)$ hereafter). 
They performed a polynomial fit to the inner part of the disc+bulge component of the rotation curve ($V_{\rm 3.6\mu m}$) 
and took the linear term as an estimate of the inner slope, following \citet[][]{2013MNRAS.433L..30L}. 

Here, we use $d_{\rm R}v_{\ast}(0)$ as a proxy of the central stellar mass concentration\footnote{
\citet[][]{2016PhDT.......168D} showed that $d_{\rm R}v_{\ast}(0)$ tightly correlates with $M_{\ast}$ 
(and thus with the maximum circular velocity), $T$, and the bulge-to-total mass ratio (see their Figure 4.6). 
This was also shown by \citet[][]{2016MNRAS.458.1199E} for a subsample of 29 S$^4$G galaxies.}
\citep[e.g.][]{2016MNRAS.458.1199E,2016A&A...587A.160D,2017ApJ...835..252S}. 
We show that the pitch angle of the galaxy is not dependent on $d_{\rm R}v_{\ast}(0)$ (right panel of Fig.~\ref{bh_mass}). 
%
%
\section{Discussion}\label{Discussion}

Here we discuss the properties of spiral arms based on the study of their pitch angles and amplitudes in the S$^4$G survey, 
as presented in previous sections. 
Particular attention is given to the theoretical coupling between bars and spirals, which is assessed with observations. 
All types of spirals are included in our analysis. 
We take advantage of the wealth in late-type galaxies hosting flocculent arms in the S$^4$G. 

In this work we use near-IR (3.6~$\mu$m) imaging, which is known to be an excellent tracer of the old stellar populations. 
The pitch angle of a galaxy can vary depending on the used wavelength \citep[e.g.][]{2014ApJ...793L..19M}. 
According to \citet[][]{2016ApJ...827L...2P}, the typical difference between 
the pitch angle measured at $3.6$~$\mu$m and $7$~$\mu$m is $\sim 3.75^{\circ}$. 
Recent work by \citet[][]{2018ApJ...869...29Y} reported a mild decrease of the pitch from red (including $3.6$~$\mu$m) to blue passbands, 
which is in agreement with the expectations from the density wave theory. 

The classification of spiral types is not strongly dependent on wavelength: for instance, 
\citet[][]{2011MNRAS.414..538K} showed that there is a correspondence between galaxies being grand-design in infrared and optical passbands. 
\citet[][]{1984ApJS...54..127E} had claimed that old stellar waves could not be found in patchy spirals, 
proposing that flocculent spirals are mainly built of star-forming regions \citep[][]{1979ApJ...233...56S,1981ApJS...47..229E}. 
Recent work by \citet[][]{2011ApJ...737...32E} studied 46 galaxies in the S$^4$G and 
showed that most of the galaxies that are flocculent in optical wavelengths can also be classified as flocculent at 3.6~$\mu$m. 
They also showed that, in the S$^4$G, early-type spiral galaxies tend to be multi-armed and grand-design, 
having underlying stellar waves, whereas their late-type counterparts host flocculent spirals. 
We confirm this in Sect.~\ref{pitches_hubble} using a factor of $\sim 8$ bigger sample \citep[see also][]{2017MNRAS.471.1070B}. 
These trends in the Hubble sequence are probably linked to the report by \citet[][]{2016MNRAS.461.3663H} that 
galaxies hosting many arms - typically multi-armed and flocculent spirals - in the Galaxy Zoo 2 \citep[][]{2013MNRAS.435.2835W} 
have bluer colours than their two-armed counterparts (i.e. the grand-design).

It is well known that the majority of galaxies present spiral arms that cannot be fitted with a single pitch angle 
\citep[e.g.][]{2013MNRAS.436.1074S,2015A&A...582A..86H}. 
In fact, in Sect.~\ref{spiral_types_ttype} we showed that the internal variation of $|\phi|$ in galaxies is approximately $10^{\circ}$ on average, 
and the differences in certain cases can be larger, even by a factor of two. 
Thus, by estimating global winding angles one misses information on individual segments, 
in the present paper and in most of the work in the literature.

In Sect.~\ref{pitches_rad} we also showed that the distribution of $|\phi|$ does not change with increasing radius, in a statistical sense, 
for all types of spirals. This implies that in discs with many logarithmic spiral segments $|\phi|$ can either get larger or smaller 
with increasing radius, and thus the arms do not systematically wind up or loosen in the galaxy outskirts. 
We are in an eclectic position regarding the radial variation of $|\phi|$ discussed in the literature: 
\citet[][]{2013MNRAS.436.1074S} and \citet[][]{2015ApJ...802L..13D} showed 
that the spiral pitch angles increase with decreasing radius, 
but an opposite trend was found by \citet[][]{2014ApJ...790...87D}.

In this work we propose the average pitch angle of galaxies as a global measure of the global winding angle, 
including a weighting for the relative arc length of the logarithmic segments (Sect.~\ref{pitch_met}), 
so that we can study its dependence on global properties of the host galaxies. 
Overall, we find a good agreement between the values of the global $|\phi|$ determined with different methods (Fig.~\ref{Fourier_visual}) and 
from the literature (Fig.~\ref{fourier_comparison_literature} and Fig.~\ref{mean_vs_median_literature}). 
However, we also show scatter and outliers in the comparison, 
mainly caused by the differences in the employed methodology and its sensitivity to radial variations of $|\phi|$, 
and to asymmetrical or decoupled spiral arms.
%
%
\subsection{Observational constraints on spiral formation: pitch angle and disc properties}\label{cmc_spirals}%
%
%
The pitch angle of the spirals in the S$^4$G increases with increasing $T$ for grand-design and multi-armed galaxies, on average, 
unlike in the case of flocculent galaxies (Sect.~\ref{ttype_pitch}). The dispersion for each $T$-type is quite large: 
this is somewhat unexpected, since the winding angle of the arms is a criterion for morphological classification 
in the de Vaucouleurs Hubble-Sandage system \citep[][]{1959HDP....53..275D}, 
applied by \citet[][]{2015ApJS..217...32B} to our sample. 
The lack of strong correlation might be due to uncertainties in 
the $T-$type determination or the asymmetry of the spiral arms. 
It might also be a consequence of the larger weight given to other criteria to classify galaxies (e.g. smoothness or bulge prominence).
The weak, or missing, correlation between pitch angle and $T$-type has been previously reported in the literature, 
and is reviewed in this section. 

Early-type galaxies are expected to host more tightly wound spirals on average 
\citep[e.g.][]{1988A&AS...76..365C,1992A&AS...93..469P,1993A&AS..100..431G}, but counterexamples are known. 
Using $K$-band imaging for 45 face-on spirals, \citet[][]{1998MNRAS.299..685S} 
found that their measurements of arm pitch angle did not correlate with Hubble type (obtained from optical images). 
They argued that this was due to the fact that the morphology of the old stellar population bears little 
resemblance to the optical morphology used to classify galaxies. 
In this work we find that the relation between the global pitch angle and the $T-$type has a lot of scatter and is not very tight, 
even when 3.6 $\mu$m imaging was used for both the morphological classification and the pitch angle measurements. 

Early work by \citet[][]{1981AJ.....86.1847K} using 113 galaxies showed that the pitch angle correlates only in an average sense with $T-$type, 
and there is a large spread in the values. \citet[][]{1981AJ.....86.1847K} 
also reported that the shape of the spirals is mainly determined by kinematic parameters (absolute rotation velocity). 
A correlation of pitch angle and the fraction of light in the disc component has been predicted by theoreticians 
\citep[e.g.][]{1964ApJ...140..646L,1975ApJ...196..381R}. 
Also, \citet[][]{2004AJ....128..183B} discussed that open spirals imply high disc-to-halo mass ratios. 
Nevertheless, we checked and confirmed that the pitch angle does not depend on the total dynamical mass or halo-to-stellar mass ratio, 
derived from inclination-corrected 21 cm line widths \citep[see Fig.~11. and Fig.~C.1. in][]{2019A&A...625A.146D}.

\citet[][]{1981AJ.....86.1847K} also showed that the spiral arms might not be either purely logarithmic (associated to a density wave origin) 
or hyperbolic (linked to a tidal origin). The environment cannot explain the formation of all spirals in all disc galaxies in the local universe, 
since spiral structure in the near-IR has been found in isolated galaxies \citep[][]{2011MNRAS.414..538K}. 
However, \citet[][]{1998MNRAS.299..685S} argued that tidal effects 
from near neighbours seem to have the effect of enhancing the amplitude of spirals. 
Analysis of Fig.~11. and Fig.~C.1. in \citet[][]{2019A&A...625A.146D} reveals that the 
pitch angle in the S$^4$G does not depend on the tidal interaction strength between galaxies or 
with the projected surface density of galaxies, taken from \citet[][]{2014MNRAS.441.1992L} \citep[see also][]{2016MNRAS.461.3663H}. 
In itself this does not rule out the influence of tidal perturbations on spiral formation: 
\citet[][]{2000MNRAS.319..393S} showed in terms of $N-$body simulations of M$\,$51 that, 
excluding the outer bridge and tail, the transient spiral density waves excited by 
interaction were very similar to those the model would generate in isolation.
%
%
\begin{figure}[h]
\begin{center}
\includegraphics[width=0.5\textwidth]{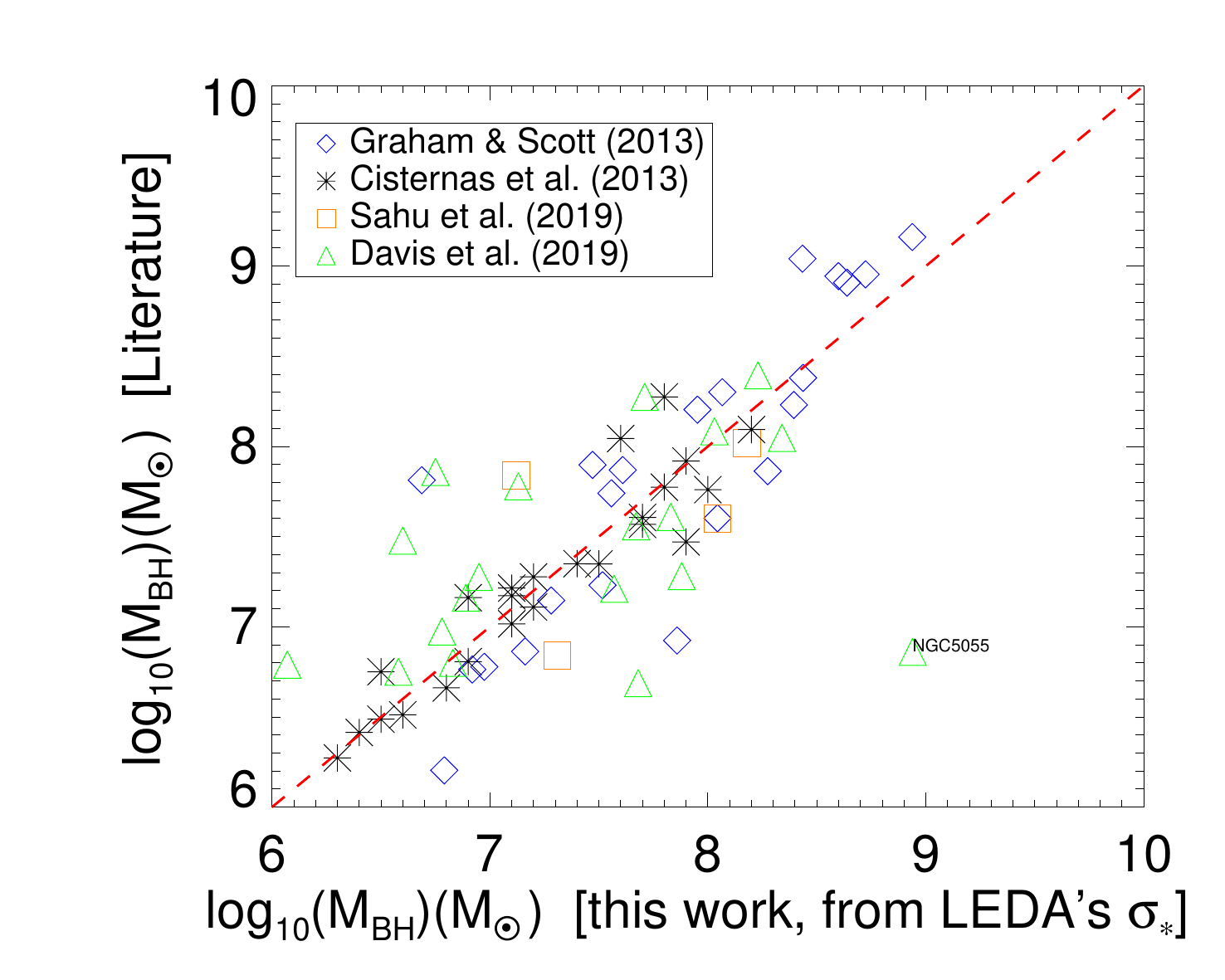}
\caption{
For a sample of S$^4$G disc galaxies with inclinations lower than 65$^{\circ}$, we show a
comparison of direct measurements of black hole masses from the compilations by 
\citet[][]{2013ApJ...764..151G,2019arXiv190304738S,2019ApJ...873...85D} -  
and also from \citet[][]{2013ApJ...776...50C} (obtained via $M_{\rm BH}-\sigma_{\ast}$ relation and using direct measurements) - with 
the ones estimated using LEDA's central velocity dispersions. In red, we show the straight line $y=x$.
}
\label{blackholescomparison}
\end{center}
\end{figure}
%
%
\begin{figure*}
\centering
\includegraphics[width=0.49\textwidth]{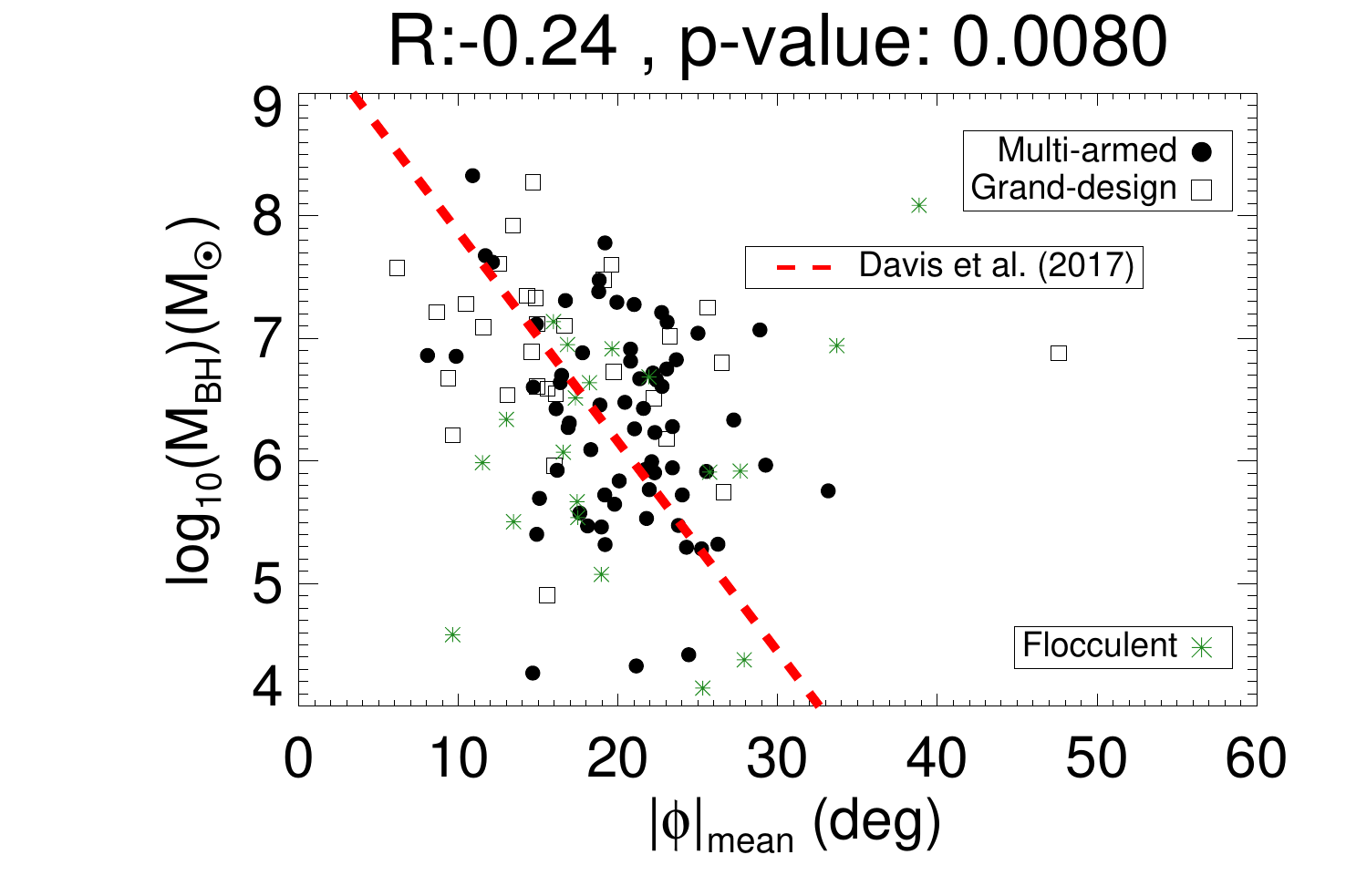}
\includegraphics[width=0.49\textwidth]{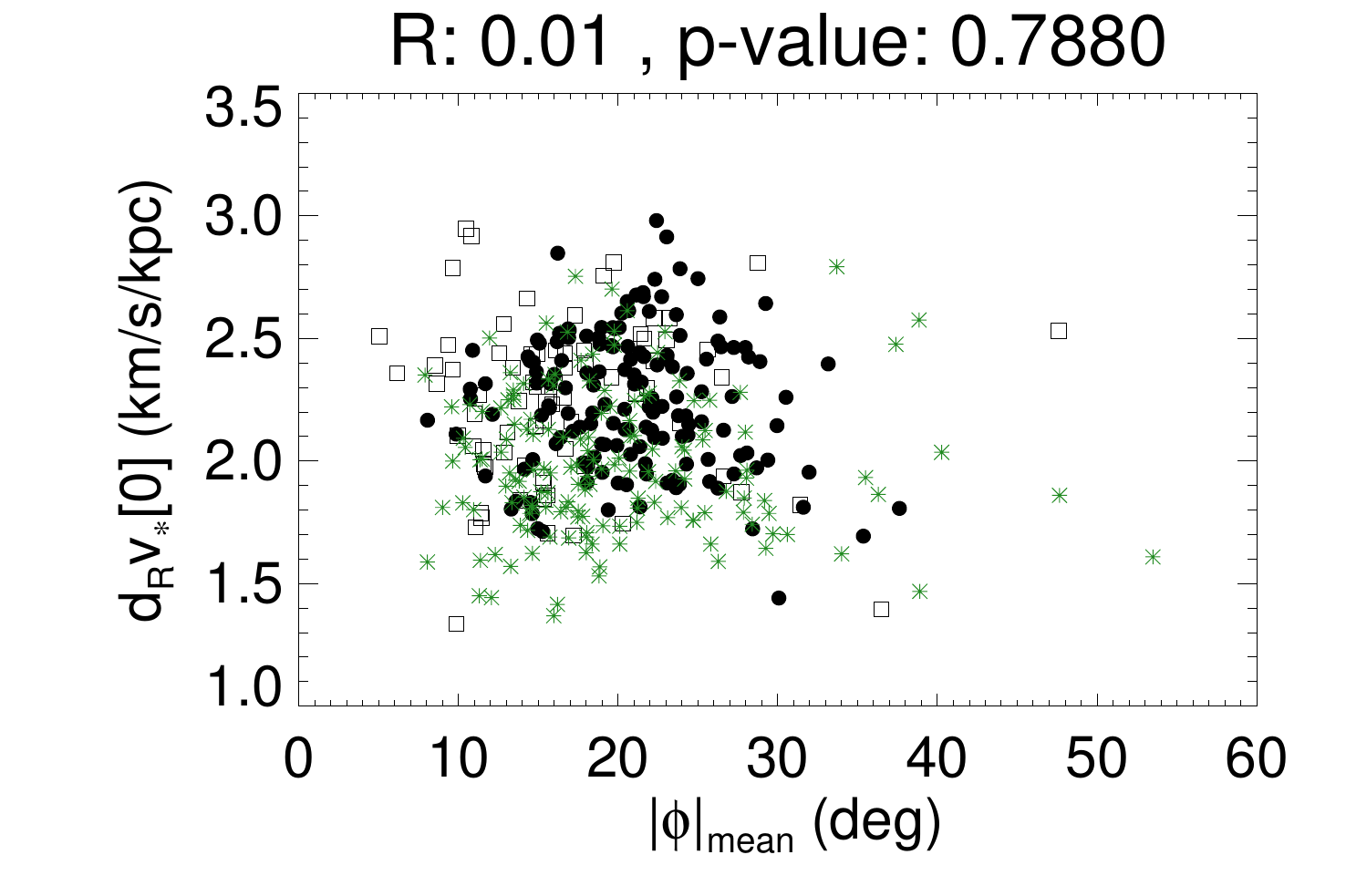}\\
\caption{
Mass of supermassive black holes (\emph{left}) (estimated from central stellar velocity dispersions) 
and inner slope of the stellar component of the rotation curve (\emph{right}) 
(tracer of central stellar mass concentration) as a function of mean pitch angle. 
The red dashed straight line (left panel) corresponds 
to the fit from the tight correlation presented by \citet[][]{2017MNRAS.471.2187D}: 
their values are consistent with ours, but we do not find such a strong dependence of the pitch angle on $M_{\rm BH}$. 
Different colours correspond to different spiral types, as indicated in the legend. 
}
\label{bh_mass}
\end{figure*}
%
%

Observations of disc galaxies at 2.1 $\mu$m by \citet[][]{1994A&A...288..365B} lead the authors to conclude that 
large-scale structure of spirals is mainly intrinsic (i.e. not caused by tidal interactions or by the environment). 
They claimed that the stellar disc temperature, the gas content, 
and the disc mass are the main agents determining the spiral morphology. 
In addition, \citet{2015ApJ...802L..13D} reported the existence of a fundamental plane of spiral structure in disc galaxies, 
determined by the pitch angle, the density of H{\sc\,i} within the disc, and the mass of the stellar bulge. 

A dependence between spiral arm pitch angles and the central mass concentration is expected from the density wave theory 
\citep[][]{1964ApJ...140..646L,1966PNAS...55..229L}. 
\citet[][]{1981AJ.....86.1847K} found that the spiral winding angle correlates only weakly with bulge-to-disc ratio. 
Recent work by \citet[][]{2019ApJ...871..194Y} using 79 galaxies from the CALIFA survey \citep[][]{2012A&A...538A...8S} 
shows a dependence of pitch angles on the prominence of bulges, the galaxy concentration, and the total stellar mass, with a large scatter. 
\citet[][]{2019MNRAS.tmp.1111M} performs the analysis of the spiral arms in the Galaxy Zoo project 
\citep[][]{2011MNRAS.410..166L,2013MNRAS.435.2835W} 
and reports a lack of strong correlation between bulge prominence and spiral arm tightness: 
galaxies with small bulges show a wide range of winding angles, while those with prominent bulges host more tightly wound arms. 
The authors speculate that the spiral arms are not static density waves, but rather re-formed structures that tend to wind-up in time. 
A link between $|\phi|$ and the relative bulge mass would also imply a correlation of the former with the disc mass fraction: 
such a correlation was not found by \citet[][]{2019MNRAS.482.5362F}, 
based on the analysis of a sample of 79 disc galaxies with estimates of the corotation radius and the pattern speed of bars and spirals.

In this paper we do not find a dependence of the pitch angle on central stellar mass concentration (Sect.~\ref{BH_mass_sec}, Fig.~\ref{bh_mass}). 
We checked and confirmed that the pitch angle does not depend on bulge-to-total mass ratio and bulge mass, 
estimated from 2-D decompositions models by \citet[][]{2015ApJS..219....4S}. 
We also reassessed the dependence of the pitch on gas fraction and gas density (Appendix~\ref{gas_pitch}), 
finding a weak correlation that holds for grand-design and multi-armed galaxies, but vanishes for flocculent spirals.

A correlation between spiral arm pitch angle and galactic shear rate has been found in $N$-body simulations by \citet[][]{2013A&A...553A..77G} 
and recently confirmed by \citet[][]{2018MNRAS.477.1451F}\footnote{
A similar dependence between local shear rate and pitch angle has also been demonstrated in collisional $N$-body simulation 
of Saturn ring's self-gravity wake structures, which arise from superposition of swing amplified noise in a disc with Toomre parameter 
$Q<2-3$ \citep[][]{2018prs..book..434S}: such structures agree very well with the stellar dynamical predictions of \citet[][]{1966ApJ...146..810J} 
and can be viewed as analogous of flocculent spiral segments.
}. Possible observational confirmation was provided by \citet[][]{2006ApJ...645.1012S}, 
who claimed that the dark matter halo profiles could be constrained using spiral arm morphology. 
Recent work by \citet[][]{2018ApJ...862...13Y} shows that the pitch angle correlates weakly, if at all, with the shear rate. 
\citet[][]{2018ApJ...862...13Y} actually reproduced the same plot from \citet[][]{2006ApJ...645.1012S} using the same sample and 
the same shear rate estimates, but their own measurements of pitch angle, finding no clear correlation. 
We note that \citet[][]{2006ApJ...645.1012S} found a connection between pitch and shear only 
when the latter was measured at a certain physical radius (10 kpc). 
The connection vanished when such radii were chosen as a function of the disc scale length. 
A correlation between spiral pitch angle and shear would imply an intimate link between spiral winding, central mass concentration, 
and dark matter haloes, which we cannot confirm in the S$^4$G. 
In Appendix~\ref{spirals_all_shear} we checked and confirmed the lack of correlation between shear and pitch angle for a subsample of 17 galaxies with 
available high-resolution rotation curves.

The whole picture for the formation of spiral arms remains unclear. 
An interpretation that disentangles the physical mechanisms that drive the formation and evolution of spiral arms 
when several observational parameters come into play is far from established. 
We conclude that, in the S$^4$G, the pitch angle of spirals does not obey any definite scaling relation involving disc fundamental properties 
such as bulge prominence, relative dark matter content, gas fraction, stellar and dynamical mass, or environmental parameters.
%
%
\subsection{Relation between the mass of the central supermassive black hole and the pitch angle of the spiral arms}\label{bhm_bulge_spirals}
%
%
Supermassive black holes are present in the centres of massive galaxies 
\citep[see e.g.][]{1971MNRAS.152..461L,1995ARA&A..33..581K,2002Natur.419..694S,2009ApJ...692.1075G}, 
with masses spanning several orders of magnitude. 
Recent pioneering work by the 
\citet[][]{2019ApJ...875L...1E} produced the first image of the event horizon 
of a supermassive black hole in the centre of M$\,87$, caught in the act of swallowing material.

\citet[][]{2008ApJ...678L..93S} found a relationship between spiral arm pitch angle and central supermassive black hole mass ($M_{\rm BH}$). 
Further evidence was provided by \citet[][]{2013ApJ...769..132B} and, more recently, by \citet{2017MNRAS.471.2187D}, who studied 44 spiral galaxies 
with measurements of $M_{\rm BH}$, and reported a surprisingly tight correlation between $M_{\rm BH}$ and $|\phi|$. 
If real, this could potentially explain previous findings in the literature of the dependence of the pitch and 
the bulge mass \citep[e.g.][]{1975A&A....44..363Y}, 
given the known scaling relation between central black holes and bulges \citep[e.g.][]{1995ARA&A..33..581K}. 
The tight correlation between $|\phi|$ and $M_{\rm BH}$ would allow for the prediction of supermassive black hole mass even in bulge-less galaxies. 

In Sect.~\ref{BH_mass_sec} we used measurements in the literature of central stellar velocity dispersions to 
estimate the mass of the supermassive black hole hosted by the spiral galaxies in our sample. 
Although these are less direct estimates of $M_{\rm BH}$, we have a factor of 2 bigger sample. 
For a subsample of galaxies, we confirmed the fairly good agreement between direct and indirect estimates of black hole masses. 
Our values are consistent with the aforementioned trend, but we find no clear link between black hole mass and pitch angle 
(only a weak correlation is found for $|\phi|_{\rm mean}$, confirmed by statistical tests). 
Recent work by \citet[][]{2019MNRAS.484..814G} reassessed the trend between $\phi$ and central velocity dispersion 
of the same sample of galaxies (see open black stars and circles in their Fig.~5), and the resulting scaling relation 
seems to be weaker.

Although the mean pitch angle correlates weakly with supermassive black hole mass (inferred from $\sigma_{\ast}$) 
we conclude that no tight scaling relation between $|\phi|$ and $M_{\rm BH}$ is found in the S$^4$G. 
This questions previous studies in the literature that were based on smaller samples \citep[e.g.][]{2008ApJ...678L..93S,2017MNRAS.471.2187D}. 
However, we note that part of the scatter in the relation might be due to the uncertainty in the determination of both parameters.
%
%
\subsection{Interplay between bars and spirals}\label{bars_spirals}
%
%
The effect of bar-induced secular evolution on mixing and reshaping discs is subtle, but its importance is becoming consolidated. 
For instance, by analysing average stellar density profiles, 
\citet[][]{2016A&A...596A..84D} showed that barred galaxies have larger disc scale-lengths and fainter extrapolated surface 
brightness ($\sim 15 \%$ difference) than their non-barred counterparts, in agreement with \citet[][]{2013MNRAS.432L..56S}. 
This proved the effect of bars pushing out the material that lies beyond corotation, 
in agreement with various simulation models \citep[e.g.][]{2002MNRAS.330...35A,2011A&A...527A.147M,2012MNRAS.426L..46A}. 
To further support this scenario, \citet[][]{2015MNRAS.451..936S} showed that the flattening of the outer stellar velocity dispersion profiles 
increases with increasing bar strength. The next question is whether bars play a role in driving the formation of spirals.

\citet[][]{2011MNRAS.414..538K} claimed that there is no evidence that bars trigger spirals 
\citep[see also][]{1998MNRAS.299..685S}, since there is no connection between a galaxy being barred and having spiral structure. 
However, we confirm that grand-design spirals are more frequently barred than non-barred \citep[in agreement with][]{1982MNRAS.201.1021E}, 
and we find that this is also the case for multi-armed and flocculent galaxies (Sect.~\ref{pitches_hubble}). 
Among late-type galaxies, most of the spiral galaxies are barred.

Nonetheless, non-axisymmetries cannot explain the formation of spirals in all the disc galaxies in the local universe, 
since spiral structure in the near-IR has been found in non-barred galaxies. 
In particular, nearly $30\%$ of the spirals in our sample are non-barred, regardless of the type. 
In these cases, it seems unlikely that the spiral pattern could have been excited by bars that dissolved afterwards, 
since bars are believed to be long-lasting 
\citep[e.g.][]{2003MNRAS.341.1179A,2004ApJ...604..614S,2010ApJ...719.1470V,2013MNRAS.429.1949A,2011MNRAS.415..709S,2015A&A...584A..90G} 
unless the disc is very gas rich \citep[e.g.][]{2002A&A...392...83B}. 

\citet[][]{2019MNRAS.tmp.1111M} reports that, at a given bulge size, strongly barred galaxies have more loosely wound arms than non-barred galaxies: 
this is linked by the authors to the effect of bars slowing the winding speed of the arms. 
In this work we do not find any difference in the distribution of the 
pitch angles for barred and non-barred galaxies when $1 \le T\le 5$ (Sect.~\ref{pitches_hubble}): 
this questions the role of bars in the formation and evolution of spirals.
%
%
\subsubsection{The hypothetical link between pitch angle and bar strength: expectations from the manifold theory}\label{manifold}
%
%
The manifold theory \citep[e.g.][]{2006A&A...453...39R,2006MNRAS.369L..56P,2006MNRAS.373..280V,2007A&A...472...63R,2009MNRAS.394...67A} 
makes specific predictions for the shape of spirals and their coupling with bars. 
The models by \citet[][]{2009MNRAS.400.1706A} show a tight correlation between the pitch angle 
and the tangential-to-radial forces evaluated at the unstable $L_1$ point (see their Fig.~5). 
Recent observational work by \citet[][]{2019MNRAS.482.5362F} does not confirm this causality. 
We find a very weak correlation between gravitational torques evaluated at 
the bar end - $Q_{\rm T}(r_{\rm bar})$ - and the pitch angle (Sect.~\ref{pitch_bar_strength}). 

We also checked that no clear dependence of $|\phi|$ on bar strength is found when other proxies of the latter are used 
-- such as the maximum bar torque ($Q_{\rm b}$), the intrinsic ellipticity, or the $m=2$ Fourier amplitude -- 
or when the dilution of bar gravitational torques by the halo contribution to the overall radial force field 
was modelled. 
For a more direct comparison, we additionally checked a possible dependence between bar strength and the average pitch angle of the innermost 
spiral segments, which are closest to the unstable Lagrangian points in barred galaxies, confirming the negative result. 
We also analysed this trend with other Fourier-based measurements of pitch angles, 
both taken from the literature or from this work, to make sure that the lack of a strong correlation cannot be attributed to the 
employed methodology to characterise the spiral winding in observations.

We conclude that the formation of the majority of the spiral arms in the S$^4$G cannot be explained with the manifold theory. 
Likewise, \citet[][]{2019A&A...625A.146D} did not confirm the manifold origin of rings in the S$^4$G from the 
predicted correlation between the ratio of outer-to-inner ring semi-major axes and the bar strength.
%
%
\subsubsection{Coupling of bars and spirals analysed from their amplitudes}\label{bars_spirals_amplitude}
%
%
The observational test of the coupling between bars and spirals has been controversial for years. 
Based on the comparison of equivalent angles of arms and bars \citep[measured as in][]{1998MNRAS.299..672S}, 
\citet[][]{2003MNRAS.342....1S} did not find evidence for the role of bars driving spiral structure in a sample of 17 inclined galaxies. 
Based on the Fourier analysis of near-IR $K_{S}$-band imaging, 
\citet[][]{2004AJ....128..183B} and \citet[][]{2009AJ....137.4487B} noted that stronger bars are associated with stronger arms. 
\citet[][]{2005AJ....130..506B} removed the spiral contribution to the bar torque \citep[following][]{2003AJ....126.1148B} 
in the OSUBSGS sample \citep[][]{2002ApJS..143...73E} and showed a correlation between the bar and spiral torques.

On the other hand, \citet[][]{2009MNRAS.397.1756D} did not find a correlation between the strengths of bars and spirals 
in their sample of isolated galaxies. \citet[][]{2009AJ....137.4487B} found the relation to be rather weak 
using $K_{S}$-band images from the Anglo-Australian Telescope (AAT) for 23 galaxies, 
and concluded that some arms might be genuine bar-driven (i.e. bars and spirals are formed from the same disc instability) 
but other factors might explain their formation as well. 

\citet{2010ApJ...715L..56S} reassessed the bar-spiral coupling in the above-referred OSUBSGS and AAT samples and 
showed a correlation between the local spiral density amplitude and the local bar-only forcing 
- the bar-induced tangential-to-radial forces after the effect of the spirals is suppressed - out to $\approx$ 1.6 bar radii. 
Within this radius, the authors interpret that spirals are a long-lasting continuation of the bar mode, 
whereas beyond the radius the spirals might be decoupled \citep[e.g.][]{1988MNRAS.231P..25S} 
or transient \citep[e.g.][]{1991MNRAS.250..278S,2002MNRAS.337.1118L}.

More recently, \citet[][]{2016A&A...596A..84D} showed that $A_{2}$ amplitudes are larger across the discs of 
barred galaxies than those of their non-barred counterparts (see their Fig.~9) when $0\le T<5$. 
Likewise, the amplitude of non-axisymmetries beyond the bar radius were more pronounced for the strongly 
barred galaxies than for the weakly barred ones: this is consistent with bars exciting the spiral pattern. 
In addition, \citet[][]{2011ApJ...737...32E} found that longer bars with larger $m=2$ density amplitudes (more evolved) 
appear together with stronger arms. 
\citet[][]{2017MNRAS.471.1070B} showed that galaxies with high bar contrasts tend to have higher arm contrasts as well. 
Lately, \citet[][]{2017A&A...603A.113S} studied the oxygen abundance of spiral arms and interarm regions 
and found small differences for barred galaxies, not detected in their unbarred counterparts, suggesting that bars may have a subtle effect on the chemical distribution of these galaxies. 

In Sect.~\ref{spiral_bar_strength_sect} we showed a clear correlation between the strength of spiral arms and bars, 
as well as a moderate dependency between spiral strength and bar length. 
In addition, the mean amplitude of the arms was found to increase with increasing pitch angle 
(this is also confirmed in Appendix~\ref{spirals_all}, for all proxies of spiral strength, using all spiral segments instead of the mean). 
Altogether, the analysis of the amplitudes of spirals and bars supports a coupling between these stellar structures, 
even though the details of this connection are far from clear.

As discussed in \citet[][and references therein]{2011ApJ...737...32E}, 
grand-design spiral arms and the symmetric inner parts of multi-armed galaxies are thought to be the result of spiral density waves, 
that could have been driven by bars. 
On the other hand, properties of flocculent galaxies and the outer parts of multi-armed galaxies can in principle be 
explained with random local gravitational instabilities in the stellar and gaseous component \citep[e.g.][]{1965MNRAS.130..125G,1966ApJ...146..810J}, 
swing amplified into spiral arms \citep[e.g.][]{1981seng.proc..111T}. 
Thus, it is meaningful that the correlation between the strength of bars and spirals is fairly similar for all types of spirals. 
Most likely, bars are not responsible for exciting flocculent spiral structure or the outer spiral modes of multi-armed galaxies galaxies. 
Thus, the correlation between the amplitudes of bars and spirals may not necessarily imply causation. 
In fact, in Appendix~\ref{Bar_strength_spiral_strength} we show that the correlation 
holds even when the mean amplitude of the two outermost spiral segments is considered 
(only slightly smaller correlation coefficients than for the innermost ones). 

We conclude that the correlation between bar and spiral amplitudes 
might simply indicate that discs that are prone to the development of bars with large amplitudes are also 
reactive to the formation of prominent spirals, as discussed in \citet[][]{2010ApJ...715L..56S}. 
Bars might play a role in creating spirals \citep[][]{1976ApJ...209...53S}, 
but mainly in strongly barred galaxies \citep[e.g.][]{1988MNRAS.231P..25S}.
%
%
\section{Summary and conclusions}\label{summarysection}
%
%
In this paper we characterise the winding angle and amplitude of the spirals arms of $391$ not-highly inclined galaxies ($i<65^{\circ}$) 
from the Spitzer Survey of Stellar Structure in Galaxies \citep[S$^4$G;][]{2010PASP..122.1397S}. 
We study barred and non-barred galaxies separately. 
We analyse all types of spirals, that is, grand-design, multi-armed, and flocculent.

We determine the pitch angle of the arms using the human-supervised measurements of spiral segments from 
\citet[][]{2015A&A...582A..86H} (Fig.~\ref{Figpitch}). 
When multiple spiral segments appear for individual galaxies ($98.5\%$ of the cases), the global pitch angle ($|\phi|$) 
is obtained from the mean ($|\phi|_{\rm mean}$), median ($|\phi|_{\rm median}$), 
the mean of innermost logarithmic segments ($|\phi|_{\rm inner}$), 
and the mean weighted by the arc length of the segments ($|\phi|_{\rm weighted}$) 
(Figs.~\ref{mean_vs_median}, \ref{fourier_comparison_literature}, \ref{mean_vs_median_literature}, and \ref{Fourier_visual}). 
Values of $|\phi|$ are listed in Table~\ref{tabulate_mean_pitches}. 
In addition, for a subsample of 32 galaxies we could reliably determine the pitch angle using 2-D Fourier analysis ($|\phi|_{\rm Fourier}$) 
(Fig.~\ref{spiral_fit_fourier}, Table~\ref{tabulate_Fourier_pitches}).

Pitch angles are analysed as a function of galaxy properties -- such as morphological type ($T$) or 
the estimated mass of the central supermassive black hole -- and the strength of non-axisymmetries. 
The latter is estimated from gravitational torques ($Q_{\rm T}$) and $m=2$ Fourier amplitudes ($A_{\rm 2}$) from \citet[][]{2016A&A...587A.160D} 
evaluated at the bar region (bar strength), and also at the radial ranges 
where the pitch angle of logarithmic segments were fitted (spiral strength) (Fig.~\ref{comparison_spiral_amp}).

The main results of this paper are the following:
\begin{itemize}
\item The distribution of grand-design, multi-armed, and flocculent spirals peaks at $T=3$ (Sb), $T=5$ (Sc), and $T=6$ (Scd), 
respectively (Fig.~\ref{histo_spiral_types_ttype}). Roughly $2/3$ of the spirals in our sample are barred 
\citep[i.e. classified as $\rm SB$, $\rm SA\underline{B}$, $\rm SAB$, or $\rm S\underline{A}B$ by][]{2015ApJS..217...32B}, 
and for $T>5$ the bar fraction is larger than $90\%$.
\item In galaxies with more than one measured spiral segment, $|\phi|$ has a dispersion of $\sigma=9.5 \pm 0.3^{\circ}$ on average, 
but the differences on individual galaxies can be $\gtrsim 15-20^{\circ}$ (Fig.~\ref{sigma_pitch_angles_fgm}). 
\item The distribution of the pitch angles as a function of galactocentric radius is flat (Fig.~\ref{pitch_radial}). 
Hence, statistically speaking, the pitch angles can either increase or decrease with radius.
\item On average, the pitch angle increases with increasing Hubble stage for grand-design and multi-armed spirals (Fig.~\ref{Fig_pitch_angles}), 
as expected by the definition of $T$, but the scatter in this relation is large. 
Among flocculent spirals, $|\phi|$ is not correlated with $T$ (Fig.~\ref{Fig_pitch_angles_fgm}).
\item We confirm that the strengths of bars and spirals are correlated (Fig.~\ref{spiral_bar_strength}). 
This either supports the role of bars driving the formation of spirals \citep[][]{1976ApJ...209...53S} 
or indicates that the discs that are prone to the formation of stronger bars are also more reactive 
to the formation of spirals of larger amplitudes \citep[see e.g.][]{2010ApJ...715L..56S}. 
The latter interpretation is favoured by the observed coupling of the amplitudes of bar and arms even in flocculent spirals, 
which are implausible to be excited by bars, or when only the outermost segments of multi-armed galaxies are analysed.
\item We confirm that long bars (in physical size or relative to the disc size) tend to be 
hosted by discs with prominent spirals 
(Fig.~\ref{spiral_bar_length}), in agreement with \citet[][]{2011ApJ...737...32E}.
\item The pitch angle of the spiral arms is not controlled by the bar-induced perturbation strength (Fig.~\ref{fig_bar_pitch}). 
This questions the manifold origin of spiral arms \citep[e.g.][]{2009MNRAS.400.1706A}. 
\item The distribution of pitch angles for barred and non-barred galaxies is roughly the same when $1 \le T\le 5$ 
(Figs.~\ref{Fig_pitch_angles}, \ref{Fig_pitch_angles_fgm}, and \ref{pitch_radial}). 
This questions the role of bars driving spiral density waves.
\item The mean pitch angle is not strongly correlated with the mass of the supermassive black hole (Fig.~\ref{bh_mass}), 
estimated from central stellar velocity dispersion (Fig.~\ref{blackholescomparison}), 
and does not depend on the central concentration of stellar mass or shear (Fig.~\ref{sparc_pitch_shear}). 
This questions previous results in the literature using smaller samples \citep[e.g.][]{1981AJ.....86.1847K,2015ApJ...802L..13D} 
and casts doubt on the validity of predicting $M_{\rm BH}$ from $|\phi|$ \citep[e.g.][]{2017MNRAS.471.2187D}. 
\end{itemize}
%
%
%
\begin{acknowledgements}
We thank the anonymous referee for comments that improved this paper. 
We acknowledge financial support from the European Union's Horizon 2020 research and innovation programme under 
Marie Sk$\l$odowska-Curie grant agreement No 721463 to the SUNDIAL ITN network, 
and from the Spanish Ministry of Economy and Competitiveness (MINECO) under grant number AYA2016-76219-P. 
S.D. acknowledges the financial support from the visitor and mobility program of the Finnish Centre for Astronomy with ESO (FINCA), 
funded by the Academy of Finland grant nr 306531. 
J.H.K. acknowledges support from the Fundación BBVA under its 2017 programme of assistance to scientific research groups, 
for the project "Using machine-learning techniques to drag galaxies from the noise in deep imaging", 
and from the Leverhulme Trust through the award of a Visiting Professorship at LJMU. 
H.S. acknowledges financial support from the Academy of Finland (grant no: 297738). 
M.H.E. thanks CONACyT for supporting his postdoctoral position through the grant No. 252531. 
We thank Sébastien Comerón for excellent comments on the manuscript. 
We thank the S$^{4}$G team for their work on the different pipelines and the data release. 
We thank Si Yue-Yu for providing their measurements of pitch angles for a subsample of S$^4$G galaxies. 
We thank Joachim Janz, Aaron Watkins, and Elena D'Onghia for useful discussions. {\it Facilities}: \emph{Spitzer} (IRAC).
%
%
\end{acknowledgements}
\bibliographystyle{aa}
\bibliography{bibliography}
\clearpage
%
%
%
\onecolumn
\begin{appendix}
%
%
\section{Tabulated pitch angles and spiral amplitudes}\label{tab_pitches}
%
%
The following data are listed in Table~\ref{tabulate_mean_pitches} for the 391 spiral galaxies in our sample:
%
\begin{itemize}
\item \emph{Column 1}: galaxy identification.
\item \emph{Column 2}: spiral types according to \citet[][]{2015ApJS..217...32B}: grand-design (G), multi-armed (M), and flocculent (F).
\item \emph{Column 3}: quality flag for the spiral fit from \citet[][]{2015A&A...582A..86H}: good=1 and acceptable=2.
\item \emph{Column 4}: number of logarithmic segments fitted by \citet[][]{2015A&A...582A..86H}.
\item \emph{Column 5}: mean pitch angle ($|\phi|_{\rm mean}$) and 
standard deviation of the mean, computed from all the fitted logarithmic segments.
\item \emph{Column 6}: standard deviation of the pitch angle of the different spiral segments ($\sigma_{|\phi|}$).
\item \emph{Column 7}: median pitch angle ($|\phi|_{\rm median}$)\footnote{Errors in the median pitch angle are not provided, 
since bootstrap resamplings are not appropriate for this work, given the small number of spiral segments per galaxy.}.
\item \emph{Column 8}: mean pitch angle of innermost spiral segments ($|\phi|_{\rm inner}$) and standard deviation of the mean.
\item \emph{Column 9}: mean pitch angle weighted by the arc length of the logarithmic segments ($|\phi|_{\rm weighted}$) 
and weighted standard deviation\footnote{The weighted standard deviation is computed as 
$\sqrt{\dfrac{\sum_{i=1}^{\rm n} s_{i} \cdot (|\phi|_{i}-|\phi|_{\rm weighted})^{2}}{\sum_{i=1}^{\rm n} s_{i}}}$, 
using the same notation as in Eqs.~\ref{eq_segm}~and~\ref{eq_segm_2}.} (in parenthesis).
\item \emph{Column 10}: mean gravitational torque of the spiral arms ($Q_{\rm T}^{\rm Spiral}$), 
followed by the uncertainty associated with the disc thickness determination \citep[for details, see][]{2016A&A...587A.160D}, and 
the internal dispersion ($\sigma$) computed from the maximum values of $Q_{\rm T}$ 
associated to the different spiral segments (when $\ge 0.01$, in parenthesis).
\item \emph{Column 11}: mean gravitational torque of the spiral arms corrected for the dark matter halo dilution 
($Q_{\rm T,halo-corr}^{\rm Spiral}$), and internal dispersion of the maximum of $Q_{\rm T}^{\rm halo-corr}$ 
associated to the different spiral segments (when $\ge 0.01$, in parenthesis).
\item \emph{Column 12}: mean $m=2$ Fourier amplitude of the spiral arms ($A_{2}^{\rm Spiral}$), 
and internal dispersion of the maximum of $A_{2}$ associated to the different spiral segments (when $\ge 0.01$, in parenthesis).
\end{itemize}
%
%
\begin{tiny}
\begin{longtable}{| c| c| c| c| c| c| c| c| c| c| c| c|}
\hline
\multicolumn{1}{|c}{\centering Galaxy} & \multicolumn{1}{|c}{\centering Type} & \multicolumn{1}{|c}{\centering Flag} & \multicolumn{1}{|c}{\centering $N$} &
\multicolumn{1}{|c}{\centering $|\phi|_{\rm mean}$} & \multicolumn{1}{|c}{\centering $\sigma_{|\phi|}$} & \multicolumn{1}{|c}{\centering $|\phi|_{\rm median}$} &
\multicolumn{1}{|c}{\centering $|\phi|_{\rm inner}$} & \multicolumn{1}{|c|}{\centering $|\phi|_{\rm weighted}$} &
\multicolumn{1}{|c}{\centering $Q_{\rm T}^{\rm Spiral}$} & \multicolumn{1}{|c}{\centering  $Q_{\rm T,halo-corr}^{\rm Spiral}$} & \multicolumn{1}{|c|}{\centering $A_{2}^{\rm Spiral}$} \tabularnewline
\multicolumn{1}{|c}{\centering } & \multicolumn{1}{|c}{\centering } & \multicolumn{1}{|c}{\centering } & \multicolumn{1}{|c}{\centering } &
\multicolumn{1}{|c}{\centering $(^{\circ})$} & \multicolumn{1}{|c}{\centering $(^{\circ})$} & \multicolumn{1}{|c}{\centering $(^{\circ})$} &
\multicolumn{1}{|c}{\centering $(^{\circ})$} & \multicolumn{1}{|c|}{\centering $(^{\circ})$} &
\multicolumn{1}{|c}{\centering } & \multicolumn{1}{|c}{\centering } & \multicolumn{1}{|c|}{\centering } \tabularnewline
\hline
\hline
ESO012-010&G&1&2& 36.5 $\pm$  4.3&  6.1& 40.8& 36.5 $\pm$  4.3& 36.8 (4.3)&0.36 $\pm$  0.03 (0.01)&0.16 (0.01)&0.35 \\
ESO026-001&F&2&2& 12.6 $\pm$  0.7&  0.9& 13.3& 12.6 $\pm$  0.7& 12.6 (0.6)&0.20 $\pm$  0.02&0.09&0.20 \\
ESO027-001&G&1&5& 17.9 $\pm$  6.6& 14.7& 15.2& 26.1 $\pm$ 10.9& 21.2 (9.2)&0.25 $\pm$  0.02 (0.05)&--&0.76 (0.34) \\
ESO054-021&F&2&2& 29.2 $\pm$  0.4&  0.5& 29.6& 29.2 $\pm$  0.4& 29.3 (0.3)&0.33 $\pm$  0.03&0.23 (0.03)&0.33 (0.03) \\
ESO238-018&F&2&2& 24.0 $\pm$ 17.0& 24.0& 40.9& 24.0 $\pm$ 17.0& 22.7 (16.9)&0.21 $\pm$  0.02&--&0.22 \\
ESO287-037&F&2&2&  8.1 $\pm$  0.1&  0.2&  8.2&  8.1 $\pm$  0.1&  8.1 (0.1)&0.25 $\pm$  0.02&0.11&0.37 \\
ESO288-013&F&2&3& 12.1 $\pm$  7.8& 13.5&  7.8& 17.5 $\pm$  9.7& 20.4 (9.6)&0.22 $\pm$  0.03&--&0.27 \\
ESO289-026&F&2&2& 34.0 $\pm$  5.2&  7.4& 39.2& 34.0 $\pm$  5.2& 34.4 (5.2)&0.39 $\pm$  0.05&0.11&0.54 \\
ESO340-042&F&2&4& 26.3 $\pm$  8.8& 17.7& 31.8& 39.1 $\pm$  7.3& 30.3 (9.7)&0.37 $\pm$  0.04 (0.12)&0.28 (0.12)&0.45 (0.01) \\
ESO342-050&M&1&4& 25.2 $\pm$  4.5&  9.0& 32.2& 32.9 $\pm$  0.7& 25.6 (7.6)&0.14 $\pm$  0.02 (0.01)&0.12 (0.01)&0.27 (0.07) \\
ESO355-026&M&2&4& 26.2 $\pm$ 11.6& 23.2& 18.0& 14.9 $\pm$  3.1& 24.3 (19.1)&0.13 $\pm$  0.02&0.09&0.18 (0.02) \\
ESO440-011&F&2&3& 25.4 $\pm$  8.5& 14.7& 19.0& 30.6 $\pm$ 11.6& 30.1 (12.6)&0.26 $\pm$  0.03 (0.11)&0.22 (0.10)&0.39 (0.09) \\
ESO443-069&F&2&4& 20.0 $\pm$  7.5& 15.0& 26.7& 23.2 $\pm$ 14.9& 23.4 (12.8)&0.26 $\pm$  0.03 (0.03)&--&0.34 (0.13) \\
ESO443-085&F&2&2& 18.8 $\pm$  0.9&  1.3& 19.8& 18.8 $\pm$  0.9& 18.9 (0.9)&0.32 $\pm$  0.02&0.17&0.27 \\
ESO485-021&F&2&4& 30.6 $\pm$ 10.3& 20.6& 46.5& 48.4 $\pm$  1.9& 32.5 (17.7)&0.32 $\pm$  0.02 (0.07)&0.04 (0.03)&0.51 (0.03) \\
ESO502-020&F&2&3& 10.3 $\pm$  4.5&  7.8& 14.6&  8.0 $\pm$  6.7& 14.3 (2.1)&0.13 $\pm$  0.02&0.04 (0.01)&0.36 (0.08) \\
ESO504-028&F&2&3& 18.4 $\pm$  4.6&  8.0& 13.8& 20.7 $\pm$  6.9& 16.7 (5.6)&0.23 $\pm$  0.03 (0.05)&0.14 (0.05)&0.20 (0.05) \\
ESO508-007&F&2&1& 38.9&-& 38.9& 38.9& 38.9&0.50 $\pm$  0.04&0.13&1.11 \\
ESO508-024&F&1&3& 28.1 $\pm$  6.3& 11.0& 31.9& 34.2 $\pm$  2.3& 31.3 (7.5)&0.29 $\pm$  0.04 (0.12)&0.18 (0.09)&0.49 (0.12) \\
ESO509-026&F&2&2& 18.4 $\pm$  6.0&  8.5& 24.4& 18.4 $\pm$  6.0& 19.9 (5.8)&0.21 $\pm$  0.03 (0.03)&--&0.63 (0.09) \\
ESO510-059&F&2&4& 26.8 $\pm$ 10.2& 20.3& 44.0& 44.3 $\pm$  0.3& 23.9 (17.0)&0.31 $\pm$  0.03 (0.03)&0.19 (0.04)&0.67 (0.15) \\
ESO532-022&F&2&2& 25.8 $\pm$  1.5&  2.1& 27.3& 25.8 $\pm$  1.5& 26.0 (1.5)&0.39 $\pm$  0.03&0.20&0.67 \\
ESO533-028&M&2&4& 27.7 $\pm$  7.4& 14.9& 32.3& 38.5 $\pm$  6.2& 30.2 (11.9)&0.07 $\pm$  0.02 (0.02)&0.05 (0.02)&0.11 (0.01) \\
ESO541-004&M&2&5& 14.6 $\pm$  2.9&  6.4& 17.4&  8.1 $\pm$  1.5& 18.5 (3.9)&0.07 $\pm$  0.01 (0.01)&0.04&0.14 (0.05) \\
ESO549-018&G&1&2& 11.1 $\pm$  1.1&  1.5& 12.1& 11.1 $\pm$  1.1& 11.1 (1.0)&0.10 $\pm$  0.01&0.06&0.14 \\
ESO576-001&G&2&4& 10.9 $\pm$  5.2& 10.3& 13.6&  9.6 $\pm$  4.0& 17.0 (6.8)&0.11 $\pm$  0.01&0.06&0.34 \\
ESO576-050&F&2&2& 15.6 $\pm$  3.4&  4.8& 19.0& 15.6 $\pm$  3.4& 16.3 (3.3)&0.20 $\pm$  0.02&0.11&0.37 \\
ESO580-030&F&2&2& 22.4 $\pm$  0.6&  0.9& 23.0& 22.4 $\pm$  0.6& 22.2 (0.6)&0.16 $\pm$  0.02&0.16&0.23 (0.02) \\
IC0163&F&2&4& 24.0 $\pm$ 10.1& 20.3& 32.2& 24.0 $\pm$ 23.4& 30.7 (11.2)&0.25 $\pm$  0.02 (0.14)&0.18 (0.11)&0.30 (0.12) \\
IC0167&G&1&4& 27.7 $\pm$ 11.8& 23.6& 41.4& 47.7 $\pm$  6.2& 42.3 (15.1)&0.60 $\pm$  0.07 (0.21)&0.30 (0.19)&1.06 (0.08) \\
IC0529&M&2&5& 19.4 $\pm$  5.0& 11.3& 14.0& 30.1 $\pm$  7.7& 21.6 (9.9)&0.16 $\pm$  0.02 (0.05)&0.12 (0.06)&0.21 (0.04) \\
IC0718&F&2&1& 53.5&-& 53.5& 53.5& 53.5&0.21 $\pm$  0.02&0.13&0.22 \\
IC0749&F&2&3& 24.2 $\pm$  7.0& 12.2& 28.7& 31.1 $\pm$  2.4& 26.7 (8.6)&0.36 $\pm$  0.03 (0.06)&0.22 (0.06)&0.49 (0.01) \\
IC0758&F&2&2& 27.9 $\pm$  4.6&  6.4& 32.4& 27.9 $\pm$  4.6& 27.7 (4.5)&0.25 $\pm$  0.02 (0.06)&0.21 (0.05)&0.24 (0.05) \\
IC0769&G&1&9& 11.6 $\pm$  3.4& 10.1&  8.2&  9.9 $\pm$  5.0& 20.5 (6.5)&0.16 $\pm$  0.01 (0.03)&0.08 (0.04)&0.41 (0.14) \\
IC0797&F&2&3& 19.0 $\pm$  4.3&  7.5& 20.3& 23.0 $\pm$  2.7& 20.8 (4.8)&0.26 $\pm$  0.03 (0.07)&0.21 (0.06)&0.27 (0.09) \\
IC1067&M&2&4& 19.6 $\pm$  1.2&  2.4& 18.6& 18.4 $\pm$  0.2& 19.6 (2.1)&0.12 $\pm$  0.01 (0.05)&0.06 (0.03)&0.21 (0.01) \\
IC1125&M&2&3& 28.1 $\pm$ 13.5& 23.4& 29.2& 40.1 $\pm$ 10.9& 38.8 (14.4)&0.27 $\pm$  0.01 (0.03)&0.22 (0.03)&0.40 \\
IC1158&F&2&2& 11.0 $\pm$  0.8&  1.2& 11.8& 11.0 $\pm$  0.8& 10.9 (0.8)&0.20 $\pm$  0.03&0.14&0.31 \\
IC1447&F&2&5& 13.2 $\pm$  5.4& 12.1&  9.8& 10.4 $\pm$  0.5& 17.2 (11.1)&0.08 $\pm$  0.01 (0.01)&0.06 (0.01)&0.10 (0.01) \\
IC1933&M&2&4& 28.7 $\pm$  8.1& 16.2& 28.7& 36.4 $\pm$ 14.6& 33.1 (14.3)&0.24 $\pm$  0.03 (0.03)&0.13 (0.01)&0.25 (0.03) \\
IC1953&M&2&3& 20.6 $\pm$  9.6& 16.6& 11.2& 11.0 $\pm$  0.2& 21.6 (13.8)&0.31 $\pm$  0.04&0.26&0.31 (0.06) \\
IC1954&G&2&4& 13.8 $\pm$  4.2&  8.3& 16.0& 19.9 $\pm$  3.9& 18.5 (6.0)&0.12 $\pm$  0.01 (0.08)&0.10 (0.08)&0.21 (0.09) \\
IC1993&M&2&3& 15.2 $\pm$  1.9&  3.2& 14.8& 13.5 $\pm$  1.3& 14.7 (2.4)&0.12 $\pm$  0.01&0.08&0.19 (0.02) \\
IC2051&F&2&2& 15.5 $\pm$  4.5&  6.4& 20.0& 15.5 $\pm$  4.5& 15.9 (4.5)&0.16 $\pm$  0.01 (0.01)&0.12 (0.01)&0.35 \\
IC2627&G&2&3& 21.6 $\pm$  5.8& 10.1& 24.5& 27.2 $\pm$  2.8& 24.9 (7.8)&0.29 $\pm$  0.03 (0.01)&--&0.59 (0.05) \\
IC3115&G&2&6& 17.2 $\pm$  5.1& 12.5& 14.8& 13.0 $\pm$  1.8& 24.6 (11.2)&0.21 $\pm$  0.02 (0.05)&0.08 (0.03)&0.50 (0.10) \\
IC3391&F&2&4& 29.5 $\pm$  9.5& 19.1& 41.2& 45.7 $\pm$  4.4& 32.0 (16.9)&0.30 $\pm$  0.04 (0.02)&--&0.20 (0.01) \\
IC4237&M&2&3& 17.2 $\pm$  2.8&  4.8& 19.0& 16.2 $\pm$  4.5& 18.6 (3.2)&0.13 $\pm$  0.01 (0.02)&0.09 (0.02)&0.26 (0.03) \\
IC4536&F&2&4& 14.1 $\pm$  1.8&  3.5& 12.6& 16.0 $\pm$  3.4& 13.5 (2.5)&0.29 $\pm$  0.03 (0.07)&0.13 (0.05)&0.31 (0.11) \\
IC4901&G&2&2&  9.9 $\pm$  1.1&  1.5& 11.0&  9.9 $\pm$  1.1&  9.9 (1.0)&0.12 $\pm$  0.01&0.11&0.45 \\
IC5069&M&2&5& 15.0 $\pm$  6.0& 13.5& 10.8& 22.8 $\pm$ 12.0& 20.5 (10.1)&0.31 $\pm$  0.03&0.27&0.44 \\
IC5273&M&2&3& 28.0 $\pm$  1.9&  3.3& 29.5& 27.2 $\pm$  3.0& 28.4 (2.6)&0.23 $\pm$  0.02&0.20&0.23 (0.01) \\
IC5321&F&2&3& 22.3 $\pm$  8.9& 15.4& 28.1& 31.0 $\pm$  2.9& 28.3 (7.5)&0.17 $\pm$  0.01 (0.01)&0.13 (0.01)&0.34 (0.02) \\
IC5325&M&2&6& 15.8 $\pm$  2.4&  6.0& 18.9& 21.5 $\pm$  2.3& 17.5 (5.0)&0.15 $\pm$  0.01 (0.12)&--&0.20 (0.10) \\
IC5332&F&2&6& 13.8 $\pm$  2.5&  6.0& 16.7& 16.8& 12.2 (5.4)&0.11 $\pm$  0.01 (0.05)&0.07 (0.02)&0.16 (0.05) \\
NGC0063&G&2&3&  9.6 $\pm$  5.7&  9.8&  5.1& 13.0 $\pm$  7.9& 13.2 (8.5)&0.05 $\pm$  0.01&0.04&0.16 \\
NGC0157&M&2&7& 15.6 $\pm$  5.8& 15.3& 11.5& 29.7 $\pm$  5.6& 26.8 (10.0)&0.41 $\pm$  0.04 (0.10)&--&0.56 \\
NGC0255&F&2&2& 19.1 $\pm$  5.1&  7.3& 24.3& 19.1 $\pm$  5.1& 20.2 (5.0)&0.22 $\pm$  0.02 (0.03)&0.12 (0.02)&0.19 (0.04) \\
NGC0298&F&2&3& 40.3 $\pm$  5.7&  9.8& 41.2& 39.8 $\pm$  9.8& 41.4 (8.3)&0.26 $\pm$  0.01&0.11 (0.02)&0.68 (0.09) \\
NGC0300&F&2&4& 20.1 $\pm$  5.3& 10.6& 23.6& 25.9 $\pm$  3.8& 23.3 (3.9)&0.17 $\pm$  0.02 (0.01)&0.09 (0.02)&0.30 (0.02) \\
NGC0406&F&2&3& 29.2 $\pm$ 13.2& 22.9& 31.1& 41.0 $\pm$ 10.0& 37.3 (12.3)&0.25 $\pm$  0.02 (0.03)&0.12 (0.05)&0.52 (0.17) \\
NGC0428&F&2&2& 27.9 $\pm$ 10.2& 14.4& 38.1& 27.9 $\pm$ 10.2& 28.8 (10.2)&0.47 $\pm$  0.05&0.20&0.68 \\
NGC0514&M&2&5& 19.2 $\pm$  4.1&  9.2& 25.3& 15.4 $\pm$  9.8& 22.7 (5.4)&0.12 $\pm$  0.01 (0.03)&0.08 (0.02)&0.21 (0.05) \\
NGC0578&G&2&3& 16.7 $\pm$  3.1&  5.3& 14.3& 18.5 $\pm$  4.2& 15.7 (3.9)&0.28 $\pm$  0.04 (0.08)&0.21 (0.08)&0.30 (0.09) \\
NGC0613&M&2&8& 22.7 $\pm$  4.9& 13.9& 29.8& 24.0 $\pm$  5.8& 27.4 (11.3)&0.36 $\pm$  0.03 (0.11)&0.27 (0.11)&0.74 (0.18) \\
NGC0628&M&2&7& 18.3 $\pm$  3.3&  8.8& 20.3& 11.8 $\pm$  0.5& 22.7 (7.8)&0.18 $\pm$  0.02 (0.03)&--&0.40 (0.14) \\
NGC0658&F&2&4& 18.1 $\pm$  6.7& 13.4& 18.5& 23.2 $\pm$ 13.8& 25.2 (12.7)&0.04&0.03 (0.01)&0.25 (0.01) \\
NGC0685&M&2&3& 18.5 $\pm$  2.4&  4.1& 19.4& 20.8 $\pm$  1.4& 18.7 (3.2)&0.29 $\pm$  0.03 (0.06)&0.22 (0.08)&0.32 (0.08) \\
NGC0772&M&1&5& 19.9 $\pm$  4.3&  9.7& 16.5& 27.5 $\pm$  9.2& 23.9 (10.7)&0.10 $\pm$  0.01 (0.01)&0.08 (0.01)&0.33 (0.10) \\
NGC0864&M&2&5& 24.4 $\pm$  2.8&  6.3& 21.6& 20.7 $\pm$  0.9& 24.8 (5.5)&0.29 $\pm$  0.02 (0.06)&0.22 (0.05)&0.38 (0.12) \\
NGC0895&G&2&3& 31.4 $\pm$  6.9& 12.0& 33.5& 30.4 $\pm$ 11.9& 29.1 (9.3)&0.34 $\pm$  0.03 (0.03)&0.20 (0.03)&0.61 (0.03) \\
NGC0908&G&2&5& 21.8 $\pm$  5.2& 11.7& 28.7& 30.5 $\pm$  1.8& 25.2 (8.3)&0.17 $\pm$  0.02 (0.01)&0.13 (0.01)&0.45 (0.23) \\
NGC0918&M&2&4& 19.0 $\pm$  2.2&  4.3& 20.6& 21.3 $\pm$  1.1& 19.5 (3.6)&0.21 $\pm$  0.03&0.16 (0.03)&0.26 (0.04) \\
NGC0986&G&2&2& 28.8 $\pm$  5.8&  8.3& 34.6& 28.8 $\pm$  5.8& 29.4 (5.8)&0.50 $\pm$  0.05 (0.01)&--&1.24 \\
NGC0991&F&2&3& 17.9 $\pm$  3.8&  6.6& 21.2& 15.8 $\pm$  5.5& 19.0 (4.8)&0.19 $\pm$  0.02 (0.07)&0.13 (0.06)&0.15 (0.02) \\
NGC1042&F&2&2& 13.2 $\pm$  0.6&  0.8& 13.8& 13.2 $\pm$  0.6& 13.2 (0.5)&0.60 $\pm$  0.06 (0.04)&0.52 (0.03)&0.74 \\
NGC1073&M&2&4& 14.7 $\pm$  4.4&  8.7& 18.9& 10.4 $\pm$  8.4& 18.7 (3.1)&0.32 $\pm$  0.04 (0.03)&--&0.35 (0.01) \\
NGC1084&F&2&3& 18.2 $\pm$  1.6&  2.8& 18.1& 16.8 $\pm$  1.3& 19.1 (1.7)&0.31 $\pm$  0.03 (0.08)&--&0.34 (0.09) \\
NGC1090&G&2&5& 11.7 $\pm$  3.7&  8.4&  8.9&  7.6 $\pm$  1.2& 14.8 (7.1)&0.14 $\pm$  0.01 (0.06)&0.10 (0.06)&0.20 (0.03) \\
NGC1179&M&2&3& 25.7 $\pm$  5.1&  8.8& 27.5& 21.9 $\pm$  5.6& 26.5 (7.4)&0.19 $\pm$  0.02 (0.01)&0.07 (0.02)&0.32 (0.02) \\
NGC1187&M&2&4& 31.4 $\pm$  8.8& 17.5& 32.4& 32.1 $\pm$  0.4& 36.4 (13.0)&0.17 $\pm$  0.02 (0.06)&0.11 (0.05)&0.43 (0.06) \\
NGC1232&M&2&7& 21.4 $\pm$  4.1& 10.9& 23.1& 11.4 $\pm$  1.1& 23.4 (8.4)&0.16 $\pm$  0.02 (0.01)&0.11&0.29 (0.02) \\
NGC1255&F&2&4& 10.4 $\pm$  1.7&  3.4& 11.3& 12.9 $\pm$  1.6& 11.5 (3.0)&0.22 $\pm$  0.02 (0.03)&0.17 (0.03)&0.28 (0.10) \\
NGC1299&F&2&3& 28.0 $\pm$  2.0&  3.4& 28.5& 26.4 $\pm$  2.1& 27.6 (3.0)&0.17 $\pm$  0.02 (0.02)&0.15 (0.02)&0.33 \\
NGC1300&G&2&4& 14.7 $\pm$  5.4& 10.7& 21.7&  5.8 $\pm$  3.3& 21.5 (5.4)&0.35 $\pm$  0.03&0.20&0.74 \\
NGC1309&M&2&5& 20.4 $\pm$  2.5&  5.5& 22.2& 24.4 $\pm$  1.2& 20.4 (4.5)&0.21 $\pm$  0.03 (0.09)&0.17 (0.08)&0.32 (0.14) \\
NGC1338&F&2&4& 19.8 $\pm$  5.9& 11.7& 25.6& 10.6 $\pm$  5.2& 26.0 (7.6)&0.27 $\pm$  0.02 (0.04)&0.24 (0.04)&0.32 (0.05) \\
NGC1347&M&2&3& 13.3 $\pm$  4.0&  6.8& 11.0& 14.4 $\pm$  6.6& 15.2 (5.7)&0.17 $\pm$  0.02 (0.01)&0.13 (0.02)&0.27 (0.07) \\
NGC1365&G&1&3& 19.1 $\pm$  5.2&  9.0& 24.2& 24.3 $\pm$  0.1& 23.0 (4.3)&0.28 $\pm$  0.03 (0.16)&--&0.96 (0.53) \\
NGC1385&F&2&3& 13.4 $\pm$  5.5&  9.6& 12.9& 18.1 $\pm$  5.2& 17.5 (6.3)&0.34 $\pm$  0.03 (0.16)&--&0.56 (0.20) \\
NGC1493&M&2&4& 20.4 $\pm$  2.6&  5.2& 21.9& 21.0 $\pm$  0.9& 21.1 (4.3)&0.14 $\pm$  0.02 (0.02)&0.02&0.20 \\
NGC1494&M&2&5& 35.4 $\pm$  8.1& 18.0& 30.9& 28.7 $\pm$ 17.5& 37.7 (13.2)&0.20 $\pm$  0.02 (0.02)&0.13 (0.02)&0.22 (0.07) \\
NGC1559&F&2&4& 21.0 $\pm$  8.3& 16.5& 20.0& 25.5 $\pm$ 19.0& 23.0 (13.1)&0.60 $\pm$  0.06 (0.07)&--&0.59 (0.12) \\
NGC1566&G&1&4& 26.5 $\pm$  7.6& 15.1& 38.7& 13.5 $\pm$  2.8& 36.5 (8.0)&0.26 $\pm$  0.05 (0.09)&0.24 (0.08)&0.75 (0.42) \\
NGC1637&M&2&4& 16.9 $\pm$  3.1&  6.1& 20.2& 18.7 $\pm$  4.9& 17.3 (4.6)&0.13 $\pm$  0.01 (0.04)&0.11 (0.04)&0.26 (0.05) \\
NGC1703&M&2&5& 20.6 $\pm$  3.7&  8.2& 22.3& 12.6 $\pm$  2.0& 22.0 (7.1)&0.17 $\pm$  0.02 (0.04)&--&0.22 (0.04) \\
NGC1792&M&2&6& 23.1 $\pm$  5.3& 13.0& 23.1& 37.7 $\pm$  3.3& 25.4 (11.4)&0.31 $\pm$  0.03 (0.11)&--&0.41 (0.05) \\
NGC2460&G&2&3&  8.5 $\pm$  3.7&  6.4& 10.9& 12.1 $\pm$  1.2& 11.2 (2.1)&0.09 $\pm$  0.03&0.08&0.41 \\
NGC2537&F&2&3& 11.5 $\pm$  2.8&  4.9&  9.2& 13.1 $\pm$  4.0& 12.4 (4.1)&0.63 $\pm$  0.08&--&0.47 \\
NGC2543&G&1&4& 14.9 $\pm$  5.2& 10.3& 23.2&  6.0 $\pm$  0.6& 22.1 (5.5)&0.25 $\pm$  0.03 (0.06)&0.18 (0.05)&0.66 (0.07) \\
NGC2608&M&2&4& 26.4 $\pm$ 11.8& 23.6& 44.6& 46.8 $\pm$  2.2& 41.0 (14.2)&0.49 $\pm$  0.05&0.39 (0.01)&0.76 (0.07) \\
NGC2633&G&1&4& 19.7 $\pm$  9.3& 18.6& 31.7& 35.5 $\pm$  3.8& 27.9 (13.0)&0.10 $\pm$  0.01 (0.03)&--&0.47 (0.16) \\
NGC2701&M&2&7& 30.5 $\pm$  7.4& 19.7& 24.6& 55.4 $\pm$ 13.6& 31.6 (20.0)&0.28 $\pm$  0.03 (0.03)&0.22 (0.03)&0.27 (0.04) \\
NGC2710&G&1&3& 20.3 $\pm$  8.3& 14.4& 24.9& 18.0 $\pm$ 13.8& 29.1 (4.4)&0.29 $\pm$  0.02&0.18&0.71 \\
NGC2712&G&1&2&  9.4 $\pm$  1.2&  1.6& 10.5&  9.4 $\pm$  1.2&  9.5 (1.1)&0.29 $\pm$  0.03&0.22&0.58 (0.09) \\
NGC2742&M&2&7& 22.1 $\pm$  4.6& 12.1& 25.3&  8.0 $\pm$  5.1& 28.2 (6.8)&0.18 $\pm$  0.02 (0.06)&0.15 (0.05)&0.15 (0.04) \\
NGC2743&F&2&1& 18.1&-& 18.1& 18.1& 18.1&0.09 $\pm$  0.01&0.09&0.23 \\
NGC2750&M&1&4& 19.8 $\pm$  8.4& 16.9& 18.7& 29.7 $\pm$ 12.3& 27.8 (12.2)&0.32 $\pm$  0.04&0.25&0.73 \\
NGC2776&M&2&6& 25.5 $\pm$  4.0&  9.7& 29.3& 15.5 $\pm$  1.4& 25.2 (8.4)&0.13 $\pm$  0.02 (0.04)&0.04 (0.02)&0.33 (0.10) \\
NGC2854&G&2&4& 16.7 $\pm$  6.3& 12.5& 27.1& 17.2 $\pm$ 10.6& 22.1 (9.2)&0.08 $\pm$  0.01 (0.02)&0.06 (0.02)&0.44 (0.12) \\
NGC2903&M&2&3& 23.1 $\pm$  1.9&  3.4& 23.1& 23.0 $\pm$  3.3& 23.3 (2.7)&0.17 $\pm$  0.02 (0.07)&0.15 (0.07)&0.47 \\
NGC2906&M&2&4& 18.0 $\pm$  6.1& 12.1& 20.3& 27.2 $\pm$  7.0& 20.7 (10.1)&0.13 $\pm$  0.02 (0.01)&0.10 (0.01)&0.20 (0.02) \\
NGC2964&F&2&3& 33.7 $\pm$  7.2& 12.5& 27.1& 37.0 $\pm$ 11.1& 30.2 (8.2)&0.30 $\pm$  0.03 (0.02)&--&0.40 (0.01) \\
NGC2967&F&2&4& 16.0 $\pm$  4.1&  8.2& 21.6&  8.9 $\pm$  0.6& 19.3 (6.5)&0.22 $\pm$  0.03 (0.03)&0.18 (0.03)&0.31 (0.01) \\
NGC2978&G&1&3& 15.9 $\pm$  1.6&  2.7& 15.3& 17.0 $\pm$  1.7& 15.4 (1.9)&0.16 $\pm$  0.02 (0.01)&0.15 (0.02)&0.45 \\
NGC3020&F&2&3& 18.0 $\pm$  5.2&  9.0& 13.5& 12.9 $\pm$  0.7& 23.1 (7.3)&0.28 $\pm$  0.03 (0.15)&0.12&0.39 (0.29) \\
NGC3021&M&2&6& 22.0 $\pm$  6.1& 15.0& 19.0& 28.0 $\pm$  9.0& 28.1 (13.6)&0.20 $\pm$  0.01 (0.01)&--&0.38 (0.12) \\
NGC3031&G&1&2& 12.5 $\pm$  0.7&  0.9& 13.2& 12.5 $\pm$  0.7& 12.6 (0.6)&0.07 $\pm$  0.01&0.05&0.39 \\
NGC3041&M&2&4& 20.8 $\pm$  4.7&  9.4& 23.3& 28.0 $\pm$  4.7& 22.4 (8.0)&0.19 $\pm$  0.02 (0.01)&0.13 (0.01)&0.33 \\
NGC3055&M&1&4& 16.3 $\pm$  3.7&  7.5& 17.0& 21.8 $\pm$  4.8& 18.2 (6.7)&0.17 $\pm$  0.01 (0.06)&0.15 (0.06)&0.31 (0.11) \\
NGC3057&F&2&3& 11.4 $\pm$  3.3&  5.7& 12.1& 11.0 $\pm$  5.7& 13.0 (4.1)&0.22 $\pm$  0.02 (0.04)&0.11 (0.02)&0.27 (0.10) \\
NGC3061&F&2&2& 14.7 $\pm$  6.1&  8.6& 20.8& 14.7 $\pm$  6.1& 17.2 (5.6)&0.25 $\pm$  0.02&0.15 (0.01)&0.46 (0.02) \\
NGC3147&M&2&7& 10.9 $\pm$  1.5&  3.9&  9.5&  9.3 $\pm$  0.2& 11.8 (3.3)&0.05 $\pm$  0.01 (0.01)&0.04 (0.01)&0.15 (0.03) \\
NGC3153&F&2&3& 24.7 $\pm$  7.0& 12.2& 27.3& 31.4 $\pm$  4.1& 25.0 (8.9)&0.13 $\pm$  0.01 (0.01)&0.08 (0.02)&0.28 \\
NGC3155&G&1&4& 12.9 $\pm$  3.7&  7.4& 15.9&  7.0 $\pm$  2.0& 17.0 (5.3)&0.11 $\pm$  0.01&0.08&0.38 \\
NGC3162&M&2&4& 23.4 $\pm$  5.2& 10.5& 31.8& 14.4 $\pm$  1.3& 26.7 (8.6)&0.23 $\pm$  0.02 (0.05)&0.20 (0.05)&0.39 (0.02) \\
NGC3177&G&2&3& 10.8 $\pm$  4.3&  7.5& 11.8& 14.8 $\pm$  3.0& 12.8 (4.2)&0.08 $\pm$  0.01 (0.01)&--&0.24 (0.05) \\
NGC3184&M&2&5& 24.3 $\pm$  3.4&  7.5& 23.1& 19.6 $\pm$  3.5& 23.0 (6.6)&0.20 $\pm$  0.03 (0.07)&0.12 (0.05)&0.38 (0.09) \\
NGC3206&M&2&3& 23.1 $\pm$  1.7&  3.0& 24.8& 22.2 $\pm$  2.6& 23.2 (2.4)&0.28 $\pm$  0.04 (0.01)&0.10&0.45 (0.13) \\
NGC3225&F&2&3& 28.0 $\pm$  7.8& 13.5& 25.3& 20.7 $\pm$  4.6& 31.8 (11.1)&0.12 $\pm$  0.01&0.07&0.13 (0.04) \\
NGC3227&G&1&4& 10.5 $\pm$  4.8&  9.7& 17.4& 18.8 $\pm$  1.4& 16.5 (6.0)&0.11 $\pm$  0.01 (0.01)&--&0.22 (0.15) \\
NGC3246&F&2&4& 16.8 $\pm$  3.7&  7.3& 20.5& 21.5 $\pm$  0.9& 18.6 (4.8)&0.16 $\pm$  0.02 (0.01)&0.09&0.14 \\
NGC3294&M&2&6& 22.3 $\pm$  4.2& 10.2& 26.9& 26.4 $\pm$  3.9& 24.9 (6.4)&0.20 $\pm$  0.02 (0.04)&0.16 (0.05)&0.35 (0.04) \\
NGC3310&G&2&2& 47.6 $\pm$  7.1& 10.0& 54.7& 47.6 $\pm$  7.1& 47.3 (7.1)&0.20 $\pm$  0.02&--&0.57 \\
NGC3319&G&2&3& 15.6 $\pm$  7.3& 12.7& 15.4&  9.2 $\pm$  6.2& 18.9 (6.7)&0.43 $\pm$  0.04 (0.02)&0.16&0.76 (0.04) \\
NGC3320&M&1&4& 30.0 $\pm$  7.3& 14.7& 38.5& 41.0 $\pm$  2.5& 36.0 (10.2)&0.16 $\pm$  0.02&0.11&0.22 (0.09) \\
NGC3321&F&2&2& 13.4 $\pm$  0.8&  1.1& 14.1& 13.4 $\pm$  0.8& 13.4 (0.7)&0.25 $\pm$  0.03&0.12&0.34 (0.03) \\
NGC3338&G&1&2& 11.6 $\pm$  0.4&  0.6& 12.0& 11.6 $\pm$  0.4& 11.6 (0.4)&0.14 $\pm$  0.01 (0.01)&0.11 (0.01)&0.63 \\
NGC3344&M&2&6& 16.9 $\pm$  4.3& 10.5& 21.8& 13.7 $\pm$  8.1& 22.7 (8.5)&0.10 $\pm$  0.01 (0.02)&0.05 (0.01)&0.22 (0.04) \\
NGC3346&F&2&4& 13.0 $\pm$  3.6&  7.2& 18.8& 19.0 $\pm$  0.2& 15.3 (5.5)&0.17 $\pm$  0.03 (0.04)&0.14 (0.04)&0.21 (0.09) \\
NGC3359&G&1&2& 26.6 $\pm$  8.4& 11.9& 35.0& 26.6 $\pm$  8.4& 27.8 (8.3)&0.30 $\pm$  0.02&0.21&0.73 (0.01) \\
NGC3364&M&2&5& 23.7 $\pm$  5.9& 13.2& 24.3& 22.6 $\pm$ 15.6& 28.4 (10.0)&0.12 $\pm$  0.02&0.10 (0.01)&0.18 \\
NGC3370&M&2&7& 16.5 $\pm$  3.1&  8.2& 13.4&  7.7 $\pm$  0.2& 20.2 (6.3)&0.17 $\pm$  0.01 (0.05)&0.13 (0.04)&0.30 (0.11) \\
NGC3381&F&2&2& 22.0 $\pm$  0.1&  0.1& 22.0& 22.0 $\pm$  0.1& 21.9 (0.0)&0.23 $\pm$  0.02 (0.04)&--&0.36 \\
NGC3423&M&2&6& 24.0 $\pm$  6.6& 16.3& 28.1& 30.3 $\pm$  2.2& 27.9 (13.5)&0.22 $\pm$  0.04 (0.02)&0.13 (0.03)&0.26 (0.04) \\
NGC3430&M&2&4& 17.6 $\pm$  4.7&  9.4& 21.9& 25.2 $\pm$  3.4& 20.9 (7.1)&0.14 $\pm$  0.01 (0.06)&0.11 (0.06)&0.26 (0.08) \\
NGC3433&M&2&8& 16.1 $\pm$  4.1& 11.7& 13.2&  6.2 $\pm$  1.4& 24.3 (9.0)&0.22 $\pm$  0.02 (0.10)&0.07 (0.04)&0.41 (0.19) \\
NGC3455&M&2&5& 25.6 $\pm$  5.7& 12.7& 24.2& 23.4 $\pm$ 15.1& 30.5 (8.9)&0.22 $\pm$  0.01&0.17&0.34 (0.08) \\
NGC3485&M&2&4& 24.3 $\pm$  7.9& 15.8& 31.7& 37.4 $\pm$  5.7& 28.8 (12.7)&0.25 $\pm$  0.02 (0.03)&0.13 (0.02)&0.62 (0.04) \\
NGC3486&F&2&9& 25.7 $\pm$  3.8& 11.5& 25.2& 15.2 $\pm$  3.0& 28.6 (10.4)&0.08 $\pm$  0.01 (0.01)&0.06 (0.02)&0.37 (0.06) \\
NGC3488&F&2&5& 11.5 $\pm$  2.9&  6.4& 13.9&  9.0 $\pm$  4.9& 15.4 (4.9)&0.18 $\pm$  0.02 (0.10)&0.15 (0.10)&0.17 (0.07) \\
NGC3507&G&1&4& 16.1 $\pm$  2.7&  5.3& 19.8& 17.8 $\pm$  2.8& 16.9 (4.8)&0.11 $\pm$  0.01 (0.04)&0.10 (0.04)&0.34 (0.04) \\
NGC3512&M&2&5& 19.0 $\pm$  5.0& 11.2& 14.9& 21.2 $\pm$ 12.0& 20.8 (9.8)&0.07 $\pm$  0.01 (0.02)&0.04 (0.02)&0.17 (0.08) \\
NGC3513&G&1&3& 21.1 $\pm$  7.4& 12.9& 26.3& 16.4 $\pm$  9.9& 27.6 (4.5)&0.42 $\pm$  0.03&--&0.81 \\
NGC3521&G&1&2& 14.9 $\pm$  0.8&  1.1& 15.6& 14.9 $\pm$  0.8& 14.9 (0.7)&0.10 $\pm$  0.01&0.08&0.57 \\
NGC3547&F&2&4& 15.6 $\pm$  4.5&  8.9& 21.6& 23.0 $\pm$  1.4& 18.9 (6.5)&0.18 $\pm$  0.01 (0.03)&0.16 (0.03)&0.29 (0.03) \\
NGC3583&G&1&3& 14.3 $\pm$  6.3& 10.9& 19.8& 20.5 $\pm$  0.8& 20.3 (2.8)&0.33 $\pm$  0.03 (0.04)&0.27 (0.04)&0.82 (0.03) \\
NGC3596&G&1&4& 14.6 $\pm$  5.5& 11.0& 20.1& 23.7 $\pm$  3.6& 18.2 (9.3)&0.15 $\pm$  0.02 (0.01)&0.12 (0.01)&0.30 (0.09) \\
NGC3614&M&2&6& 14.5 $\pm$  4.5& 10.9& 16.4& 10.6 $\pm$  6.7& 21.7 (8.9)&0.21 $\pm$  0.02 (0.07)&0.14 (0.06)&0.40 (0.02) \\
NGC3627&M&1&4& 21.0 $\pm$  6.7& 13.4& 30.5& 20.7 $\pm$  9.8& 30.7 (6.6)&0.19 $\pm$  0.02 (0.02)&--&0.68 (0.18) \\
NGC3629&F&2&4& 14.6 $\pm$  4.1&  8.3& 19.4& 21.3 $\pm$  1.9& 17.5 (5.9)&0.21 $\pm$  0.02&0.12 (0.01)&0.28 (0.04) \\
NGC3631&M&2&6& 26.2 $\pm$  6.5& 15.9& 25.8& 28.4 $\pm$  3.2& 31.2 (12.7)&0.27 $\pm$  0.03 (0.05)&--&0.42 (0.15) \\
NGC3659&G&2&3& 23.9 $\pm$  8.8& 15.3& 24.3& 16.3 $\pm$  7.9& 31.5 (9.4)&0.22 $\pm$  0.01&0.18&0.28 (0.12) \\
NGC3673&M&2&5& 15.7 $\pm$  6.5& 14.6&  7.0& 22.4 $\pm$ 15.4& 25.1 (12.4)&0.21 $\pm$  0.03 (0.11)&0.12 (0.09)&0.44 (0.13) \\
NGC3683A&M&2&5& 21.7 $\pm$  5.6& 12.6& 17.1& 22.9 $\pm$ 11.6& 23.6 (11.7)&0.09 $\pm$  0.01 (0.01)&0.06 (0.01)&0.22 (0.08) \\
NGC3684&M&2&4& 25.2 $\pm$  6.2& 12.3& 27.0& 16.2 $\pm$  3.8& 26.3 (12.2)&0.15 $\pm$  0.02&0.14&0.22 (0.02) \\
NGC3686&G&2&4& 19.6 $\pm$  6.0& 11.9& 21.7& 27.5 $\pm$  5.8& 20.7 (5.3)&0.14 $\pm$  0.03 (0.07)&0.12 (0.06)&0.22 (0.01) \\
NGC3687&M&2&6& 26.4 $\pm$  3.0&  7.2& 30.0& 30.2 $\pm$  0.2& 27.5 (6.5)&0.09 $\pm$  0.01 (0.01)&0.04 (0.01)&0.24 (0.07) \\
NGC3689&G&2&5& 22.2 $\pm$  6.3& 14.2& 19.9& 36.1 $\pm$  4.1& 24.1 (10.4)&0.18 $\pm$  0.02 (0.02)&0.17 (0.02)&0.36 (0.04) \\
NGC3701&M&2&4& 20.0 $\pm$  2.9&  5.9& 20.1& 16.0 $\pm$  2.3& 20.2 (5.5)&0.09 $\pm$  0.01 (0.03)&0.07 (0.04)&0.20 (0.02) \\
NGC3715&M&2&4& 23.9 $\pm$  6.5& 13.0& 28.1& 29.2 $\pm$ 10.2& 25.7 (8.6)&0.14 $\pm$  0.01&0.12&0.55 (0.09) \\
NGC3726&M&2&4& 19.0 $\pm$  4.8&  9.6& 22.6& 19.5 $\pm$  3.2& 22.0 (6.4)&0.25 $\pm$  0.03&0.17&0.39 (0.09) \\
NGC3752&F&2&3& 20.8 $\pm$  5.1&  8.9& 24.1& 25.8 $\pm$  1.7& 23.7 (5.9)&0.27 $\pm$  0.01 (0.03)&0.19 (0.03)&0.53 (0.11) \\
NGC3756&M&2&7& 18.1 $\pm$  3.1&  8.3& 18.0& 14.9 $\pm$  5.3& 20.6 (6.5)&0.09 $\pm$  0.01 (0.02)&0.06 (0.02)&0.10 (0.02) \\
NGC3780&M&1&4& 16.4 $\pm$  4.1&  8.3& 18.9& 11.2 $\pm$  6.6& 20.2 (3.4)&0.22 $\pm$  0.02&0.17&0.33 \\
NGC3810&M&1&7& 18.9 $\pm$  2.3&  6.2& 15.6& 14.6 $\pm$  0.4& 19.0 (6.0)&0.11 $\pm$  0.01 (0.02)&0.09&0.28 (0.08) \\
NGC3887&M&2&4& 21.5 $\pm$  6.3& 12.7& 23.0& 30.8 $\pm$  7.8& 24.2 (11.3)&0.31 $\pm$  0.03&0.24 (0.01)&0.40 (0.02) \\
NGC3888&M&1&5& 20.7 $\pm$  7.2& 16.1& 18.1& 34.5 $\pm$ 12.8& 25.3 (16.2)&0.10 $\pm$  0.01 (0.07)&--&0.23 (0.11) \\
NGC3893&M&1&7& 14.7 $\pm$  3.3&  8.7& 14.6& 17.8 $\pm$  1.1& 18.8 (7.3)&0.28 $\pm$  0.02 (0.05)&--&0.67 (0.20) \\
NGC3930&F&2&3& 18.0 $\pm$  7.2& 12.5& 17.9& 11.8 $\pm$  6.2& 23.3 (8.1)&0.16 $\pm$  0.02 (0.02)&0.09&0.31 (0.04) \\
NGC3938&M&2&6& 14.9 $\pm$  3.3&  8.0& 13.7& 12.4 $\pm$  1.3& 17.3 (7.4)&0.11 $\pm$  0.01 (0.02)&--&0.29 (0.10) \\
NGC3949&M&2&5& 33.2 $\pm$  8.1& 18.0& 42.3& 13.9 $\pm$  3.7& 40.1 (12.6)&0.26 $\pm$  0.03 (0.08)&0.25 (0.07)&0.24 (0.05) \\
NGC3953&M&1&6& 16.7 $\pm$  3.3&  8.1& 18.4& 21.8 $\pm$  3.4& 19.3 (5.8)&0.09 $\pm$  0.01 (0.01)&0.06 (0.01)&0.27 (0.04) \\
NGC3982&M&2&4& 22.3 $\pm$  5.0&  9.9& 29.1& 21.7 $\pm$  7.4& 24.8 (8.4)&0.11 $\pm$  0.01 (0.05)&0.09 (0.04)&0.20 \\
NGC3985&F&2&2& 11.9 $\pm$  9.6& 13.6& 21.6& 11.9 $\pm$  9.6& 16.1 (8.7)&0.21 $\pm$  0.01&0.11&0.30 \\
NGC3992&M&1&7& 12.1 $\pm$  3.5&  9.1& 11.0& 10.0 $\pm$  4.7& 17.6 (9.1)&0.11 $\pm$  0.01 (0.04)&0.08 (0.03)&0.27 (0.09) \\
NGC4030&M&2&8& 14.4 $\pm$  3.2&  9.2& 14.5& 14.5 $\pm$  7.7& 20.0 (6.7)&0.08 $\pm$  0.01 (0.03)&--&0.25 (0.10) \\
NGC4035&F&2&2& 14.5 $\pm$  6.3&  8.8& 20.7& 14.5 $\pm$  6.3& 16.7 (5.8)&0.30 $\pm$  0.09 (0.29)&--&0.34 (0.13) \\
NGC4041&M&2&7& 21.6 $\pm$  4.3& 11.3& 19.9& 32.8 $\pm$  1.3& 25.5 (9.2)&0.07 $\pm$  0.01 (0.05)&0.06 (0.05)&0.25 (0.07) \\
NGC4051&M&2&6& 22.4 $\pm$  6.4& 15.6& 33.1& 34.1 $\pm$  1.0& 30.1 (8.5)&0.28 $\pm$  0.03 (0.11)&0.18 (0.09)&0.62 (0.26) \\
NGC4067&M&2&5& 14.9 $\pm$  5.1& 11.4& 21.0& 24.2& 21.4 (6.0)&0.05&0.04 (0.01)&0.17 (0.04) \\
NGC4080&F&2&2& 10.3 $\pm$  2.4&  3.4& 12.7& 10.3 $\pm$  2.4& 10.7 (2.4)&0.16 $\pm$  0.01&0.04&0.32 (0.05) \\
NGC4108B&F&2&2& 34.0 $\pm$ 18.9& 26.7& 52.9& 34.0 $\pm$ 18.9& 25.3 (16.8)&0.21 $\pm$  0.02&0.01&0.16 \\
NGC4123&M&2&3& 21.1 $\pm$  3.7&  6.4& 21.3& 21.0 $\pm$  6.3& 21.0 (5.4)&0.24 $\pm$  0.02 (0.09)&0.20 (0.08)&0.45 (0.15) \\
NGC4133&M&2&8& 16.2 $\pm$  3.7& 10.6& 10.9& 21.0 $\pm$ 12.9& 18.8 (10.6)&0.09 $\pm$  0.01 (0.05)&0.06 (0.04)&0.30 (0.05) \\
NGC4136&F&2&4& 19.0 $\pm$  4.5&  9.1& 16.7& 24.4 $\pm$  7.7& 19.7 (8.1)&0.18 $\pm$  0.02&0.12 (0.03)&0.29 (0.01) \\
NGC4145&F&2&3& 21.2 $\pm$  2.7&  4.7& 23.8& 23.9 $\pm$  0.1& 22.5 (3.2)&0.19 $\pm$  0.02 (0.06)&0.15 (0.08)&0.30 \\
NGC4152&M&2&5& 29.2 $\pm$  5.7& 12.8& 29.9& 36.2 $\pm$  6.3& 34.8 (9.2)&0.25 $\pm$  0.02 (0.03)&--&0.56 (0.11) \\
NGC4162&M&2&5& 27.2 $\pm$  4.5& 10.1& 26.7& 29.5 $\pm$  9.5& 26.5 (8.1)&0.12 $\pm$  0.01 (0.03)&0.11 (0.03)&0.16 (0.03) \\
NGC4165&G&2&6& 14.2 $\pm$  3.8&  9.2& 11.7& 17.5 $\pm$  8.5& 20.1 (8.1)&0.19 $\pm$  0.01 (0.03)&0.10 (0.02)&0.50 (0.04) \\
NGC4189&M&2&5& 16.0 $\pm$  4.5& 10.0& 18.6& 22.5 $\pm$  3.9& 21.8 (6.4)&0.22 $\pm$  0.02 (0.08)&0.14 (0.06)&0.31 (0.11) \\
NGC4193&G&2&2& 11.0 $\pm$  0.1&  0.1& 11.1& 11.0 $\pm$  0.1& 11.0 (0.1)&0.11 $\pm$  0.02&0.09&0.21 \\
NGC4204&F&2&2& 35.5 $\pm$  1.9&  2.7& 37.4& 35.5 $\pm$  1.9& 35.3 (1.9)&0.47 $\pm$  0.04 (0.07)&0.11 (0.02)&0.68 (0.03) \\
NGC4210&M&2&4& 10.7 $\pm$  4.0&  7.9&  9.5& 15.9 $\pm$  6.4& 12.2 (7.3)&0.05 $\pm$  0.01 (0.02)&0.04 (0.02)&0.08 (0.01) \\
NGC4212&M&2&5& 16.2 $\pm$  4.1&  9.2& 11.2& 21.5 $\pm$ 10.3& 18.4 (9.3)&0.20 $\pm$  0.01 (0.07)&0.18 (0.07)&0.33 (0.06) \\
NGC4237&G&2&4& 16.0 $\pm$  2.7&  5.4& 14.4& 17.6 $\pm$  6.2& 15.9 (4.1)&0.08 $\pm$  0.01&--&0.11 (0.04) \\
NGC4254&M&1&5& 20.8 $\pm$  6.8& 15.1& 21.9& 17.1 $\pm$ 10.0& 29.3 (10.6)&0.23 $\pm$  0.02 (0.05)&--&0.49 (0.15) \\
NGC4273&M&2&3& 20.1 $\pm$  4.2&  7.3& 17.7& 23.0 $\pm$  5.3& 20.1 (5.4)&0.11 $\pm$  0.01&--&0.36 \\
NGC4276&F&2&2& 24.2 $\pm$  4.4&  6.3& 28.6& 24.2 $\pm$  4.4& 25.6 (4.2)&0.33 $\pm$  0.02 (0.01)&0.31 (0.01)&0.37 (0.03) \\
NGC4294&M&2&4& 29.4 $\pm$  9.5& 19.1& 26.6& 41.5 $\pm$ 14.9& 30.3 (13.3)&0.26 $\pm$  0.02 (0.08)&0.19 (0.08)&0.35 (0.04) \\
NGC4303&M&2&5& 23.0 $\pm$  6.5& 14.4& 16.5& 38.6 $\pm$  1.1& 24.9 (12.6)&0.30 $\pm$  0.03 (0.16)&--&0.51 (0.04) \\
NGC4314&G&2&4& 23.2 $\pm$ 12.5& 25.0& 44.4& 44.8 $\pm$  0.4& 34.7 (18.2)&0.21 $\pm$  0.03 (0.07)&0.09 (0.04)&0.60 (0.18) \\
NGC4319&G&2&4& 18.0 $\pm$  6.0& 11.9& 19.7& 22.1 $\pm$ 12.0& 20.1 (10.8)&0.14 $\pm$  0.02&--&0.64 \\
NGC4321&M&2&4& 21.4 $\pm$  5.0& 10.1& 26.5& 29.8 $\pm$  3.3& 26.0 (8.1)&0.23 $\pm$  0.02 (0.10)&0.20 (0.11)&0.64 (0.09) \\
NGC4390&G&2&6& 17.9 $\pm$  5.7& 14.0& 16.7& 26.6 $\pm$ 18.3& 23.9 (14.0)&0.23 $\pm$  0.02 (0.02)&0.18 (0.01)&0.21 (0.05) \\
NGC4411A&G&2&3& 11.4 $\pm$  5.1&  8.8& 11.0& 15.7 $\pm$  4.7& 13.8 (5.5)&0.29 $\pm$  0.02 (0.01)&0.13 (0.01)&0.60 (0.01) \\
NGC4411B&M&2&6& 14.1 $\pm$  2.9&  7.2& 17.7& 12.2 $\pm$  7.8& 17.1 (4.7)&0.15 $\pm$  0.01 (0.02)&0.12 (0.03)&0.23 (0.12) \\
NGC4412&G&2&3& 17.3 $\pm$  8.0& 13.9& 11.8&  9.4 $\pm$  2.4& 22.4 (12.2)&0.34 $\pm$  0.04 (0.12)&0.28 (0.10)&0.37 (0.04) \\
NGC4413&G&2&2& 11.3 $\pm$  0.4&  0.6& 11.7& 11.3 $\pm$  0.4& 11.3 (0.4)&0.14 $\pm$  0.01 (0.01)&0.11 (0.01)&0.36 (0.03) \\
NGC4414&M&2&5& 28.9 $\pm$  6.0& 13.4& 30.5& 32.3 $\pm$  1.9& 31.4 (7.8)&0.06 $\pm$  0.01 (0.01)&--&0.09 (0.04) \\
NGC4428&M&2&6& 21.6 $\pm$  3.9&  9.5& 20.8& 15.0 $\pm$  0.9& 22.1 (7.4)&0.12 $\pm$  0.01 (0.02)&0.10 (0.01)&0.29 (0.10) \\
NGC4430&F&1&3&  9.6 $\pm$  4.3&  7.5&  7.5& 12.7 $\pm$  5.2& 15.4 (4.7)&0.20 $\pm$  0.02&0.14&0.27 \\
NGC4480&M&2&5& 18.4 $\pm$  4.8& 10.8& 16.8& 17.2 $\pm$  0.4& 20.8 (11.3)&0.14 $\pm$  0.01 (0.01)&0.11 (0.02)&0.21 (0.06) \\
NGC4487&M&2&5& 32.0 $\pm$  4.2&  9.5& 27.1& 41.5 $\pm$  5.1& 30.1 (7.7)&0.20 $\pm$  0.03 (0.01)&0.16 (0.02)&0.25 \\
NGC4501&M&2&8& 19.2 $\pm$  3.9& 10.9& 18.5& 15.8 $\pm$  2.7& 22.3 (9.2)&0.07 $\pm$  0.01&0.06 (0.01)&0.24 (0.11) \\
NGC4504&G&2&4& 15.4 $\pm$  3.7&  7.5& 19.3& 15.8 $\pm$  8.1& 16.5 (6.2)&0.15 $\pm$  0.02 (0.01)&0.08 (0.02)&0.21 (0.03) \\
NGC4519&M&2&4& 21.8 $\pm$  6.5& 13.0& 22.3& 29.1 $\pm$  8.9& 25.3 (10.3)&0.29 $\pm$  0.03 (0.07)&0.23 (0.08)&0.37 (0.18) \\
NGC4535&M&2&6& 17.8 $\pm$  4.8& 11.7& 17.7& 14.6 $\pm$  3.2& 23.8 (9.1)&0.24 $\pm$  0.03 (0.10)&0.12 (0.08)&0.53 (0.14) \\
NGC4536&M&2&4& 25.0 $\pm$  5.6& 11.2& 25.5& 32.8 $\pm$  7.2& 27.5 (9.0)&0.21 $\pm$  0.02 (0.05)&--&0.72 (0.10) \\
NGC4545&M&2&4& 22.8 $\pm$  5.0& 10.1& 19.3& 27.2 $\pm$ 10.5& 22.7 (9.0)&0.22 $\pm$  0.02&0.17&0.26 \\
NGC4548&G&2&5&  8.6 $\pm$  2.2&  5.0&  9.1&  8.8 $\pm$  5.9& 11.3 (3.3)&0.14 $\pm$  0.02 (0.07)&0.11 (0.06)&0.28 (0.17) \\
NGC4559&M&2&4& 21.8 $\pm$  5.2& 10.5& 19.1& 18.3 $\pm$  0.8& 20.8 (8.6)&0.19 $\pm$  0.02 (0.01)&0.13 (0.04)&0.42 (0.09) \\
NGC4561&F&2&2& 22.1 $\pm$  5.1&  7.3& 27.3& 22.1 $\pm$  5.1& 22.8 (5.1)&0.21 $\pm$  0.02 (0.12)&0.06 (0.05)&0.26 (0.06) \\
NGC4567&F&2&3& 16.6 $\pm$  4.8&  8.4& 14.0& 20.0 $\pm$  5.9& 17.1 (6.3)&0.25 $\pm$  0.03 (0.05)&0.22 (0.04)&0.32 (0.02) \\
NGC4568&M&2&4& 22.8 $\pm$  5.1& 10.1& 24.9& 22.5 $\pm$ 12.2& 26.0 (7.1)&0.18 $\pm$  0.02 (0.03)&--&0.41 (0.17) \\
NGC4571&F&2&3& 12.7 $\pm$  2.9&  5.0& 12.3& 10.1 $\pm$  2.2& 13.5 (3.2)&0.16 $\pm$  0.02 (0.02)&0.09 (0.03)&0.21 (0.01) \\
NGC4647&F&2&4& 17.5 $\pm$  2.2&  4.4& 21.0& 21.2 $\pm$  0.2& 16.9 (3.9)&0.15 $\pm$  0.03 (0.05)&--&0.31 (0.04) \\
NGC4651&M&2&4&  9.8 $\pm$  1.8&  3.6& 12.0& 10.4 $\pm$  1.6& 11.2 (2.4)&0.06&0.05&0.22 (0.05) \\
NGC4653&G&2&7& 15.3 $\pm$  4.1& 10.8& 12.7& 17.3 $\pm$ 11.2& 21.2 (7.1)&0.32 $\pm$  0.03 (0.13)&0.14 (0.02)&0.48 (0.20) \\
NGC4654&M&2&4& 23.8 $\pm$  7.3& 14.6& 33.8& 28.6 $\pm$  6.9& 29.3 (7.6)&0.37 $\pm$  0.05&0.29&0.65 \\
NGC4658&F&2&2& 24.8 $\pm$  5.1&  7.3& 29.9& 24.8 $\pm$  5.1& 25.4 (5.1)&0.24 $\pm$  0.02 (0.05)&0.22 (0.05)&0.48 \\
NGC4680&G&2&1& 21.4&-& 21.4& 21.4& 21.4&0.28 $\pm$  0.02&--&0.55 \\
NGC4682&M&2&7& 13.6 $\pm$  3.4&  9.0& 13.5&  9.6 $\pm$  3.9& 18.1 (6.4)&0.07 $\pm$  0.01 (0.02)&0.06 (0.02)&0.12 (0.03) \\
NGC4701&F&2&3& 17.7 $\pm$  7.0& 12.1& 12.1& 21.8 $\pm$  9.7& 21.2 (10.1)&0.09 $\pm$  0.01&--&0.16 \\
NGC4713&F&2&4& 25.3 $\pm$  7.5& 15.0& 33.4& 12.8 $\pm$  1.6& 30.6 (12.4)&0.31 $\pm$  0.03 (0.03)&0.26 (0.01)&0.35 (0.12) \\
NGC4731&F&2&2& 29.7 $\pm$  1.1&  1.6& 30.8& 29.7 $\pm$  1.1& 29.5 (1.1)&0.83 $\pm$  0.10 (0.01)&0.43 (0.01)&1.02 \\
NGC4775&F&2&4& 21.0 $\pm$  4.0&  8.1& 23.2& 21.5 $\pm$  1.7& 21.5 (6.3)&0.25 $\pm$  0.03 (0.04)&--&0.27 (0.08) \\
NGC4779&M&2&5& 14.5 $\pm$  2.5&  5.5& 16.8& 12.9 $\pm$  4.8& 16.1 (4.4)&0.40 $\pm$  0.04 (0.11)&0.28 (0.10)&0.59 (0.13) \\
NGC4781&M&2&6& 16.9 $\pm$  4.2& 10.2& 22.3& 16.4 $\pm$  7.3& 21.3 (7.8)&0.25 $\pm$  0.02 (0.02)&0.22 (0.04)&0.39 (0.14) \\
NGC4800&F&2&6& 19.6 $\pm$  3.2&  7.8& 21.9& 23.5 $\pm$  6.4& 21.3 (5.3)&0.07 $\pm$  0.01&0.06&0.27 (0.01) \\
NGC4806&F&2&5& 15.4 $\pm$  3.7&  8.3& 17.4& 18.0 $\pm$  0.7& 18.1 (6.6)&0.16 $\pm$  0.03 (0.02)&0.14 (0.03)&0.27 (0.10) \\
NGC4814&F&2&4& 15.6 $\pm$  4.6&  9.2& 23.3& 23.6 $\pm$  0.3& 19.1 (7.1)&0.09&0.03 (0.01)&0.58 (0.05) \\
NGC4897&F&2&2&  7.9 $\pm$  0.5&  0.7&  8.4&  7.9 $\pm$  0.5&  7.9 (0.5)&0.05&0.04 (0.02)&0.22 \\
NGC4899&F&2&4& 14.1 $\pm$  4.3&  8.7& 20.5& 21.2 $\pm$  0.7& 16.7 (5.9)&0.14 $\pm$  0.02 (0.04)&0.13 (0.04)&0.24 \\
NGC4902&M&1&6& 19.7 $\pm$  4.5& 11.1& 19.9& 23.0 $\pm$  3.8& 23.8 (9.4)&0.12 $\pm$  0.01 (0.05)&0.08 (0.04)&0.25 (0.09) \\
NGC4904&F&2&5& 23.8 $\pm$  7.7& 17.2& 17.1& 29.2 $\pm$ 12.1& 28.7 (14.2)&0.33 $\pm$  0.02 (0.17)&0.28 (0.16)&0.41 (0.22) \\
NGC4928&F&2&3& 21.7 $\pm$  9.5& 16.5& 14.2& 25.4 $\pm$ 15.2& 30.5 (13.3)&0.26 $\pm$  0.02 (0.06)&--&0.57 (0.18) \\
NGC4965&F&2&2& 19.0 $\pm$  3.7&  5.2& 22.7& 19.0 $\pm$  3.7& 19.4 (3.6)&0.31 $\pm$  0.04&0.25&0.36 \\
NGC4981&M&2&6& 20.6 $\pm$  6.0& 14.8& 15.9&  9.3 $\pm$  6.6& 25.6 (13.0)&0.11 $\pm$  0.01 (0.01)&0.10 (0.01)&0.33 (0.20) \\
NGC4995&M&2&6& 11.7 $\pm$  4.3& 10.5& 13.0& 13.0 $\pm$  6.7& 18.2 (7.6)&0.27 $\pm$  0.02 (0.05)&0.26 (0.05)&0.39 (0.04) \\
NGC5012&M&2&7& 18.8 $\pm$  3.4&  9.0& 17.8& 26.1 $\pm$ 11.0& 23.5 (10.3)&0.12 $\pm$  0.01 (0.02)&0.09 (0.03)&0.34 (0.06) \\
NGC5016&F&2&3& 10.7 $\pm$  2.3&  4.0&  9.1& 12.1 $\pm$  3.0&  9.2 (2.0)&0.09 $\pm$  0.01 (0.02)&0.08 (0.02)&0.17 (0.05) \\
NGC5033&M&2&7& 18.8 $\pm$  3.7&  9.8& 22.3& 25.1 $\pm$  0.1& 23.6 (6.0)&0.12 $\pm$  0.01 (0.03)&0.08 (0.04)&0.54 (0.21) \\
NGC5042&F&2&3& 13.6 $\pm$  6.1& 10.6& 11.8& 18.4 $\pm$  6.6& 17.2 (7.3)&0.21 $\pm$  0.02 (0.07)&0.14 (0.07)&0.24 (0.14) \\
NGC5054&M&1&7& 19.7 $\pm$  5.3& 14.0& 17.7& 31.9 $\pm$  2.1& 28.3 (8.5)&0.11 $\pm$  0.01 (0.05)&0.07 (0.04)&0.27 (0.11) \\
NGC5055&M&2&7&  8.1 $\pm$  2.6&  7.0&  5.5& 16.9 $\pm$  3.5& 11.6 (6.5)&0.05 $\pm$  0.01 (0.02)&0.04 (0.02)&0.15 (0.05) \\
NGC5085&G&2&6& 15.3 $\pm$  3.6&  8.8& 17.6& 22.0 $\pm$  4.4& 18.3 (5.8)&0.13 $\pm$  0.01 (0.01)&0.09 (0.03)&0.34 (0.13) \\
NGC5105&F&2&5& 19.8 $\pm$  4.8& 10.7& 20.3& 11.5 $\pm$  7.8& 24.8 (7.1)&0.26 $\pm$  0.02 (0.02)&0.12 (0.02)&0.46 (0.07) \\
NGC5112&M&2&5& 37.6 $\pm$  5.3& 11.8& 36.5& 42.3 $\pm$  5.8& 38.8 (9.4)&0.46 $\pm$  0.05&0.25&0.44 \\
NGC5194&G&1&6& 14.9 $\pm$  2.5&  6.2& 15.8& 15.6 $\pm$  0.2& 17.0 (4.1)&1.18 $\pm$  0.19 (1.62)&--&0.87 (0.68) \\
NGC5205&G&2&2& 15.5 $\pm$  1.1&  1.6& 16.6& 15.5 $\pm$  1.1& 15.9 (1.0)&0.23 $\pm$  0.01 (0.01)&0.16 (0.01)&0.47 (0.01) \\
NGC5236&M&2&4& 18.1 $\pm$  4.9&  9.8& 21.3& 10.6 $\pm$  3.3& 20.6 (6.9)&0.31 $\pm$  0.03 (0.11)&0.23 (0.10)&0.42 \\
NGC5240&F&2&5& 15.5 $\pm$  4.5& 10.0& 14.5& 12.4 $\pm$  2.1& 18.7 (8.2)&0.09 $\pm$  0.01 (0.04)&0.07 (0.03)&0.15 (0.07) \\
NGC5247&M&1&5& 24.2 $\pm$  6.7& 15.1& 25.1& 39.1 $\pm$  0.4& 30.7 (10.9)&0.55 $\pm$  0.06 (0.05)&--&0.75 (0.06) \\
NGC5248&M&2&7& 23.6 $\pm$  6.1& 16.2& 15.7& 25.8 $\pm$ 19.8& 34.1 (12.1)&0.33 $\pm$  0.04 (0.15)&0.24 (0.11)&0.97 (0.33) \\
NGC5254&M&2&6& 15.3 $\pm$  5.5& 13.5& 14.6& 13.6 $\pm$  1.0& 21.7 (12.6)&0.24 $\pm$  0.02 (0.03)&0.17 (0.03)&0.26 (0.06) \\
NGC5300&M&2&4& 17.8 $\pm$  1.8&  3.6& 17.7& 15.8 $\pm$  1.9& 18.4 (3.2)&0.15 $\pm$  0.02 (0.03)&0.10 (0.02)&0.14 (0.03) \\
NGC5313&M&2&4& 20.2 $\pm$  9.6& 19.2& 34.6& 20.9 $\pm$ 13.7& 32.5 (10.0)&0.17 $\pm$  0.01&0.15&0.36 (0.05) \\
NGC5320&F&2&4& 20.7 $\pm$  5.0& 10.1& 24.8& 21.8 $\pm$  2.9& 22.7 (7.0)&0.17 $\pm$  0.01 (0.05)&0.11 (0.04)&0.26 (0.04) \\
NGC5334&F&2&3& 13.3 $\pm$  3.4&  6.0& 16.2& 16.7 $\pm$  0.5& 15.5 (3.2)&0.09 $\pm$  0.02 (0.02)&0.06 (0.03)&0.22 (0.14) \\
NGC5339&F&2&3& 13.5 $\pm$  4.2&  7.3& 15.5& 17.5 $\pm$  2.0& 15.0 (5.0)&0.39 $\pm$  0.05 (0.21)&0.33 (0.18)&0.56 (0.08) \\
NGC5350&F&2&4& 25.5 $\pm$  7.8& 15.7& 27.4& 36.8 $\pm$  9.4& 24.1 (11.5)&0.22 $\pm$  0.02 (0.06)&0.18 (0.06)&0.31 (0.06) \\
NGC5364&G&1&4& 15.5 $\pm$  4.1&  8.2& 14.1& 20.9 $\pm$  6.8& 11.4 (4.1)&0.12 $\pm$  0.02 (0.01)&0.10 (0.02)&0.41 (0.29) \\
NGC5371&G&1&5& 13.4 $\pm$  7.0& 15.8&  5.9& 11.6 $\pm$  5.7& 23.1 (13.2)&0.17 $\pm$  0.02 (0.04)&0.10 (0.04)&0.60 (0.21) \\
NGC5375&M&2&5& 18.5 $\pm$  6.0& 13.4& 15.0& 26.8 $\pm$ 11.8& 23.4 (10.3)&0.17 $\pm$  0.02 (0.02)&0.11 (0.02)&0.45 \\
NGC5376&M&2&3& 10.7 $\pm$  1.1&  1.9& 10.2&  9.7 $\pm$  0.5& 10.5 (1.5)&0.03&0.03 (0.01)&0.09 (0.01) \\
NGC5426&M&2&4& 19.7 $\pm$  3.9&  7.8& 25.4& 26.4 $\pm$  1.0& 22.0 (6.3)&0.25 $\pm$  0.05 (0.10)&0.19 (0.09)&0.50 (0.09) \\
NGC5427&G&1&7& 23.0 $\pm$  6.0& 16.0& 21.1& 42.5 $\pm$  8.5& 21.2 (11.8)&0.15 $\pm$  0.02 (0.06)&0.08 (0.01)&0.46 (0.12) \\
NGC5430&G&1&2& 12.9 $\pm$  2.0&  2.8& 14.8& 12.9 $\pm$  2.0& 12.9 (1.9)&0.15 $\pm$  0.01 (0.04)&--&0.38 (0.04) \\
NGC5452&F&1&3& 20.1 $\pm$  7.4& 12.8& 15.5& 25.0 $\pm$  9.5& 18.8 (9.0)&0.16 $\pm$  0.01&0.10&0.20 (0.02) \\
NGC5457&M&2&6& 19.2 $\pm$  3.7&  9.0& 24.2& 18.5 $\pm$  8.3& 20.1 (7.1)&0.22 $\pm$  0.01 (0.03)&0.15 (0.02)&0.47 (0.26) \\
NGC5468&M&2&7& 28.2 $\pm$  5.9& 15.5& 32.5& 42.8 $\pm$  3.5& 33.3 (8.5)&0.30 $\pm$  0.03 (0.05)&--&0.32 (0.04) \\
NGC5480&F&2&2& 38.8 $\pm$  0.1&  0.1& 38.9& 38.8 $\pm$  0.1& 38.9 (0.0)&0.26 $\pm$  0.02&0.20&0.66 (0.07) \\
NGC5584&M&2&6& 23.6 $\pm$  3.7&  9.0& 21.3& 34.4 $\pm$  2.5& 23.5 (7.5)&0.38 $\pm$  0.05 (0.10)&0.28 (0.08)&0.28 (0.11) \\
NGC5595&M&2&5& 22.4 $\pm$  4.2&  9.3& 26.7& 20.0 $\pm$ 11.2& 25.5 (6.1)&0.30 $\pm$  0.03 (0.07)&0.28 (0.07)&0.43 (0.05) \\
NGC5600&F&2&2& 16.1 $\pm$ 10.7& 15.1& 26.7& 16.1 $\pm$ 10.7& 19.3 (10.1)&0.16 $\pm$  0.01 (0.02)&--&0.33 (0.05) \\
NGC5660&F&2&4& 27.6 $\pm$  7.4& 14.7& 38.2& 15.2 $\pm$  3.9& 35.7 (9.2)&0.31 $\pm$  0.04 (0.21)&--&0.33 (0.16) \\
NGC5668&F&2&3& 17.4 $\pm$  5.6&  9.7& 18.3& 22.5 $\pm$  4.2& 21.2 (6.5)&0.17 $\pm$  0.02 (0.04)&--&0.45 (0.01) \\
NGC5669&F&2&3& 18.1 $\pm$  7.3& 12.6& 21.0& 25.0 $\pm$  4.0& 24.2 (6.1)&0.35 $\pm$  0.05 (0.13)&0.22 (0.11)&0.47 (0.02) \\
NGC5676&G&2&5& 16.6 $\pm$  3.5&  7.9& 18.0& 16.1 $\pm$  2.7& 19.1 (6.4)&0.20 $\pm$  0.02 (0.05)&0.18 (0.05)&0.36 (0.05) \\
NGC5693&F&2&2& 21.9 $\pm$  3.7&  5.2& 25.6& 21.9 $\pm$  3.7& 23.8 (3.2)&0.22 $\pm$  0.02&0.19&0.26 \\
NGC5768&F&2&2& 19.8 $\pm$  4.9&  6.9& 24.6& 19.8 $\pm$  4.9& 19.4 (4.8)&0.12 $\pm$  0.01 (0.02)&0.06 (0.02)&0.34 \\
NGC5774&F&2&4& 17.8 $\pm$  2.8&  5.6& 20.0& 15.1 $\pm$  4.9& 19.1 (4.2)&0.16 $\pm$  0.02 (0.03)&0.13 (0.05)&0.28 (0.07) \\
NGC5783&F&2&4& 23.1 $\pm$  5.1& 10.1& 30.8& 31.9 $\pm$  1.1& 22.7 (8.6)&0.33 $\pm$  0.02 (0.01)&0.21 (0.02)&0.32 (0.01) \\
NGC5832&F&2&1&  9.0&-&  9.0&  9.0&  9.0&0.21 $\pm$  0.01&0.08&0.39 \\
NGC5861&G&2&5& 15.7 $\pm$  3.6&  8.1& 16.0& 17.0 $\pm$  1.0& 18.5 (5.2)&0.29 $\pm$  0.03 (0.02)&0.21 (0.02)&0.51 (0.27) \\
NGC5885&M&2&5& 20.4 $\pm$  4.3&  9.6& 21.6& 29.4 $\pm$  3.0& 23.2 (7.2)&0.15 $\pm$  0.01&0.14 (0.01)&0.22 (0.05) \\
NGC5892&F&2&5& 17.0 $\pm$  3.2&  7.2& 15.6& 24.2 $\pm$  1.0& 15.4 (5.3)&0.20 $\pm$  0.02 (0.02)&0.16 (0.03)&0.27 (0.02) \\
NGC5915&F&2&3& 20.6 $\pm$ 10.4& 18.0& 20.1& 29.5 $\pm$  9.3& 25.1 (11.8)&0.35 $\pm$  0.02 (0.13)&--&0.85 (0.10) \\
NGC5921&F&1&4& 17.3 $\pm$  4.3&  8.6& 20.4& 23.8 $\pm$  3.4& 22.4 (6.4)&0.26 $\pm$  0.03 (0.13)&--&0.47 (0.28) \\
NGC5937&F&2&3& 22.9 $\pm$  4.1&  7.1& 27.0& 27.0 $\pm$  0.1& 25.4 (4.2)&0.12 $\pm$  0.01 (0.07)&--&0.30 (0.04) \\
NGC5950&G&2&2& 15.3 $\pm$  2.2&  3.1& 17.5& 15.3 $\pm$  2.2& 15.9 (2.1)&0.40 $\pm$  0.05 (0.06)&0.32 (0.05)&0.49 (0.03) \\
NGC5956&M&2&4& 23.9 $\pm$  6.0& 11.9& 26.5& 30.0 $\pm$  7.9& 26.6 (8.7)&0.07 $\pm$  0.01 (0.02)&0.04 (0.02)&0.17 (0.03) \\
NGC5957&M&2&5& 21.9 $\pm$  6.6& 14.7& 21.1& 36.0 $\pm$  3.8& 27.0 (10.6)&0.15 $\pm$  0.02 (0.06)&--&0.36 (0.12) \\
NGC5962&F&2&4& 16.8 $\pm$  4.2&  8.3& 23.2& 24.0 $\pm$  0.8& 19.5 (6.6)&0.16 $\pm$  0.01 (0.08)&0.15 (0.08)&0.28 (0.14) \\
NGC5963&F&2&4& 14.5 $\pm$  7.4& 14.8& 27.1& 27.3 $\pm$  0.2& 24.1 (8.2)&0.11 $\pm$  0.01&0.09&0.26 \\
NGC5964&M&2&4& 28.4 $\pm$  4.4&  8.8& 31.8& 31.0 $\pm$  0.8& 29.7 (7.1)&0.29 $\pm$  0.02 (0.11)&0.23 (0.11)&0.46 (0.03) \\
NGC5970&M&2&5& 14.9 $\pm$  2.1&  4.7& 15.8& 16.3 $\pm$  0.5& 15.8 (3.4)&0.12 $\pm$  0.01 (0.04)&0.10 (0.04)&0.27 (0.02) \\
NGC5985&M&2&6& 11.7 $\pm$  2.1&  5.0& 10.5&  7.6 $\pm$  0.1& 14.2 (4.9)&0.09 $\pm$  0.01 (0.04)&0.06 (0.02)&0.17 (0.07) \\
NGC6063&G&2&4& 13.9 $\pm$  2.4&  4.9& 16.8& 18.0 $\pm$  1.2& 13.2 (4.1)&0.14 $\pm$  0.01 (0.03)&0.10 (0.03)&0.18 (0.03) \\
NGC6070&M&2&5& 22.2 $\pm$  2.7&  6.1& 20.7& 28.5 $\pm$  0.3& 23.3 (5.3)&0.18 $\pm$  0.02 (0.06)&0.14 (0.06)&0.28 (0.08) \\
NGC6106&F&2&3& 14.7 $\pm$  2.7&  4.7& 16.0& 14.0 $\pm$  4.6& 16.1 (3.5)&0.21 $\pm$  0.02 (0.03)&0.16 (0.06)&0.29 (0.02) \\
NGC6181&G&2&4& 25.6 $\pm$  5.6& 11.2& 26.4& 33.0 $\pm$  6.6& 28.6 (9.7)&0.24 $\pm$  0.02 (0.09)&--&0.55 (0.09) \\
NGC6207&F&2&3& 21.9 $\pm$  5.4&  9.4& 24.7& 27.2 $\pm$  2.4& 24.3 (7.1)&0.24 $\pm$  0.02&0.21 (0.01)&0.39 (0.06) \\
NGC6339&M&2&3& 23.4 $\pm$  6.4& 11.1& 24.6& 29.2 $\pm$  4.7& 27.0 (8.2)&0.35 $\pm$  0.03 (0.11)&0.24 (0.09)&0.41 (0.09) \\
NGC6412&F&2&2& 13.5 $\pm$  1.2&  1.6& 14.6& 13.5 $\pm$  1.2& 13.5 (1.1)&0.33 $\pm$  0.03&0.19&0.41 \\
NGC6923&M&2&5& 21.6 $\pm$  5.7& 12.7& 25.7& 22.0 $\pm$  3.7& 27.6 (4.9)&0.13 $\pm$  0.01&0.10 (0.01)&0.41 (0.18) \\
NGC7070&F&2&4& 20.7 $\pm$  8.7& 17.3& 33.0& 35.6 $\pm$  2.6& 30.6 (11.9)&0.19 $\pm$  0.02 (0.04)&0.08 (0.04)&0.32 (0.04) \\
NGC7163&G&2&2&  5.1 $\pm$  2.5&  3.5&  7.5&  5.1 $\pm$  2.5&  5.7 (2.4)&0.12 $\pm$  0.02 (0.01)&0.09 (0.01)&0.34 (0.02) \\
NGC7167&G&2&2& 16.6 $\pm$  5.0&  7.1& 21.6& 16.6 $\pm$  5.0& 17.4 (4.9)&0.17 $\pm$  0.02&0.13&0.23 (0.02) \\
NGC7171&G&2&4& 13.0 $\pm$  2.9&  5.7& 11.5& 16.5 $\pm$  5.0& 12.2 (4.3)&0.15 $\pm$  0.01 (0.04)&0.11 (0.03)&0.30 (0.05) \\
NGC7205&F&2&2& 13.1 $\pm$  1.0&  1.4& 14.1& 13.1 $\pm$  1.0& 13.2 (1.0)&0.17 $\pm$  0.02&0.14&0.26 \\
NGC7247&M&1&7& 15.1 $\pm$  3.6&  9.5& 10.1&  7.6 $\pm$  2.5& 21.3 (9.2)&0.09 $\pm$  0.01 (0.03)&--&0.18 (0.06) \\
NGC7254&G&2&2& 17.0 $\pm$  4.7&  6.6& 21.7& 17.0 $\pm$  4.7& 18.8 (4.3)&--&--&-- \\
NGC7371&G&2&2&  6.2 $\pm$  0.8&  1.2&  7.0&  6.2 $\pm$  0.8&  6.3 (0.8)&0.06 $\pm$  0.01 (0.01)&0.04&0.12 \\
NGC7412&M&1&7& 22.1 $\pm$  3.9& 10.3& 21.3& 34.8 $\pm$  4.9& 24.8 (8.7)&0.26 $\pm$  0.02 (0.07)&0.18 (0.06)&0.54 (0.05) \\
NGC7418&F&2&4& 19.5 $\pm$  5.5& 10.9& 25.1& 24.5 $\pm$  3.8& 23.5 (4.9)&0.18 $\pm$  0.02 (0.08)&0.15 (0.08)&0.26 (0.08) \\
NGC7424&F&2&2& 15.8 $\pm$  1.2&  1.8& 17.0& 15.8 $\pm$  1.2& 15.6 (1.2)&0.24 $\pm$  0.03&0.10&0.45 \\
NGC7437&F&2&2& 15.1 $\pm$  1.6&  2.3& 16.8& 15.1 $\pm$  1.6& 14.9 (1.6)&0.26 $\pm$  0.03&0.18&0.20 \\
NGC7448&M&2&4& 21.0 $\pm$  4.7&  9.4& 21.1& 15.9 $\pm$  5.2& 22.5 (6.6)&0.26 $\pm$  0.04 (0.03)&--&0.26 (0.05) \\
NGC7531&M&2&3& 18.0 $\pm$  1.8&  3.1& 16.9& 19.2 $\pm$  2.3& 18.5 (2.6)&0.15 $\pm$  0.02 (0.02)&0.12 (0.02)&0.38 (0.08) \\
NGC7661&F&2&4& 29.3 $\pm$ 15.8& 31.6& 54.9& 56.7 $\pm$  1.8& 49.0 (19.0)&0.24 $\pm$  0.02 (0.02)&0.13 (0.02)&0.31 \\
NGC7689&M&2&3& 26.6 $\pm$  7.4& 12.9& 21.0& 19.2 $\pm$  1.8& 26.7 (10.5)&0.25 $\pm$  0.03&0.19&0.26 \\
NGC7723&F&2&3& 22.5 $\pm$  8.0& 13.8& 19.3& 28.4 $\pm$  9.1& 21.2 (10.0)&0.14 $\pm$  0.01 (0.05)&0.11 (0.05)&0.20 (0.04) \\
NGC7724&G&2&3& 11.4 $\pm$  2.6&  4.5& 13.5&  9.9 $\pm$  3.7& 11.8 (3.3)&0.12 $\pm$  0.02 (0.06)&0.09 (0.05)&0.20 (0.07) \\
NGC7741&F&2&3&  9.6 $\pm$  4.4&  7.6&  9.3& 13.4 $\pm$  4.1& 13.3 (4.5)&0.33 $\pm$  0.03 (0.11)&0.20 (0.10)&0.39 (0.15) \\
NGC7743&G&2&6& 22.2 $\pm$  4.0&  9.7& 20.4& 34.2 $\pm$  2.2& 23.1 (9.0)&0.17 $\pm$  0.03&--&0.46 (0.01) \\
NGC7757&M&2&5& 23.8 $\pm$  4.6& 10.4& 24.1& 31.3 $\pm$  7.2& 27.5 (8.4)&0.26 $\pm$  0.02 (0.04)&0.25 (0.05)&0.52 (0.07) \\
NGC7764&F&2&3& 23.6 $\pm$  7.7& 13.3& 20.4& 29.3 $\pm$  8.9& 25.2 (10.4)&0.41 $\pm$  0.04 (0.27)&0.38 (0.25)&0.39 (0.17) \\
NGC7793&F&2&2& 11.4 $\pm$  1.2&  1.6& 12.5& 11.4 $\pm$  1.2& 11.5 (1.1)&0.15 $\pm$  0.02 (0.01)&0.12 (0.02)&0.24 \\
NGC7798&M&2&5& 16.9 $\pm$  5.2& 11.6& 16.0& 13.6 $\pm$  3.0& 19.4 (10.6)&0.14 $\pm$  0.01 (0.02)&--&0.41 (0.12) \\
PGC011367&F&2&2& 18.5 $\pm$  0.2&  0.2& 18.6& 18.5 $\pm$  0.2& 18.4 (0.1)&0.19 $\pm$  0.03&0.09&0.26 \\
PGC012664&F&2&3& 14.3 $\pm$  3.4&  5.9& 11.8& 11.0 $\pm$  0.8& 18.9 (4.2)&0.25 $\pm$  0.03 (0.07)&0.14 (0.05)&0.19 (0.05) \\
PGC027616&F&2&5& 17.5 $\pm$  3.1&  6.9& 15.3& 19.2 $\pm$  3.9& 18.4 (5.8)&0.26 $\pm$  0.02 (0.05)&0.08 (0.02)&0.33 (0.03) \\
PGC028380&F&2&2& 16.9 $\pm$  1.5&  2.1& 18.4& 16.9 $\pm$  1.5& 16.7 (1.5)&0.27 $\pm$  0.03 (0.01)&0.13 (0.01)&0.37 (0.01) \\
PGC032091&M&2&3& 21.4 $\pm$  9.2& 16.0& 24.7& 30.1 $\pm$  5.4& 27.1 (8.7)&0.19 $\pm$  0.01&0.09&0.22 \\
PGC043345&F&2&2& 18.8 $\pm$  2.2&  3.1& 21.0& 18.8 $\pm$  2.2& 18.8 (2.2)&0.21 $\pm$  0.02&0.11&0.32 (0.02) \\
PGC043458&F&2&2& 28.4 $\pm$  1.1&  1.6& 29.5& 28.4 $\pm$  1.1& 28.4 (1.1)&0.31 $\pm$  0.03&0.17&0.38 \\
PGC045958&G&2&2&  9.9 $\pm$  0.0&  0.1&  9.9&  9.9&  9.8 (0.0)&0.15 $\pm$  0.02&--&0.17 \\
PGC047721&G&2&3&  9.6 $\pm$  5.3&  9.2&  7.0& 13.4 $\pm$  6.5& 11.3 (7.1)&0.11 $\pm$  0.01 (0.03)&0.01&0.46 (0.16) \\
PGC048087&F&2&2& 21.2 $\pm$  2.2&  3.2& 23.5& 21.2 $\pm$  2.2& 21.2 (2.2)&0.18 $\pm$  0.02&0.09&0.20 \\
PGC048179&F&2&3& 14.6 $\pm$  5.0&  8.7& 15.8& 19.2 $\pm$  3.4& 16.0 (5.1)&0.30 $\pm$  0.03 (0.07)&0.25 (0.06)&0.32 (0.11) \\
PGC069448&M&2&4& 27.2 $\pm$  4.5&  9.0& 33.5& 34.5 $\pm$  1.0& 28.8 (7.3)&0.31 $\pm$  0.04 (0.05)&0.23 (0.05)&0.34 (0.11) \\
UGC02081&F&2&2& 12.3 $\pm$  0.9&  1.3& 13.2& 12.3 $\pm$  0.9& 12.3 (0.9)&0.13 $\pm$  0.02&0.07&0.17 \\
UGC02443&F&2&5& 16.4 $\pm$  5.2& 11.6& 21.0& 23.3 $\pm$  2.3& 22.9 (7.5)&0.16 $\pm$  0.03&0.11&0.16 \\
UGC04151&M&2&4& 24.4 $\pm$  2.5&  5.0& 23.6& 26.9 $\pm$  4.8& 24.7 (4.3)&0.13 $\pm$  0.03&0.08&0.16 \\
UGC04169&M&2&5& 31.6 $\pm$  9.7& 21.6& 20.1& 27.3 $\pm$  8.8& 31.1 (16.8)&0.20 $\pm$  0.02 (0.01)&0.10 (0.01)&0.35 (0.05) \\
UGC04549&F&2&3& 13.5 $\pm$  4.0&  7.0& 16.1& 17.5 $\pm$  1.3& 15.4 (4.2)&0.13 $\pm$  0.01 (0.04)&0.08 (0.03)&0.18 (0.03) \\
UGC04841&F&2&2& 16.2 $\pm$  0.8&  1.1& 17.0& 16.2 $\pm$  0.8& 16.2 (0.8)&0.22 $\pm$  0.02&0.17 (0.01)&0.32 \\
UGC04867&F&2&4& 15.7 $\pm$  7.2& 14.4& 27.1& 28.2 $\pm$  1.1& 25.5 (7.9)&0.37 $\pm$  0.02&0.10 (0.02)&0.57 (0.04) \\
UGC05358&F&2&2& 11.3 $\pm$  6.8&  9.6& 18.1& 11.3 $\pm$  6.8& 15.3 (5.5)&0.33 $\pm$  0.04&0.11&0.60 \\
UGC05707&F&2&3& 13.0 $\pm$  5.1&  8.8& 13.1& 17.4 $\pm$  4.3& 16.4 (5.1)&0.27 $\pm$  0.03&0.10 (0.03)&0.62 \\
UGC06023&F&2&2& 14.3 $\pm$  2.2&  3.1& 16.5& 14.3 $\pm$  2.2& 14.6 (2.2)&0.15 $\pm$  0.01&0.12&0.27 \\
UGC06194&F&2&3& 18.2 $\pm$  5.8& 10.1& 18.3& 18.2 $\pm$ 10.1& 23.0 (7.2)&0.18 $\pm$  0.02 (0.03)&0.14 (0.04)&0.17 (0.04) \\
UGC06335&F&2&4& 13.9 $\pm$  4.1&  8.2& 18.2& 19.7 $\pm$  1.5& 16.3 (3.7)&0.15 $\pm$  0.01 (0.01)&0.15 (0.01)&0.23 \\
UGC06903&F&2&3& 16.9 $\pm$  5.9& 10.2& 14.0& 21.1 $\pm$  7.1& 15.3 (6.5)&0.19 $\pm$  0.02 (0.07)&0.09 (0.05)&0.26 (0.13) \\
UGC06930&F&2&2& 47.7 $\pm$  9.6& 13.6& 57.3& 47.7 $\pm$  9.6& 48.0 (9.6)&0.27 $\pm$  0.03&0.17&0.34 \\
UGC07133&M&2&3& 30.1 $\pm$  4.6&  8.0& 33.5& 28.3 $\pm$  7.4& 30.5 (6.3)&0.19 $\pm$  0.02 (0.02)&0.09 (0.04)&0.18 \\
UGC07848&F&2&3& 16.0 $\pm$  4.8&  8.3& 11.2& 18.4 $\pm$  7.2& 12.9 (4.7)&0.13 $\pm$  0.01 (0.02)&0.07 (0.01)&0.12 (0.01) \\
UGC08041&M&2&4& 20.5 $\pm$  5.1& 10.3& 20.4& 14.4 $\pm$  6.0& 22.9 (7.8)&0.33 $\pm$  0.06 (0.05)&0.14 (0.03)&0.29 (0.03) \\
UGC08516&F&2&2& 17.6 $\pm$  3.8&  5.3& 21.4& 17.6 $\pm$  3.8& 18.1 (3.7)&0.25 $\pm$  0.02&--&0.20 \\
UGC08658&F&2&4& 14.6 $\pm$  3.9&  7.9& 19.8& 21.2 $\pm$  1.4& 15.1 (6.7)&0.12 $\pm$  0.01 (0.03)&0.06 (0.02)&0.25 (0.03) \\
UGC08909&F&2&2& 21.2 $\pm$  1.4&  2.0& 22.6& 21.2 $\pm$  1.4& 21.2 (1.4)&0.09 $\pm$  0.01&0.02&0.14 \\
UGC09569&F&2&2& 17.5 $\pm$  0.3&  0.4& 17.8& 17.5 $\pm$  0.3& 17.5 (0.3)&0.27 $\pm$  0.02 (0.03)&0.17 (0.02)&0.63 \\
UGC09730&F&2&2& 24.7 $\pm$  2.8&  4.0& 27.5& 24.7 $\pm$  2.8& 24.4 (2.8)&0.25 $\pm$  0.02 (0.08)&0.08 (0.04)&0.34 \\
UGC09837&M&2&3& 18.1 $\pm$  1.0&  1.8& 17.5& 17.0 $\pm$  0.4& 18.2 (1.5)&0.16 $\pm$  0.03&0.09&0.20 \\
UGC10020&F&2&3& 36.3 $\pm$ 10.5& 18.2& 37.2& 27.5 $\pm$  9.8& 41.7 (10.0)&0.08 $\pm$  0.01 (0.02)&0.07 (0.02)&0.17 (0.12) \\
UGC10445&F&2&2& 18.0 $\pm$  3.7&  5.2& 21.7& 18.0 $\pm$  3.7& 19.6 (3.3)&0.24 $\pm$  0.03&0.11&0.30 (0.13) \\
UGC10721&M&2&5& 27.1 $\pm$  6.1& 13.5& 31.2& 34.6 $\pm$  3.4& 31.1 (6.2)&0.25 $\pm$  0.01&0.19&0.59 (0.08) \\
UGC10803&F&2&1& 37.4&-& 37.4& 37.4& 37.4&0.09 $\pm$  0.02&0.07&0.15 \\
\hline
\caption{
Average pitch angles and spiral amplitudes.
}
\label{tabulate_mean_pitches}
\end{longtable}
\end{tiny}

\clearpage
%
%
%
\section{Bar strength versus spiral strength: extended analysis}\label{Bar_strength_spiral_strength}
\begin{figure}[h]
\begin{center}
\includegraphics[width=0.99\textwidth]{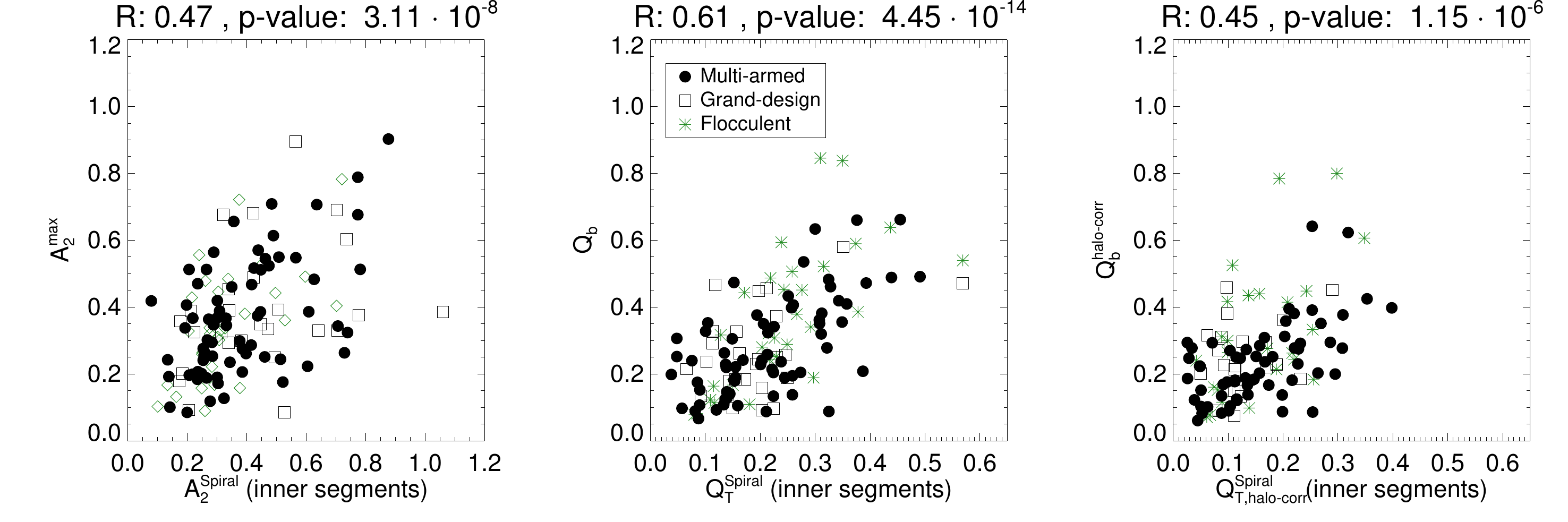}
\includegraphics[width=0.99\textwidth]{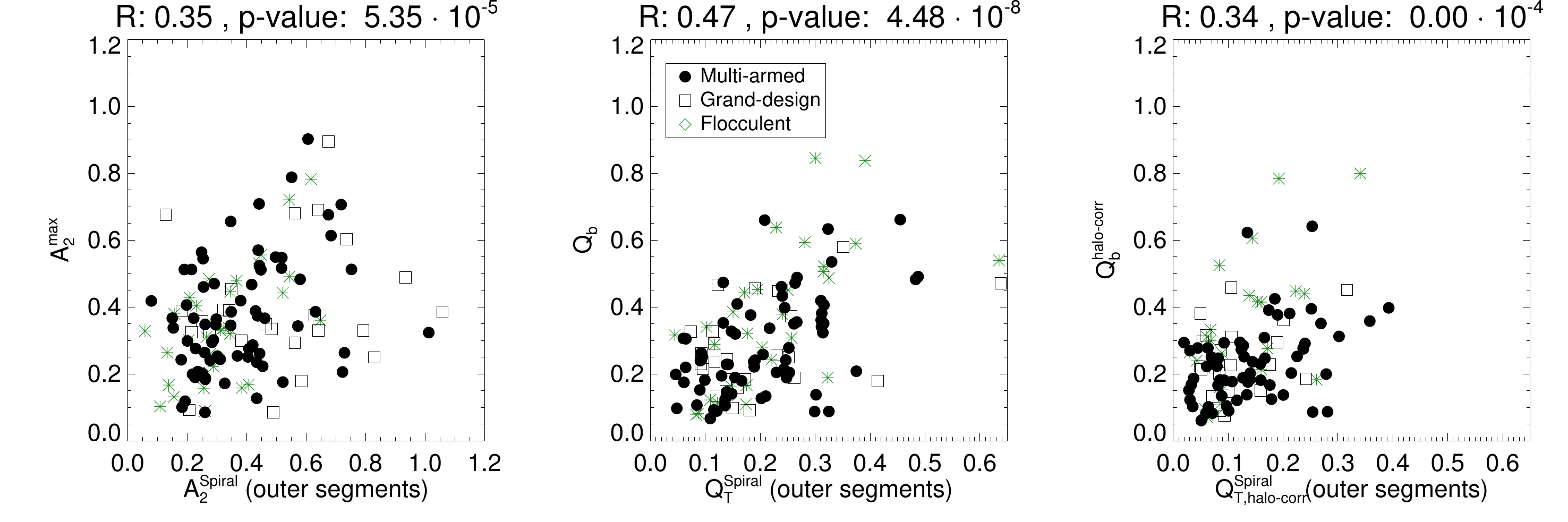}
\caption{
As in Fig.~\ref{spiral_bar_strength}, but taking the mean of the maximum amplitude 
of the two innermost (\emph{upper panels}) and outermost (\emph{lower panels}) spiral logarithmic segments.
}
\label{pitch_vs_spiral_strength_inner_outer}
\end{center}
\end{figure}

%
%
In Sect.~\ref{spiral_bar_strength_sect} we showed a correlation between the bar strength and the mean strength of the spirals. 
Here we test this dependence when only the mean of the amplitude the two innermost and the two outermost spiral segments are considered 
(see Fig.~\ref{pitch_vs_spiral_strength_inner_outer}). We only use galaxies with more than four measured spiral segments. 
The strength of the spiral segments is measured from the $m=2$ Fourier density amplitudes (left), 
tangential-to-radial forces (centre), and also torques corrected for the halo-dilution (right). 

\begin{figure}[h]
\begin{center}
\includegraphics[width=0.32\textwidth]{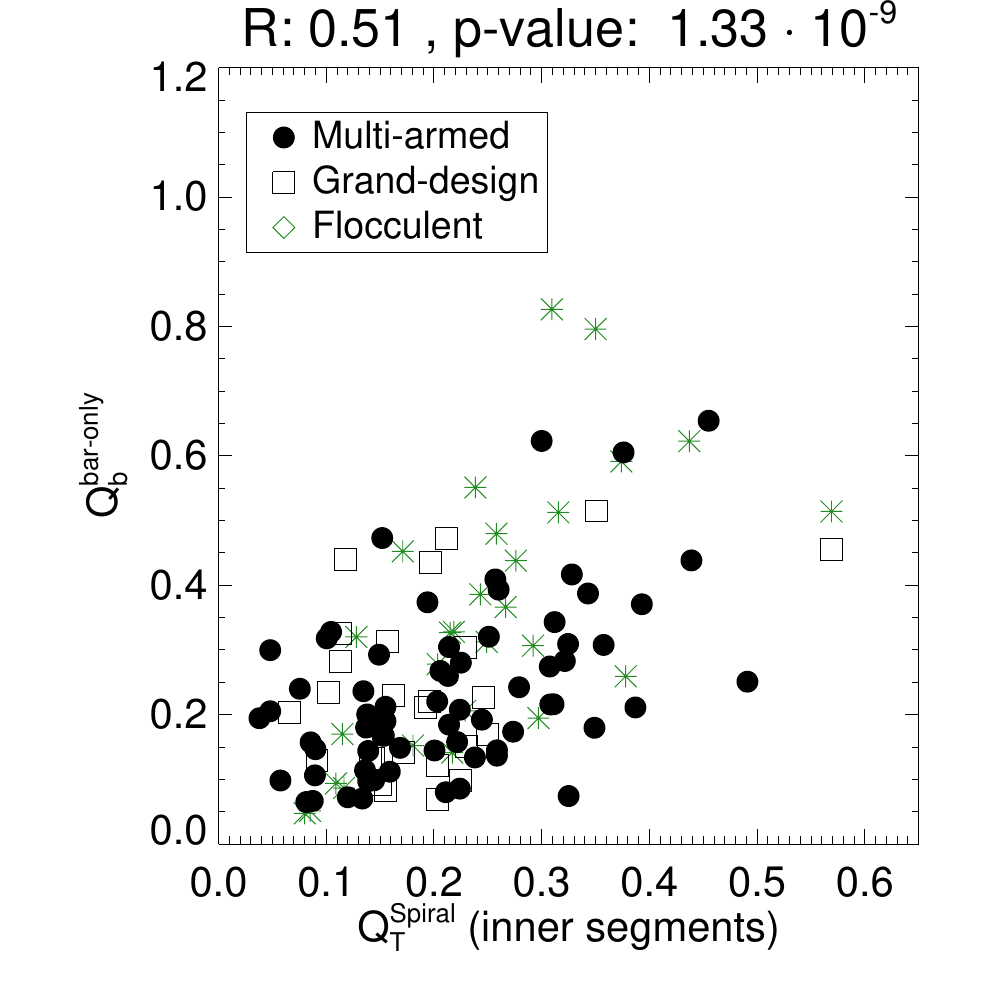}
\includegraphics[width=0.32\textwidth]{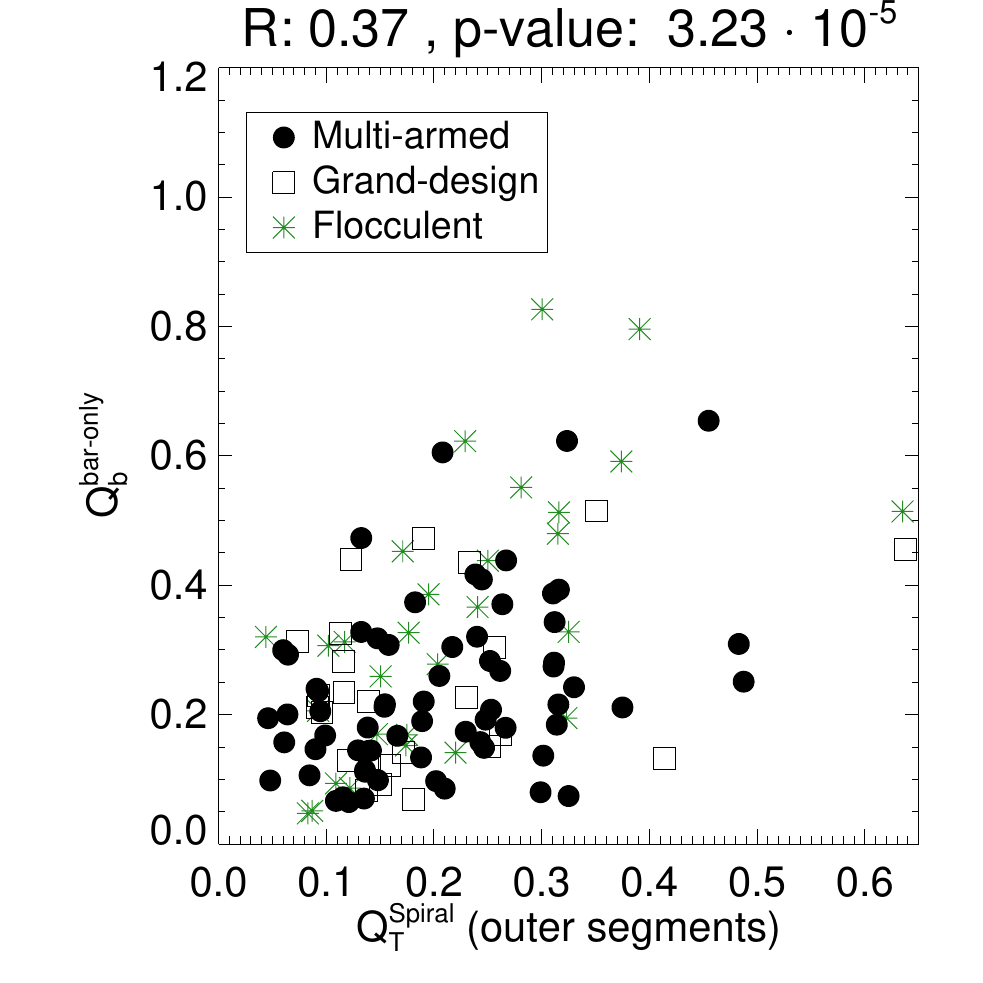}
\caption{
As in Fig.~\ref{pitch_vs_spiral_strength_inner_outer}, 
but eliminating the spiral contribution to the bar gravitational torque.
}
\label{pitch_vs_spiral_strength_inner_outer_BAR_ONLY}
\end{center}
\end{figure}

In Fig.~\ref{pitch_vs_spiral_strength_inner_outer_BAR_ONLY} we confirm the same trends when using 
bar-only forces (see Sect.~\ref{bar_strength_methods}). 
Interestingly, the correlation between the amplitude of the bar and 
that of the spiral modes holds in these extreme radial ranges.
%
%
%
%
%
%
\clearpage
%
%
\section{Pitch angle as a function of the relative mass of cold gas}\label{gas_pitch}
\begin{figure}
\centering
\includegraphics[width=0.49\textwidth]{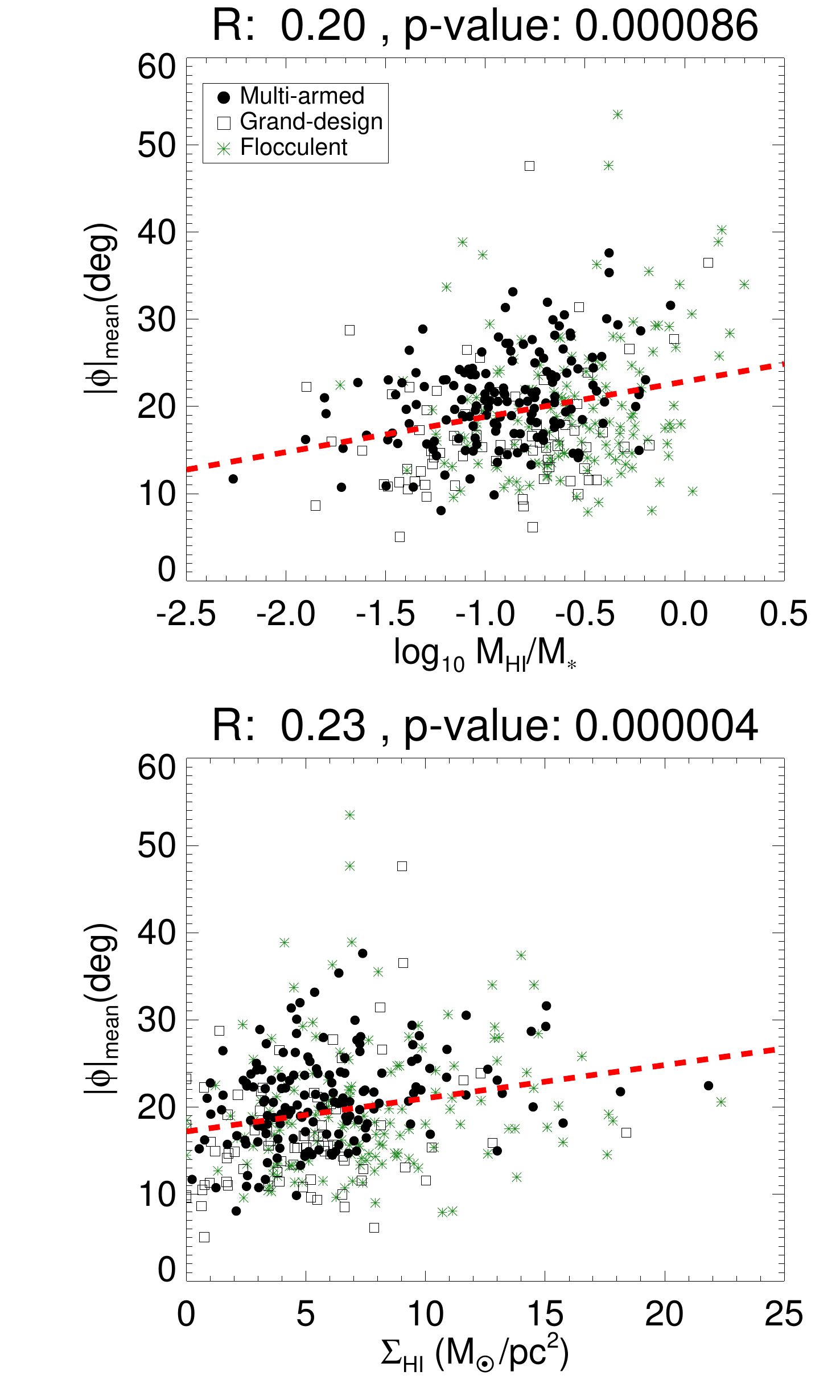}
\includegraphics[width=0.49\textwidth]{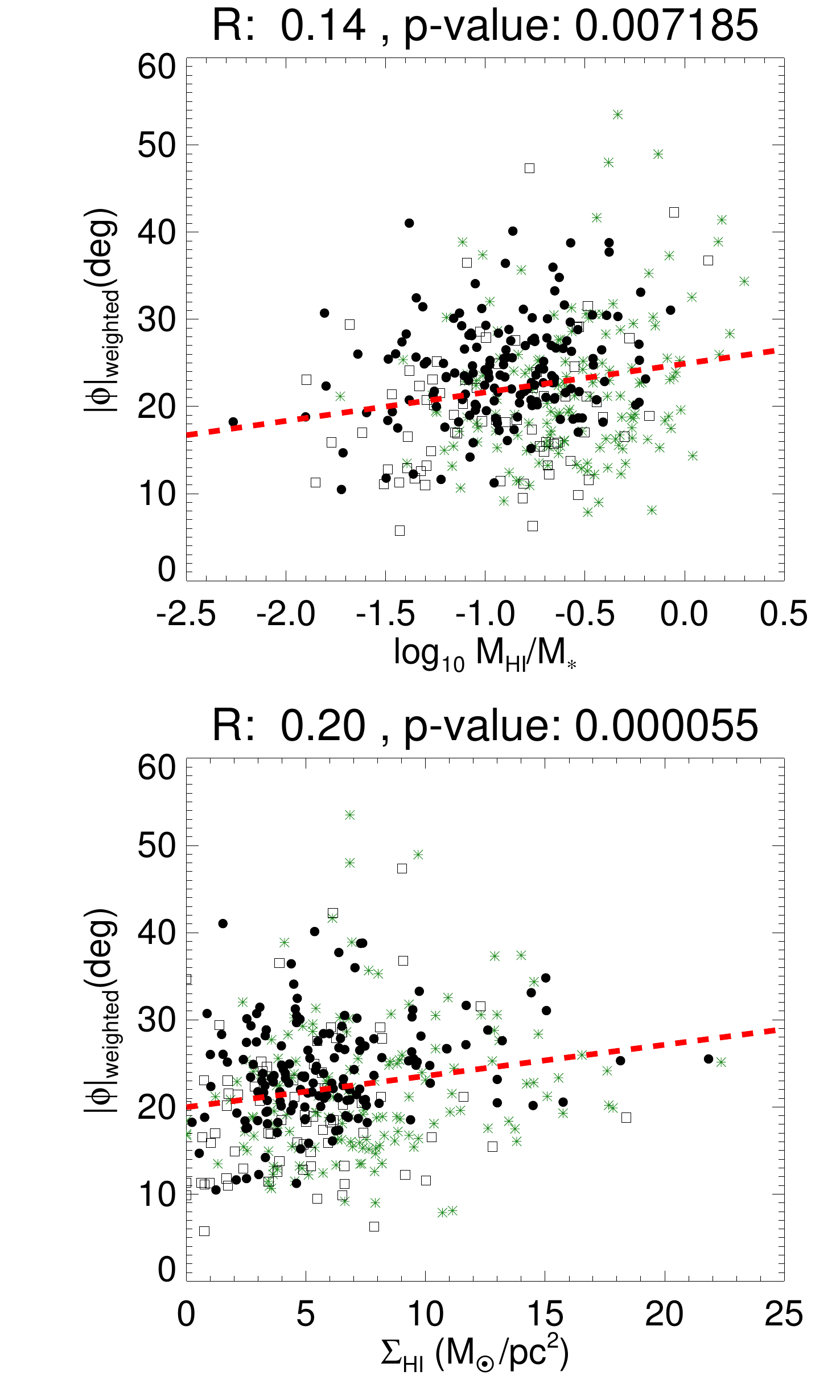}
\caption{
Pitch angle versus H{\sc\,i} gas fraction (\emph{upper panels}) and H{\sc\,i} surface density (\emph{lower panels}) 
for S$^4$G galaxies with inclination $< 65 ^{\circ}$, 
including barred and non-barred and all types of spirals (multi-armed, flocculent, and grand-design). 
The Spearman's correlation coefficient and significance are indicated on top of the plots. 
Different spiral types are shown with different colours and symbols, as indicated in the legend. 
The red line shows a linear fit to the cloud of points.
}
\label{Fig_gas}
\end{figure}

%
%
Here we assess a possible dependence between $|\phi|$ and the relative content of gas \citep[e.g.][]{2015ApJ...802L..13D}. 
Atomic gas masses are estimated as \citep[e.g.][]{1988gera.book..522G,2018MNRAS.474.5372E}:
\begin{equation}\label{gasfrac}
M_{\rm HI}=2.356 \cdot 10^5 \cdot D^2 \cdot 10^{0.4 \cdot (17.4-m21c)},
\end{equation}
where $m21c$ is the corrected 21-cm line flux in magnitude from HyperLEDA, 
$D$ is the distance to the galaxy (in megaparsecs) adopted by \citet[][]{2015ApJS..219....3M}. 
H{\sc\,i} surface gas densities are obtained as follows:
\begin{equation}
\Sigma_{\rm HI}=M_{\rm HI}/(\pi \cdot R_{25.5}^2),
\end{equation}
%
%
where $R_{25.5}$ indicates the isophotal radii at the surface brightness 25.5~mag~arcsec$^{-2}$ in the $3.6\,\mu$m images.
%
%
The pitch angles of the spirals in the S$^4$G are weakly correlated with gas fraction or density (Fig.~\ref{Fig_gas}). 
A slightly tighter trend is seen when we take only multi-armed and grand-design spiral galaxies. 
%
%
%
%
\clearpage
%
%
%
\section{Pitch angle as a function of galaxy mass}\label{gal_mass}
%
%
We test a possible dependence of the spiral pitch angle on the total stellar mass of the galaxies, 
estimated by \citet[][]{2015ApJS..219....3M}, 
and on dynamical mass ($M_{\rm dyn}$) and (dark) halo-to-stellar mass ratio ($M_{\rm halo}/M_{\ast}$). 
\citet[][]{2016A&A...587A.160D} derived a first-order estimate of $M_{\rm halo}/M_{\ast}$ 
within the optical radius ($R_{\rm opt} \sim 3.2 \cdot h_{\rm R}$), which we use here\footnote{
Under the assumption that the halo within the optical disc 
contributes approximately a constant fraction of the total halo mass ($\sim 4\%$), 
\citet[][]{2016A&A...587A.160D,2016arXiv161101844D} found that the trend of the $M_{\rm halo}/M_{\ast}$ versus 
$M_{\ast}$ relation agreed with the prediction of $\Lambda$CDM models, fitted at $z\approx 0$ 
based on abundance matching and halo occupation distribution methods \citep[e.g.][]{2010ApJ...710..903M}.
}:
\begin{equation}\label{dmeq}
M_{\rm halo}/M_{\ast}(<R_{\rm opt})\approx 1.34 \cdot\bigg(\frac{(V_{\rm HI}^{\rm max})^{2}}{V_{3.6 \mu \rm m}^{2}(R_{\rm opt})}-1\bigg),
\end{equation}
where $V_{\rm HI}^{\rm max}$ refers to the inclination-corrected H{\sc\,i} velocity amplitude from HyperLEDA 
and $V_{3.6 \mu \rm m}$ corresponds to the stellar component of the circular velocity (inferred from 3.6~$\mu$m imaging). 
We also estimate the dynamical mass within $R_{\rm opt}$ (assuming a spherical mass distribution):
\begin{equation}
M_{\rm dyn}(<R_{\rm opt})\approx \frac{(V_{\rm HI}^{\rm max})^{2} \cdot R_{\rm opt}}{G},
\label{m_dyn}
\end{equation}
where $G$ is the Newtonian constant of gravitation. 

\begin{figure*}
\centering
\includegraphics[width=0.33\textwidth]{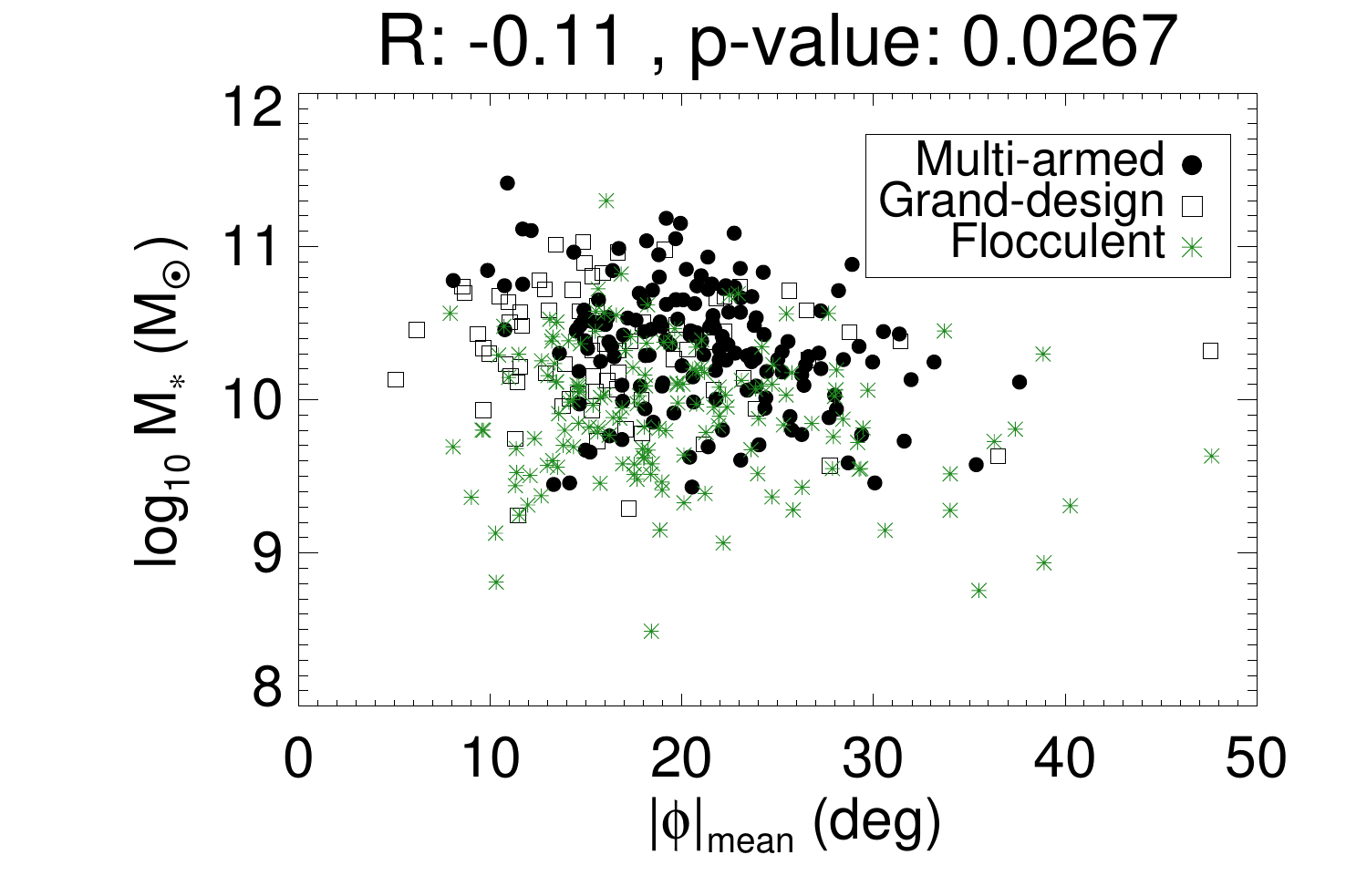}
\includegraphics[width=0.33\textwidth]{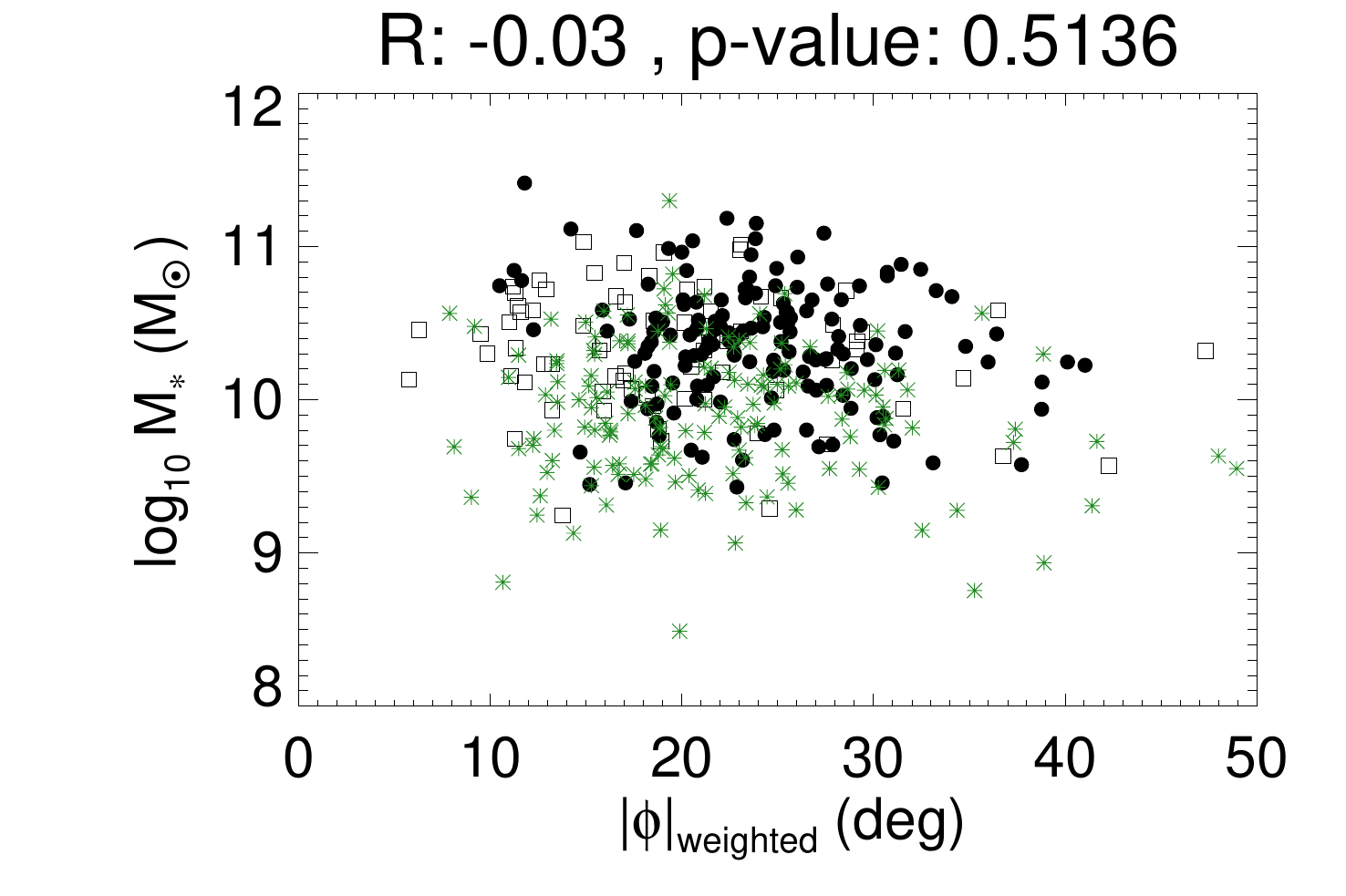}
\includegraphics[width=0.33\textwidth]{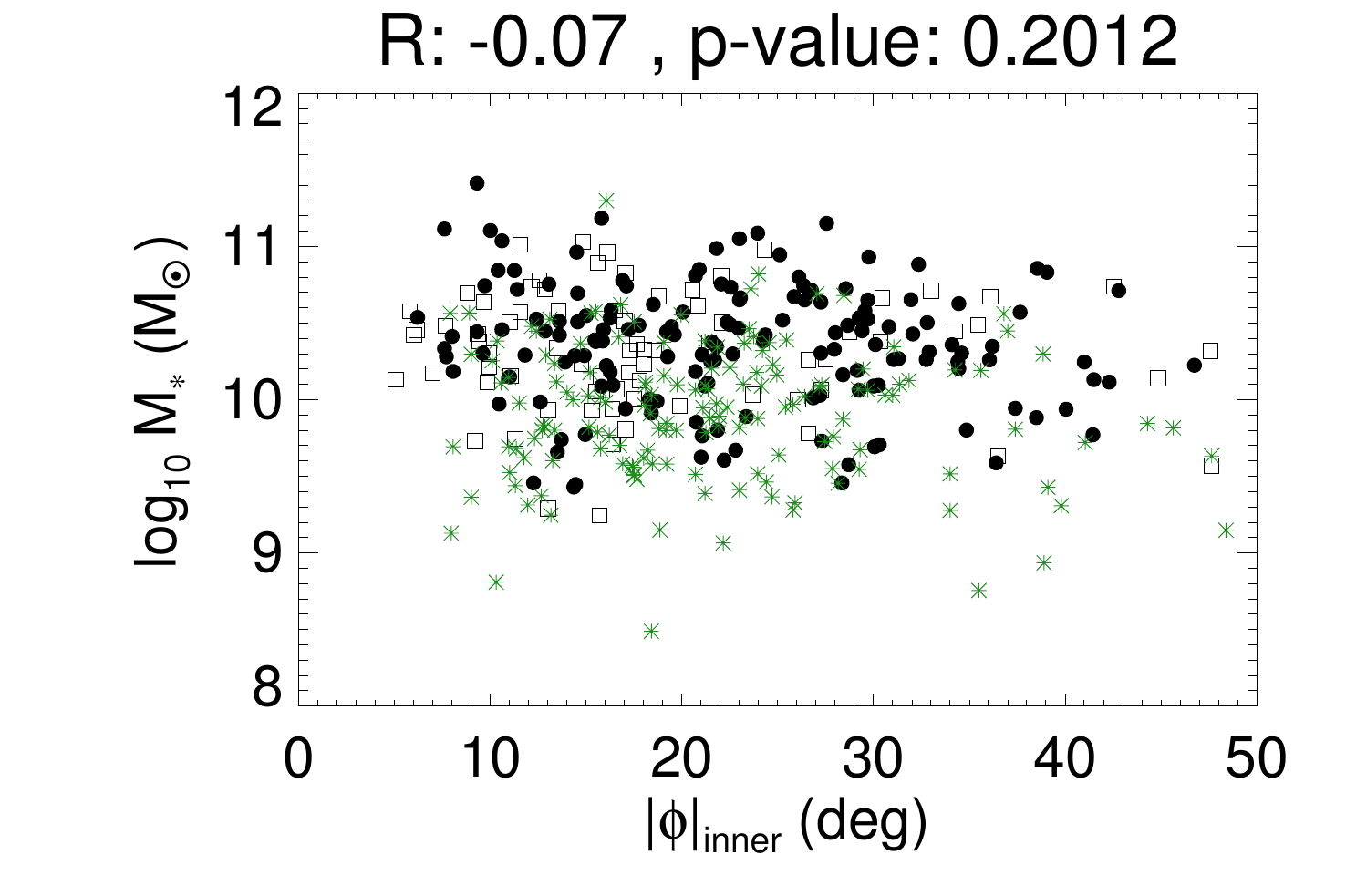}\\
\includegraphics[width=0.33\textwidth]{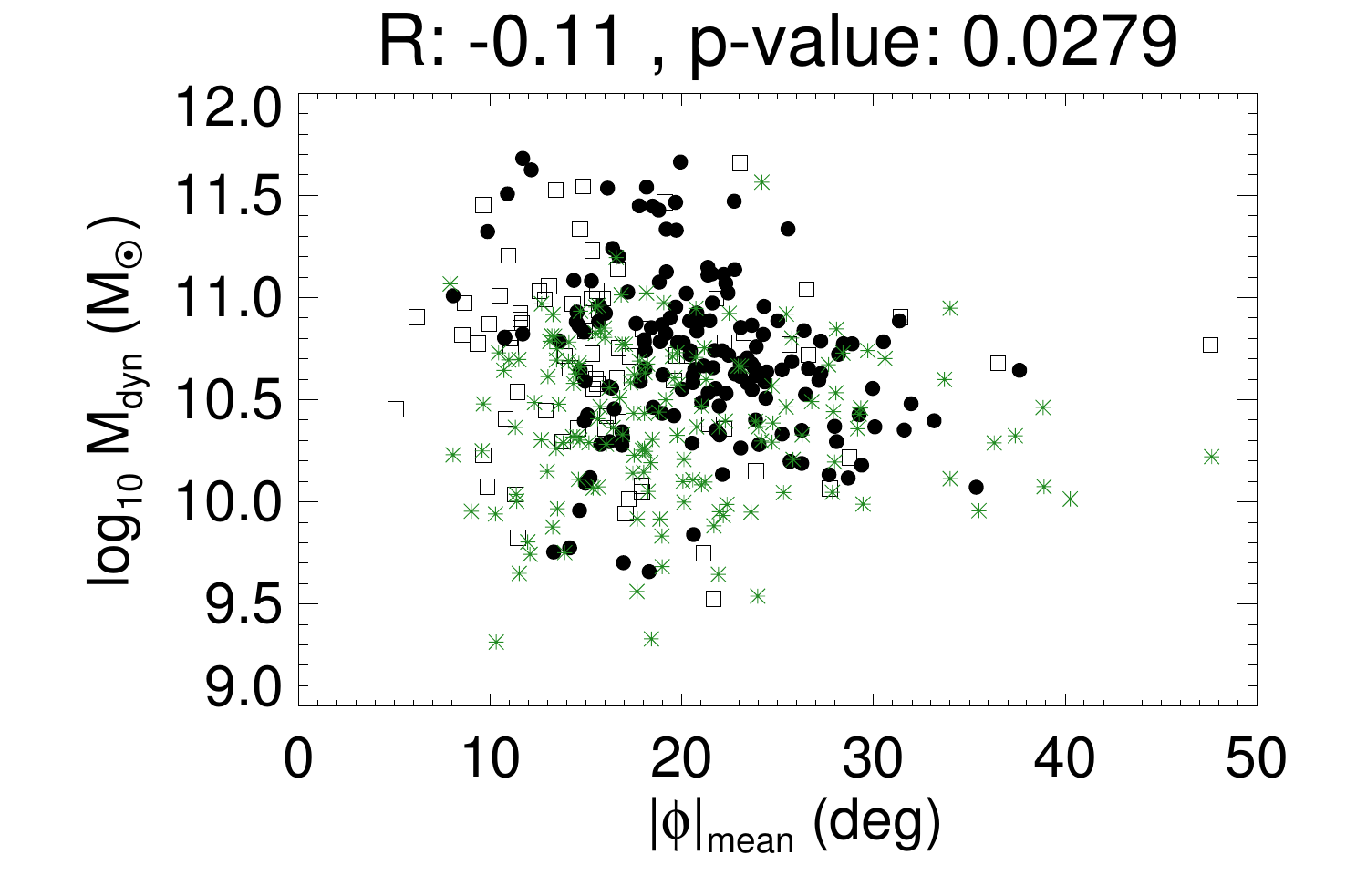}
\includegraphics[width=0.33\textwidth]{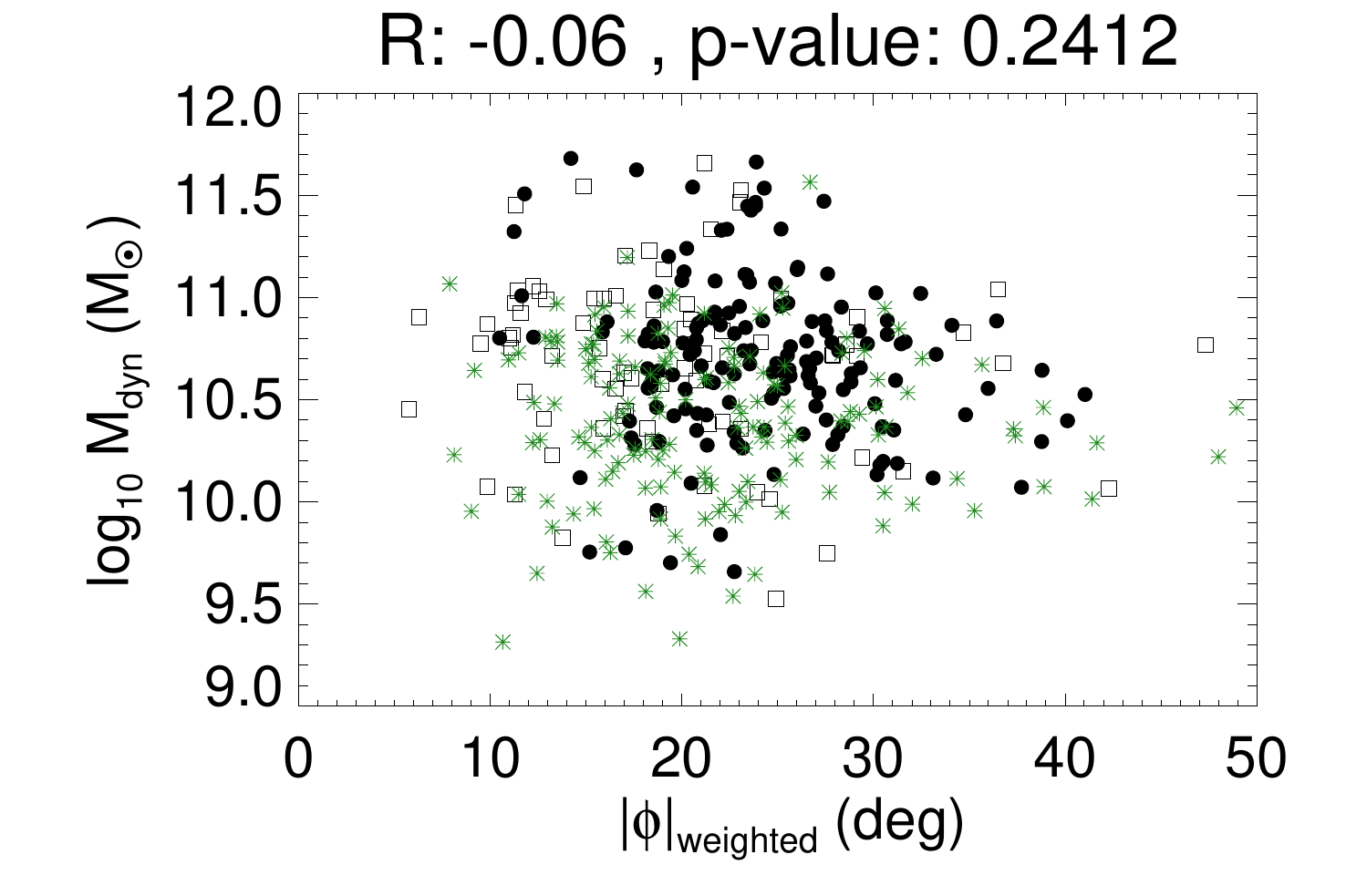}
\includegraphics[width=0.33\textwidth]{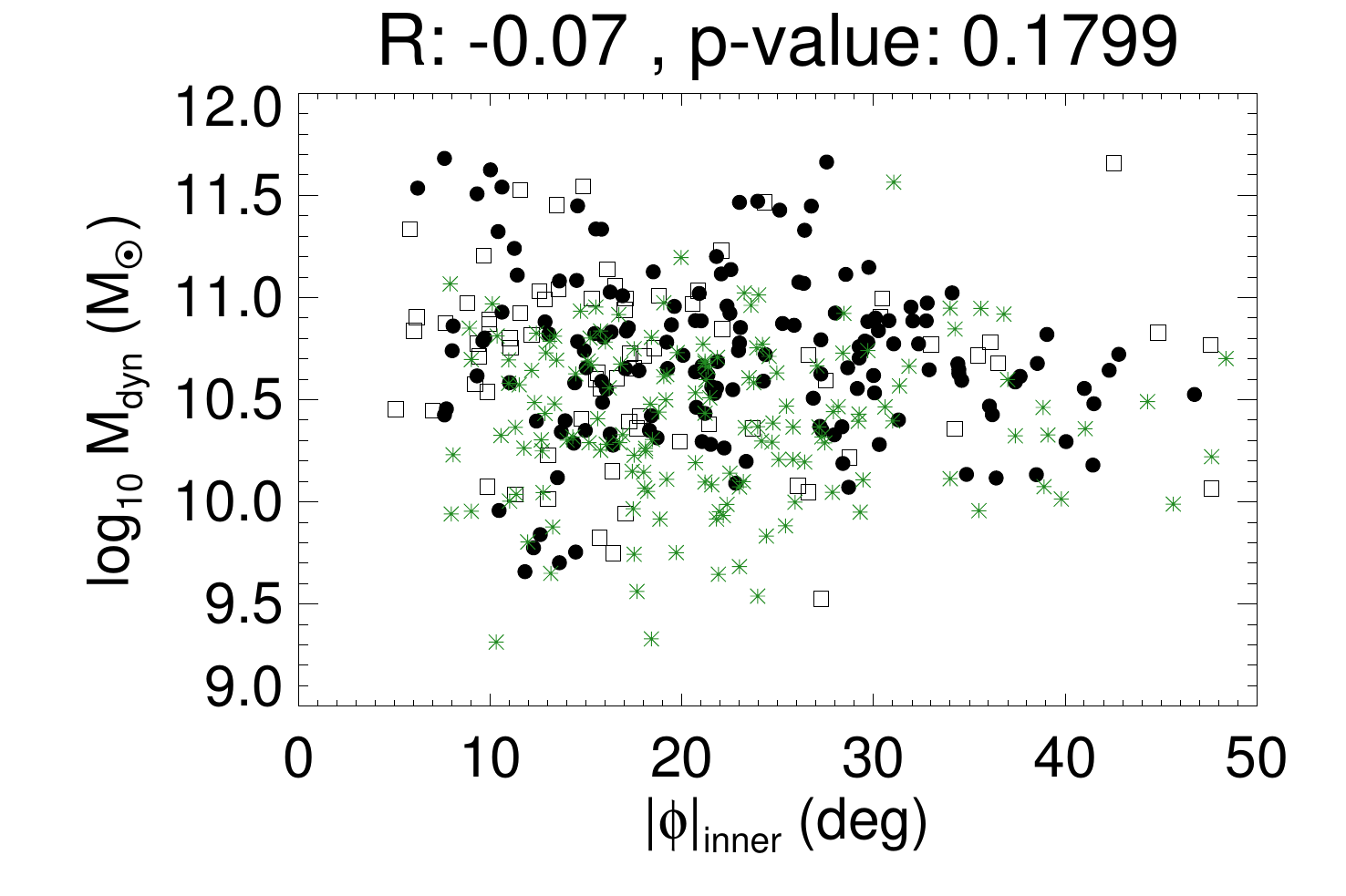}\\
\includegraphics[width=0.33\textwidth]{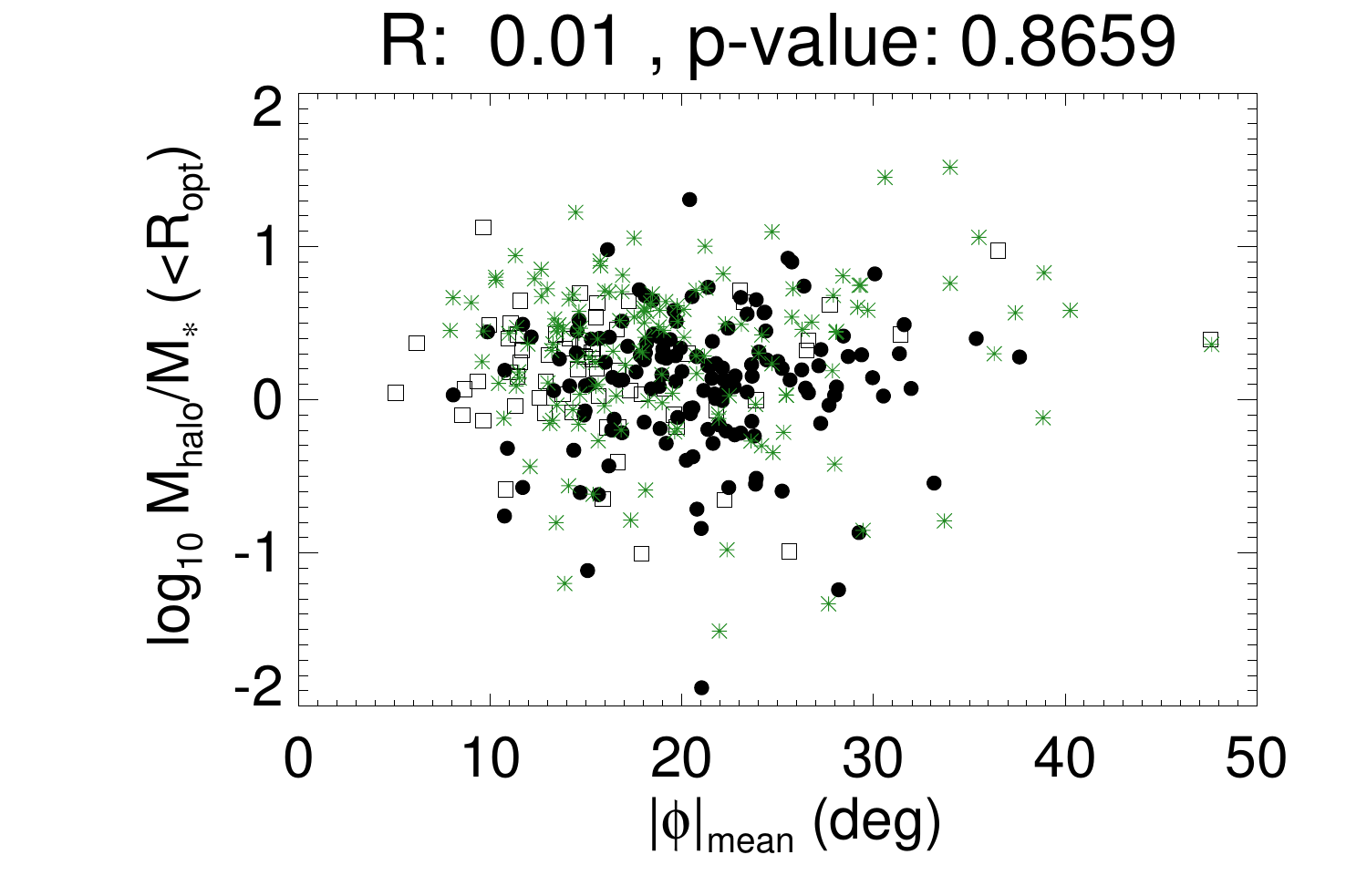}
\includegraphics[width=0.33\textwidth]{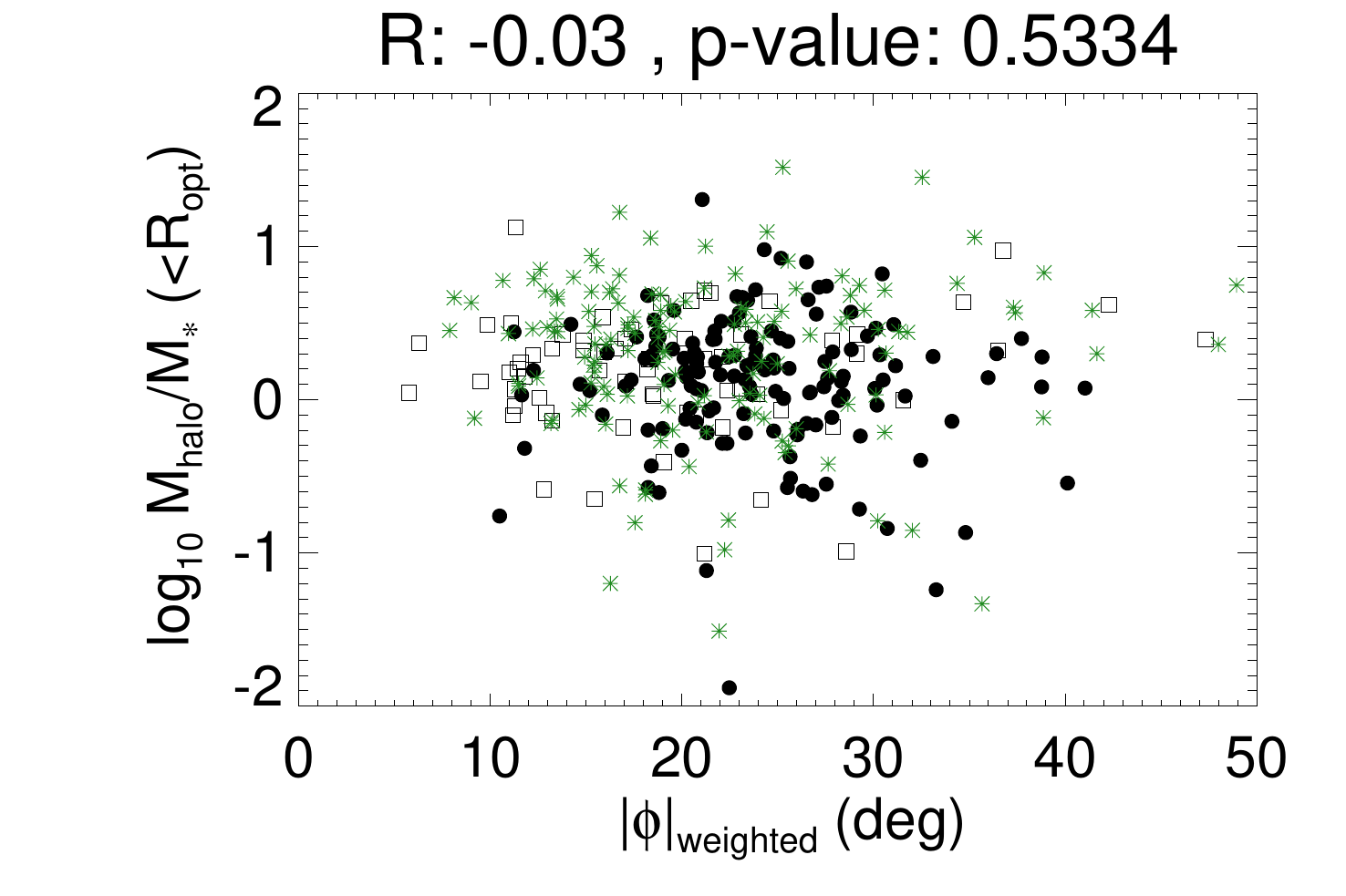}
\includegraphics[width=0.33\textwidth]{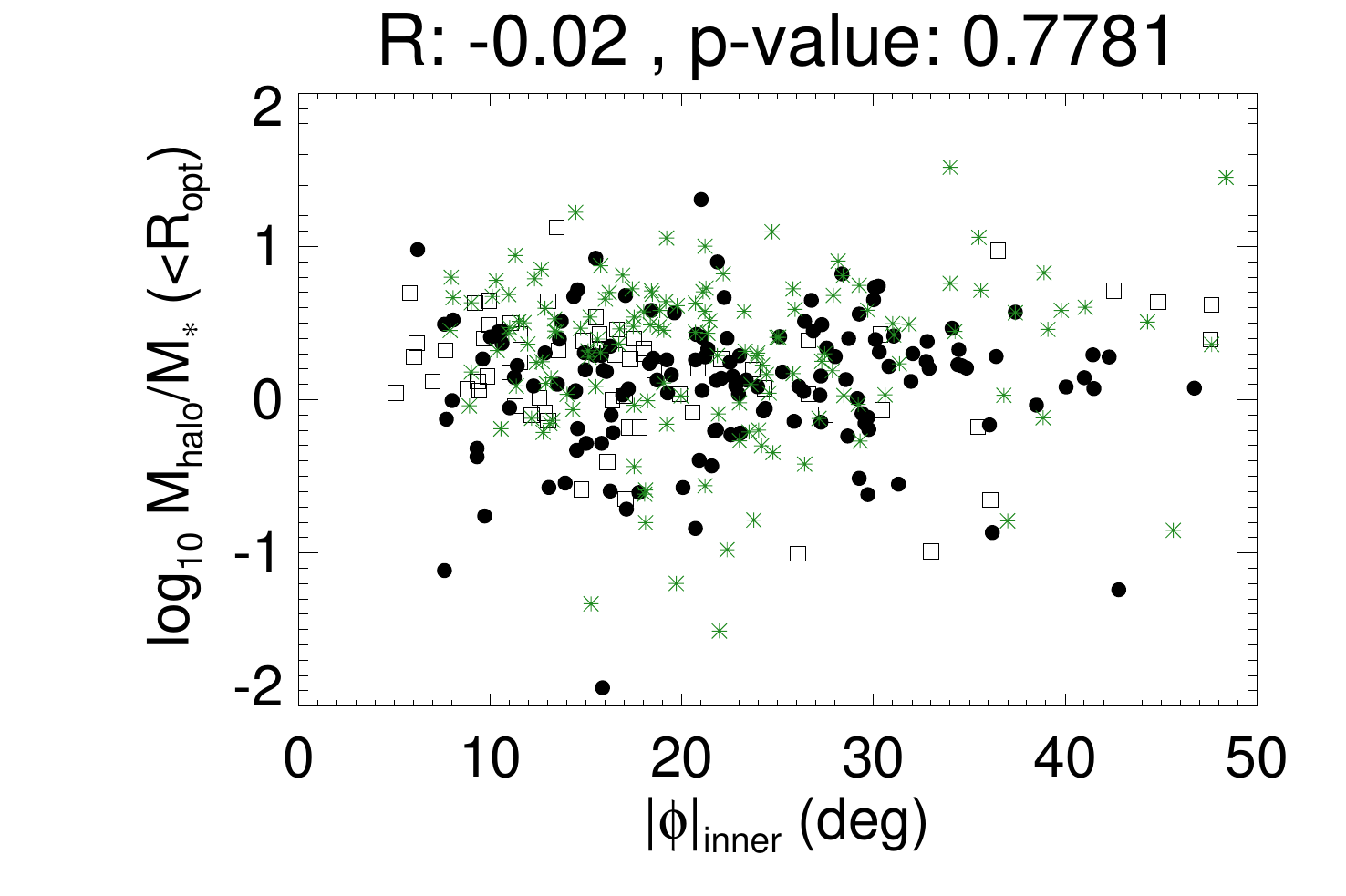}\\
\caption{
Galaxy mass versus pitch angle. We plot the total stellar mass (\emph{upper row}), dynamical mass (\emph{central row}), 
and halo-to-stellar mass ratio (\emph{lower row}) 
versus the mean pitch angle (\emph{left column}), 
mean weighted by the arc length of the arms (\emph{central column}), and mean of the innermost logarithmic segments 
(\emph{right column}). Different colours and symbols represent different types of spirals, as indicated in the legend of the upper left panel 
(as in Fig~\ref{spiral_bar_strength}).
}
\label{galaxy_mass_fig}
\end{figure*}

No dependence of the global spiral pitch angle on total stellar mass, dynamical mass, 
or halo-to-stellar mass ratio is found in the S$^4$G (Fig.~\ref{galaxy_mass_fig}). 
This questions the effect of the global mass distribution in determining the winding angle of the arms 
\citep[e.g.][]{1981AJ.....86.1847K,1982ApJ...253..101K,1998MNRAS.299..685S,2006ApJ...645.1012S,2013A&A...553A..77G}. 
%
%
%
%
\clearpage
%
%
%
\section{Strength versus pitch angle of spiral segments}\label{spirals_all}
\begin{figure}[h]
\begin{center}
\includegraphics[width=0.66\textwidth]{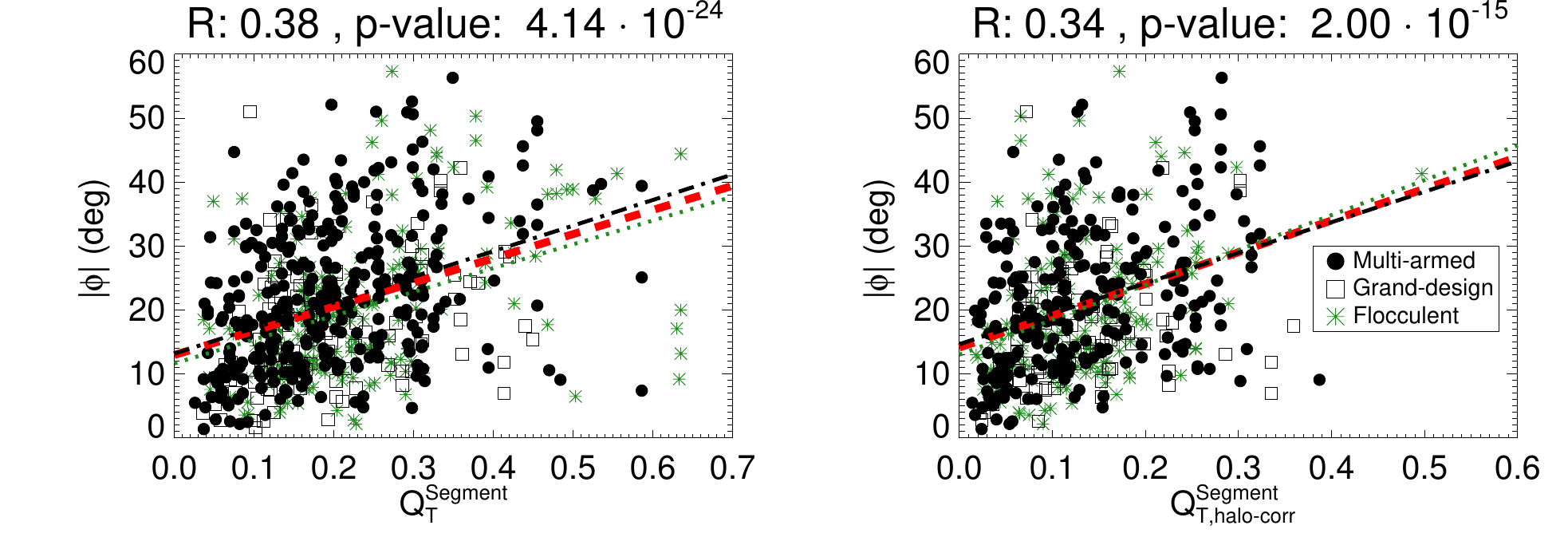}
\includegraphics[width=0.33\textwidth]{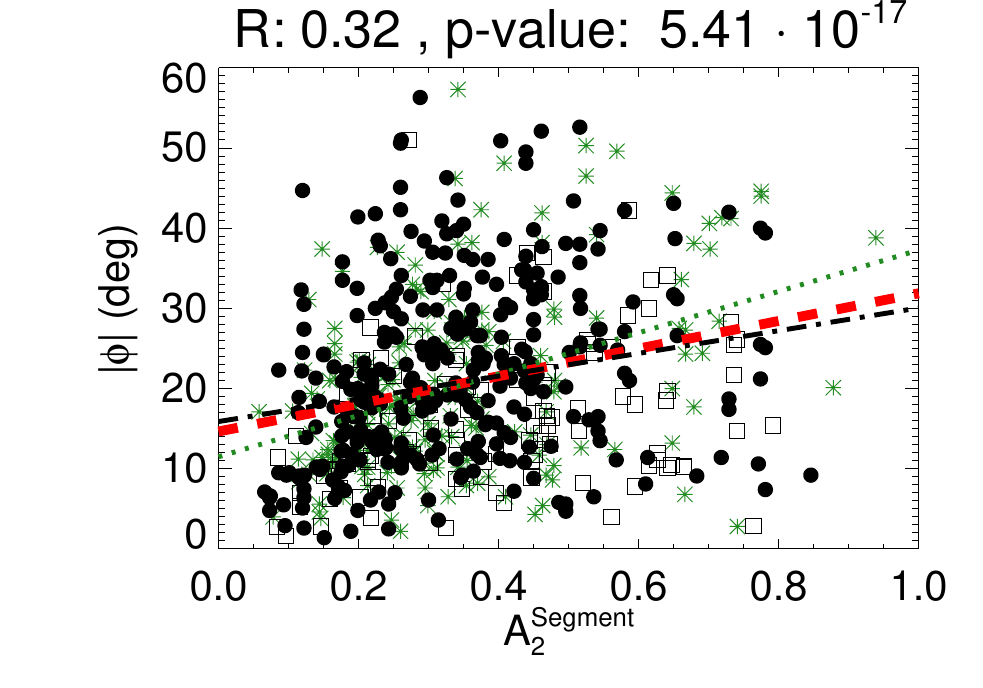}
\caption{
As in Fig.~\ref{pitch_vs_spiral_strength} but using the strength of all the spiral segments, instead of the mean, 
and showing also the $m=2$ Fourier density amplitude of the spirals (\emph{right panel}).
}
\label{pitch_vs_spiral_strength_mean_all_segments}
\end{center}
\end{figure}

%
%
In Sect.~\ref{spiral_bar_strength_sect} we showed a correlation between the mean pitch and the mean strength of the spirals 
(see Fig.~\ref{pitch_vs_spiral_strength}). 
Here we confirm this dependence when all the individual logarithmic segments (often several per galaxy) are used 
(Fig.~\ref{pitch_vs_spiral_strength_mean_all_segments}).
%
%
%
%
\clearpage
%
%
\section{Pitch angle versus shear}\label{spirals_all_shear}
%
%
We calculate the shear ($\Gamma$) from H{\sc\,i} and H$_{\alpha}$ rotation curves -- $V(r)$ -- 
from the SPARC compilation \citep[][]{2016AJ....152..157L} as follows \citep[e.g.][]{2018MNRAS.477.1451F}: 
\begin{equation}
\Gamma=-d {\rm ln} \Omega/d {\rm ln} r,
\end{equation}
where $\Omega=V/r$ is the angular velocity. 
The shear is obtained for the 17 SPARC galaxies with reliable measurements of the pitch angle in \citet[][]{2015A&A...582A..86H}. 
Specifically, $\Gamma$ was calculated in the same radial ranges where $|\phi|$ was fitted, 
and thus several measurements per galaxy are provided. We do not find any correlation between $\Gamma$ and $|\phi|$.
%
%
\begin{figure}[h]
\begin{center}
\includegraphics[width=0.99\textwidth]{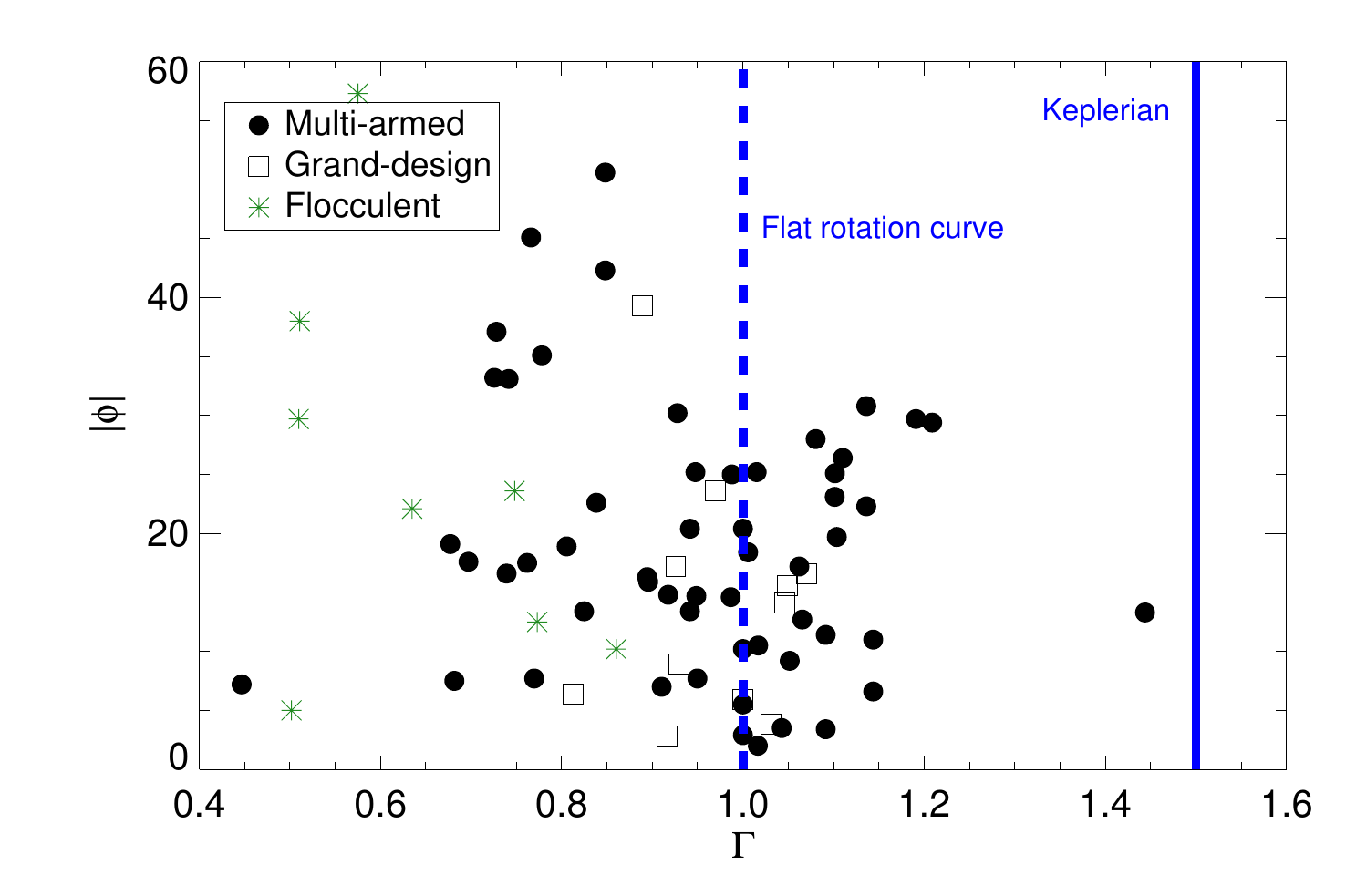}
\caption{
Pitch angle versus shear for a sample of 17 galaxies. 
Every point (72) corresponds to a logarithmic segment. 
In green we show the flocculent spirals, while black squares and filled circles 
correspond to grand-design and multi-armed galaxies, respectively (see legend). 
The dashed (solid) lines show the expected value of $\Gamma$ for a flat (Keplerian) rotation curve.
}
\label{sparc_pitch_shear}
\end{center}
\end{figure}
\end{appendix}
%
%
\end{document}